\documentstyle[preprint,aps,epsfig,amsmath,amssymb]{revtex}

 \def\lp{\left. }
 \def\rp{\right. }
 \def\lr{\left( }
 \def\rr{\right) }
 \def\le{\left[ }
 \def\re{\right] }

 \def\li{\mbox{Li}_2}

 \newcommand{\ms}{m_{\tilde{q}}}
 \newcommand{\mc}{m_{\tilde{\chi}}}
 \newcommand{\mg}{m_{\tilde{g}}}
 \newcommand{\nonu}{\nonumber \\}

\def\bea{\begin{eqnarray}}
\def\eea{\end{eqnarray}}

\def\ra{\rightarrow}

\def\chiplus{{\tilde{\chi}}^{+}}
\def\chiminus{{\tilde{\chi}}^{-}}
\def\chizero{{\tilde{\chi}}^{0}}
\def\chargino{{\tilde{\chi}}^{\pm}}
\def\neutralino{{\tilde{\chi}}^{0}}
\def\gaugino{{\tilde{\chi}}}
\def\gluino{\tilde{g}}
\def\bino{\tilde{B}}
\def\w3ino{\tilde{W_3}}
\def\h1ino{\tilde{H_1}}
\def\hino2{\tilde{H_2}}
\def\wpino{\tilde{W}^{+}}
\def\hpino{\tilde{H}^{+}}

\def\alphas{\alpha_{S}}
\def\alphash{\hat{\alpha}_{S}}
\def\MS{$\overline{\rm MS}$\,\,}

\def\u6d{u_{6 \Delta}}
\def\ud7{u_{7 \Delta}}
\def\s4d{s_{4 \Delta}}
\def\sd3{s_{3 \Delta}}
\def\delu{\Delta_{u}}
\def\delt{\Delta_{t}}

\newcommand{\msqu}[1]{m_{\tilde{q}_{#1}}}
\newcommand{\ih}[2]{\,\hat{I}\left( \frac{#1}{#2} \right)}

\begin{document}
\draft
\preprint{
\vbox{
\halign{&##\hfil\cr
	& hep-ph/0005196 \cr
	& ANL-HEP-PR-00-045 \cr
	& DESY 00-069 \cr}}
}
\title{Next-to-Leading Order SUSY-QCD Predictions \\ 
	for Associated Production of Gauginos and Gluinos}
\author{Edmond~L.~Berger$^a$, Michael~Klasen$^b$, and Tim~M.~P.~Tait$^a$}
\address{$^a$High Energy Physics Division, Argonne National Laboratory \\
             Argonne, Illinois 60439 \\
         $^b$II.~Institut f\"ur Theoretische Physik, Universit\"at Hamburg \\
             Hamburg, Germany}
\date{\today}
\maketitle

\begin{abstract}
We present complete results of a next-to-leading order 
calculation of the production of gaugino-like charginos ($\chargino$) and 
neutralinos ($\neutralino$) in association with gluinos ($\gluino$) at 
hadron colliders, including the strong corrections from the exchange of 
colored particles and sparticles.  Adopting a variety of models for 
the sparticle mass spectrum, including typical supergravity (SUGRA) models 
and a light gluino model, we provide predictions for total and differential 
cross sections at the energies of the Fermilab Tevatron and CERN Large Hadron 
Collider (LHC).

\end{abstract} 
\vspace{0.2in}
\section{Introduction}
\label{intro}

Weak scale supersymmetry (SUSY) \cite{susy} is a theoretically attractive
extension of the Standard Model of particle physics.  Supersymmetric 
theories can solve the Higgs hierarchy puzzle \cite{hierarchy},
break electroweak symmetry radiatively at low energies
\cite{radbreak}, and explain the unification of the gauge
couplings at a high energy scale \cite{unif}.  SUSY introduces
a superpartner for each Standard Model particle with the same
quantum numbers, but for a difference in spin of 1/2.  If supersymmetry 
were exact, these superparticles would be degenerate in mass
with their Standard Model partners.  However, SUSY can be broken
softly in such a way that its attractive features survive,
while the superpartners become heavy enough to evade current limits from
collider searches \cite{searches}.  A powerful and general parametrization 
for the soft SUSY-breaking terms is provided by the Minimal Supersymmetric 
Standard Model (MSSM) \cite{mssm}. Soft breaking mechanisms require that 
the superpartners remain lighter than a few TeV, and thus there is reason to
expect that high energy investigations, such as those at LEP II, Run II of 
the Fermilab Tevatron, and the CERN Large Hadron Collider (LHC) will 
discover supersymmetric particles or place strong constraints on 
supersymmetric models.

The search for direct experimental evidence of supersymmetry at 
colliders requires a good understanding of 
theoretical predictions of the cross sections for production of the
superparticles.  In the case of hadron colliders,
where collisions of strongly interacting hadrons are studied,
the large strong coupling strength ($\alphas$) leads to potentially
large contributions beyond the leading order (LO) in a perturbation
series expansion of the cross section.  To have accurate
theoretical estimates of production rates and differential 
cross sections, it is necessary to 
include corrections at next-to-leading order (NLO) or beyond.  
Next-to-leading order calculations of the hadroproduction of gluinos 
and squarks \cite{squarkgluino}, top squarks \cite{stop}, 
sleptons \cite{slepton,gaugino}, and gauginos \cite{gaugino} have 
been published, including our brief report on associated production 
of gauginos and gluinos \cite{sletter}.  In this paper, we provide a 
detailed exposition of our calculation of associated production, and 
we present new predictions of total and differential cross sections 
for a variety of assumptions about the superpartner mass spectrum.  

Associated production of a gaugino ($\gaugino$) with a gluino 
($\gluino$) or with a squark ($\tilde{q}$) is potentially a very 
important production mechanism.  Associated production processes are 
semi-weak in that they involve one somewhat smaller coupling constant 
than the production of a pair of colored sparticles.  However, in popular 
models of SUSY breaking~\cite{sugra,susybreak}, the mass spectrum favors much 
lighter masses for the color-neutral, low-lying neutralinos and charginos 
than for the colored squarks and gluinos.  Their lighter mass means that 
the phase space for production of neutralinos and charginos, and the 
relevant partonic 
luminosities, will be greater than that for gluinos and squarks.  This 
effect is potentially decisive at a collider with limited energy, such as 
the Tevatron.  Indeed, as our numerical results show, the extra phase space 
may more than offset the smaller coupling, and gauginos may be produced more 
copiously than squarks at the Tevatron.  Another point in favor of associated 
production is the relative simplicity of the final state.  For example, the 
lowest lying neutralino is the (stable) lightest supersymmetric 
particle (LSP) in supergravity (SUGRA) models~\cite{sugra}, manifest only 
as missing energy in the events, and it is the second lightest in 
gauge-mediated models\cite{susybreak}.  The charginos and 
higher mass neutralinos may decay leptonically leaving a lepton signature plus 
missing transverse energy; relatively clean events ensue.  Furthermore, 
associated production may be the best channel for measurement of the gluino 
mass \cite{lo2}.   

In this paper we present our complete NLO calculation in SUSY-QCD of the 
hadroproduction of a gluino in association with a gaugino, including 
contributions from virtual loops of colored sparticles and particles, and 
three-particle final states in which light quarks or gluons are emitted.
We extract the ultraviolet, infrared, and collinear divergences by use of 
dimensional 
regularization and illustrate how they may be absorbed by the usual 
renormalization and mass factorization procedures. In computing the 
virtual contributions, we encountered divergent four-point functions 
that had not been evaluated previously.  We use a combined analytic and 
numerical phase space slicing method to treat the contributions from real 
emission of light particles.  Associated production was calculated at LO 
some years ago \cite{lo2,lo1}. Our reason to focus first on the $\gaugino$ 
plus $\gluino$ final state, rather than the $\gaugino$ plus $\tilde{q}$ 
final state, is that the LO cross sections for $\gaugino + \gluino$ are 
3 to 6 times greater than those for $\gaugino + \tilde{q}$ at the energy 
of the Tevatron when the mass $m_{\gluino} = m_{\tilde{q}} = 300$ GeV, and 
6 to 15 times greater when $m_{\gluino} = m_{\tilde{q}} = 600$ GeV.  These 
comparisons are pertinent for the lighter mass neutralinos 
$\neutralino_{1,2}$ and chargino $\chargino_1$.  In obtaining 
the $\tilde{q}$ cross sections, we sum over five flavors of degenerate 
squarks and antisquarks.  

Our analysis is general in that it is not tied to a particular 
SUSY breaking model.  We can provide cross sections for arbitrary 
gluino and gaugino masses, and, indeed, the values of the cross sections at 
Tevatron and LHC energies depend crucially on the sparticle masses.  Mass 
generation in supersymmetry is accomplished in a hidden sector and transmitted 
to the MSSM fields.  In SUGRA models~\cite{sugra}, transmission is through 
gravitational interactions, while in gauge-mediated~\cite{susybreak} and 
gaugino-mediated~\cite{gaugino1,gaugino2,gaugino3} models, it occurs 
through gauge interactions.  Anomaly mediated SUSY breaking, 
also gravitational in 
origin, is a fourth possibility \cite{amsb}.  In the presentation of 
predictions for cross sections, we consider illustrative mass spectra 
typical of each scenario and consistent with bounds established from current 
data~\cite{Abel:2000,lepdata}.  We also examine the 
phenomenologically open case of a gluino with mass light compared to 
the SUSY scale, $m_{\tilde g} \simeq 30$ GeV.  This possibility arises in some 
models of gauge-mediated SUSY breaking \cite{raby,raby2}.  

As is shown in detail below, the NLO corrections to associated production 
are generally positive, but they can be modest in size, ranging in the SUGRA 
model from a few percent at the energy of the Fermilab Tevatron to 100\% at 
the energy of the LHC, depending on the sparticle masses.  In the light-gluino 
case, NLO contributions increase the cross section by factors of 1.3 to 1.4 at 
the energy of the Tevatron and by factors of 2 to 3.5 at the energy of the 
LHC.  Owing to these enhancements, collider searches for signatures of associated 
production will generally discover or exclude sparticles with 
masses larger than one would estimate based on LO production rates alone.  
More significant from the viewpoint of reliability, the renormalization and 
factorization scale dependence of the cross sections is reduced by a 
factor of more than two when NLO contributions are included.  

At Run II of the Fermilab Tevatron, for an integrated luminosity of 
2 $\rm{fb}^{-1}$, we expect that 10 or more events could be produced in 
each of the lighter gaugino channels of the SUGRA model, 
$\gluino \neutralino_1$, 
$\gluino \neutralino_2$, and $\gluino \chargino_1$, provided that the 
gluino mass $m_{\gluino}$ is less than 450 GeV.  The cross sections 
for the three heavier gaugino channels, $\gluino \neutralino_3$, 
$\gluino \neutralino_4$, and $\gluino \chargino_2$, are smaller by 
an order of magnitude or more than those of the lighter gaugino 
channels.  In the light gluino model that we consider, more than 100
events could be produced in the three lighter gaugino channels provided 
that the common GUT-scale fermion mass $m_{1/2}$ is less than 400 GeV, 
and as many as 10 events in the three heavier gaugino channels as long 
as $m_{1/2}$ is less than 200 GeV.  At the higher energy and luminosity 
of the LHC, at least a few events should be produced in every channel in 
the SUGRA model and many more in the light gluino model.    

The shapes of the rapidity distributions of the gauginos are not altered 
appreciably by NLO contributions, but the locations of the maximum cross 
section in transverse momentum ($p_T$) are shifted to smaller values 
by NLO contributions.  At LHC energies where the contribution of the 
$qg$ initial state is important, modifications of the $p_T$ spectra 
can be pronounced.  

We begin in Sec.~\ref{leading} with a brief review of the LO calculation in 
order to introduce our notation.  In this section, we also introduce the SUSY 
breaking models we adopt and summarize salient aspects of their predicted mass 
spectra.  This discussion is followed in Sec.~\ref{nlocorr} by a detailed 
presentation of our NLO ${\cal O}(\alpha\alphas^2)$ calculation.  We present 
partonic scaling functions in Sec.~\ref{results} as well as predictions for 
inclusive and differential cross sections at Tevatron and LHC energies.  Our 
conclusions may be found in Sec.~\ref{conclusions}.  In 
Appendices~\ref{virtapp}~--~\ref{qapp}, we present detailed analytic results
related to the NLO calculation.

\section{Leading Order Production of Gauginos and Gluinos}
\label{leading}

We begin with the Born level cross sections for the partonic processes 
\bea
 q\, \bar{q} \ra \gluino\, \neutralino_k &,& q\, \bar{q} \ra 
 \gluino\, \chargino_k ,
\eea
derived first in \cite{lo2,lo1}.  In anticipation of the renormalization
and mass factorization of the NLO contributions, we proceed in the 
$n = 4 - 2 \epsilon$ dimensions of standard dimensional regularization.  
We assume that there is no mixing between squarks of
different generations and that the squark mass eigenstates are aligned with
the squark chirality states, equivalent to the assumption that
the two squarks of a given flavor are degenerate in mass.  We ignore
the $n_f = 5$ light quark masses in all of the kinematics and couplings,
and thus study the production of gaugino-like charginos and neutralinos,
but not the production of Higgsino-like ones \cite{foot1}. We assume 
further that the entries in the chargino and neutralino mass matrices are 
real, and thus that the unitary transformations from the 
($\bino$, $\w3ino$, $\h1ino$, $\hino2$) and ($\wpino$, $\hpino$) bases
to the ($\neutralino_1$, $\neutralino_2$, $\neutralino_3$,
$\neutralino_4$) and ($\chiplus_1$, $\chiplus_2$) bases are given by
orthogonal matrices.  A result of this convention is that it is possible
for the mass of one or more neutralinos to be negative inside a 
polarization sum.  The chargino masses are chosen to be positive as 
may be done for Dirac fermions.  

The Dirac matrix $\gamma_5$ (or equivalently the projectors
$P_{L (R)} = (1 \pm \gamma_5) / 2$)
appearing in the
gaugino and gluino couplings is treated in the `naive' scheme in which
it anti-commutes with all of the other $\gamma_{\mu}$ matrices.  This
scheme is acceptable at the one loop level for calculations free from
anomaly.  In evaluating the Feynman diagrams involving 
Majorana and explicitly charge-conjugated fermions, we have followed 
the approach described in Ref.~\cite{denner}.

We express our leading order results in terms of the Mandelstam
variables 
\begin{alignat}{3}
  s & = (p_a +p_b)^2 \\
  t & = (p_b -p_2)^2 &\qquad t_1 & = (p_b -p_2)^2 - {m_1}^2
  &\qquad  t_2 &= (p_b -p_2)^2 - {m_2}^2 \notag\\
  u & = (p_a - p_2)^2 &\qquad u_1 & = (p_a -p_2)^2 - {m_1}^2
  &\qquad  u_2 &= (p_a -p_2)^2 - {m_2}^2, \notag
\end{alignat}
where $p_a$, $p_b$, $p_1$, and $p_2$ refer to the four-momenta of the
incoming quark, the incoming anti-quark, the produced gluino, and the
produced gaugino, respectively.  Variable $m_1$ denotes the mass of the 
gluino and $m_2$ that of the gaugino.  The incoming
partons are treated as massless.  The momenta are on mass-shell,
$p_a^2 = p_b^2 = 0$, $p_1^2 = m_1^2$, and $p_2^2 = m_2^2$.
The invariants obey the relation $s + t + u = m_1^2 + m_2^2$.

After the $n$-dimensional phase space integration we obtain the lowest
order partonic differential cross section,
\begin{eqnarray}
   \frac{d^2\hat{\sigma}^B_{ij}}{dt_2\,du_2} &=& \frac{\pi 
    S_{\epsilon}}{s^2\,\Gamma (1 - \epsilon)} \left[\frac{ t_2\,u_2 - m^{2}_{2}
    s}{\mu^2\,s} \right]^{- \epsilon} \Theta(t_2\,u_2 - m^{2}_{2}\,s)
    \label{bornph} \\[1mm]
 & & \times \Theta (s - (m_1 + m_2)^2)\,
   \delta (s + t + u - m^{2}_1 - m^{2}_2)\,
   \overline{|{\cal M}^B_{ij}|}^{2} , \nonumber
\end{eqnarray}
where $S_{\epsilon} = (4 \pi)^{-2 + \epsilon}$.  The
arbitrary scale $\mu$ is introduced, as usual, to provide the correct
mass dimension for the coupling in $n$ dimensions; 
$\overline{|{\cal M}^B_{ij}|}^{2}$ is the leading order matrix element 
summed over the colors and helicities of all of the outgoing particles, 
and averaged over the colors and helicities of the incoming ones.  The 
indices $(i,j)$ label the incident partons.  For neutralino production 
at leading order, the partons are quarks and antiquarks of the same 
flavor. For chargino production, the incident quarks and antiquarks 
have different flavor.  At next-to-leading order, quark gluon initial 
states contribute also.  

\begin{figure}
\centerline{\hbox{
\psfig{figure=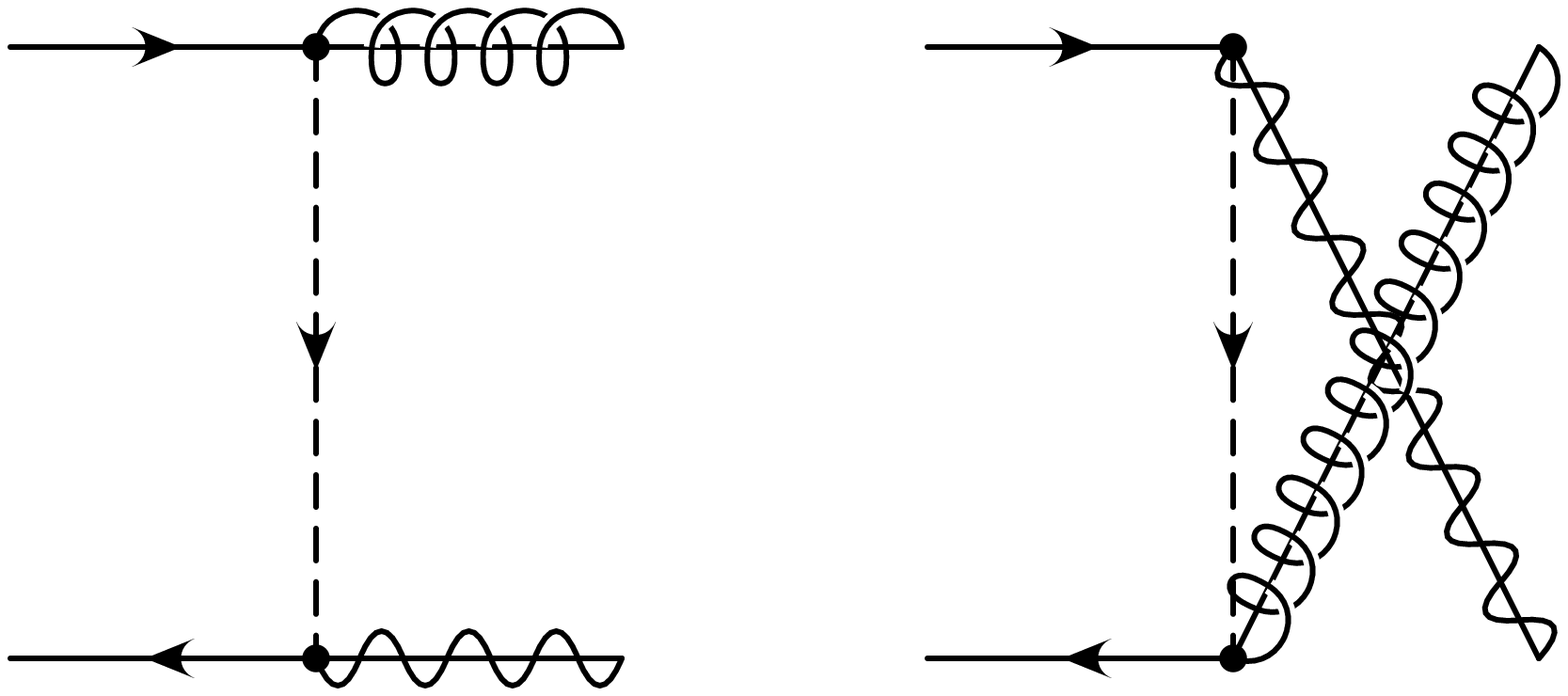,height=1.2in}}}
\vskip 0.1 in
\caption{Leading order Feynman diagrams for $q \bar{q} \ra \gluino
\gaugino$.}
\label{feyborn}
\end{figure}

As is shown in Fig.~\ref{feyborn}, the Born matrix element for
associated production of gluinos and gauginos proceeds via $t-$ or $u-$
channel exchange of a squark.  In the case of charged gauginos, only the
left-handed chiral squarks participate, whereas neutral gauginos receive
contributions from both left- and right-handed chiral squarks.  Furthermore, 
in the case of charged gaugino production, the squarks exchanged in the 
$t-$ and $u-$ channels correspond to different flavors, while in the neutral
case the $t-$ and $u-$ channel squarks have the same flavor.  Under our
assumption that the squark mass eigenstates correspond to squarks of
definite chirality, the (massless) incoming quark and anti-quark 
are forced to have a particular helicity, and thus the sets of graphs in 
which a right-handed chiral squark is exchanged cannot interfere with 
those mediated by a left-handed chiral squark.
The matrix element has the analytic form \cite{lo2} 
\begin{eqnarray}
\label{bornme}
\overline{|{\cal M}^B|}^{2} = \frac{ 8 \pi\, \alphash}{9}
\left[ \frac{X_t\, t_1\, t_2}{(t - \msqu{t}^2)^2} - 
\frac{2 \, X_{tu}\, s\, m_1\, m_2}{(t - \msqu{t}^2)(u - \msqu{u}^2)} +
\frac{X_u\, u_1\, u_2}{(u - \msqu{u}^2)^2} \right] ,
\end{eqnarray}
where $\alphash = \hat{g}^2_s / 4 \pi$ is the coupling between gluinos,
squarks, and quarks;  
$\msqu{t (u)}$ is the mass of the squark exchanged
in the $t- (u-)$ channel graph; and the $X$ represent the gaugino 
interactions with quark and squark.  

For production of a neutralino of type $\chizero_k$, 
the $X$ are \cite{chicoupling}
\bea
  X_t = X_{t u} = X_u = 2 \, \left| 
  e \, e_q \, N'_{k \, 1} +
  \frac{e}{\sin \theta_W \, \cos \theta_W}\left( \, T_q -
  e_q \, \sin^2 \theta_W \right) N'_{k \, 2} \right|^2.
\eea
In the expressions above, $e$ is
the electron charge, $\theta_W$ the weak mixing angle, $T_q$ the
third component of the weak isospin for the squark, and $e_{q}$ is the
charge of the quark in units of $e$.  For up-type quarks $e_q = 2 / 3$
and for down-type quarks $e_q = - 1 / 3$.  The matrix $N' \,$ is the
transformation from the interaction to mass eigenbasis defined in 
Ref.~\cite{chicoupling}.
The expressions for production of a positive chargino of 
type $\chiplus_k$ are
\begin{alignat}{3}
  X_t &= \frac{e^2}{\sin^2 \theta_W} |V_{k \, 1}|^2 , &\qquad 
  X_{t u} &= \frac{e^2}{\sin^2 \theta_W} \,
  \text{Re} \, (\, V_{k \, 1} \, U^*_{k \, 1}) , &\qquad 
  X_u &= \frac{e^2}{\sin^2 \theta_W} |U_{k \, 1}|^2 .
\end{alignat}
For the negative chargino $\chiminus_k$ they have the form,
\begin{alignat}{3}
  X_t &= \frac{e^2}{\sin^2 \theta_W} |U_{k \, 1}|^2 , &\qquad 
  X_{t u} &= \frac{-e^2}{\sin^2 \theta_W} \,
  \text{Re} \, (\, V^*_{k \, 1} \, U_{k \, 1}) , &\qquad 
  X_u &= \frac{e^2}{\sin^2 \theta_W} |V_{k \, 1}|^2 .
\end{alignat}
Matrices $U$ and $V$ are the chargino transformation matrices from
interaction to mass eigenstates defined in Ref.~\cite{chicoupling}.

To compute cross sections for hadroproduction,
\begin{alignat}{2}
h_a\, h_b \ra \gluino\, \neutralino_k \, X,& \qquad h_a\, h_b \ra 
\gluino\, \chargino_k \, X,
\end{alignat}
where $h_a$ and $h_b$ label the incoming hadrons, one must convolve the
partonic cross section with the parton distribution functions.
In the high energy scattering limit, one may neglect the mass of the
incoming hadrons compared with their momenta, and obtain
\begin{alignat}{3}
  S & = (P_a + P_b)^2 \\
  T & = (P_b -p_2)^2 &\qquad T_1 & = (P_b -p_2)^2 - {m_1}^2
  &\qquad  T_2 &= (P_b -p_2)^2 - {m_2}^2 \notag\\
  U & = (P_a - p_2)^2 &\qquad U_1 & = (P_a -p_2)^2 - {m_1}^2
  &\qquad  U_2 &= (P_a -p_2)^2 - {m_2}^2, \notag
\end{alignat}
in which $P_i$ indicates the momentum of hadron $h_i$.
We define $x_i$ by the relations
\begin{alignat}{3}
  s & = x_a \, x_b \, S, &\qquad t_2 & = x_b \, T_2,
  &\qquad  u_2 &= x_a \, U_2 .
\end{alignat}
The convolution with the parton distribution functions may be 
written \cite{squarkgluino} as 
\bea
  & & \frac{d^2\sigma}{dT_2 \, dU_2}(S,T,U, \mu^2_F) = 
  \label{convolution} \\
  & & \sum_{i,j= q,\bar{q}}
  \int_{x_a^-}^1 dx_a \int_{x_b^-}^1 dx_b \, x_a f_{i}^{h_a} (x_a,\mu_F^2)\,
  x_b f_{j}^{h_b} (x_b,\mu_F^2)
  \,\frac{d^2\hat{\sigma}_{ij}(s, t, u, \mu_F^2)}{dt_2\, du_2} .
  \nonumber
\eea
In this equation $\mu_F$ refers to the factorization scale, and 
$d^2\hat{\sigma}_{ij} / dt_2\, du_2$ is the hard cross section, equal 
to the Born cross section at leading order.  The 
lower limits of integration on the convolution are
\bea
x_a^- &=& \frac{-T_1}{S + U_2} ,\\
x_b^- &=& \frac{-x_a \, U_2 - m_2^2 + m_1^2}{x_a \, S + T_2} .
\eea

The differential cross section in the transverse momentum ($p_T$) 
and rapidity ($y$) of the gaugino is related to the differential
cross section in $U_2$ and $T_2$ by
\bea
\frac{d^2\sigma (S, p_T, y, \mu_F^2)}{dp_T \, dy } = 2\, p_T \, S \,
\frac{d^2\sigma (S, T, U, \mu_F^2)}{dT_2 \, dU_2},
\eea
with
\bea
 p_T^2 &=& \frac{T_2\,U_2}{S} - m_2^2=\frac{t_2\,u_2}{s} - m_2^2 ,\\
 y &=& \frac{1}{2}\log{\left(\frac{T_2}{U_2} \right)}.
\eea
The total cross
section is obtained by integrating over the full range of transverse
momentum and rapidity,
\bea
  \sigma(S,\mu_F^2) &=& 
  \int_0^{p_T^{max}(0)} dp_T \int_{-y^{max}(p_T)}^{y^{max}(p_T)} dy\, 
  \frac{d^2\sigma (S, p_T, y, \mu_F^2)}{dp_T \, dy} \\[0.2cm]
  &=& \int_{-y^{max}(0)}^{y^{max}(0)} dy \int_0^{p_T^{max}(y)} dp_T\,
  \frac{d^2\sigma (S, p_T, y, \mu_F^2)}{dp_T dy} .
\label{totalxs}
\eea
The limits of integration are 
\bea
   p_T^{max}(y) & = &
  \frac{1}{2 \, \sqrt{S}\,\text{Cosh}\,(y)}\sqrt{\left(S+m_2^2-m_1^2\right)^2
  -4 \, m_2^2 \, S \,\text{Cosh}^2 \, (y)} ,  
\eea
and 
\bea
  y^{max}(p_T) & = &
 \text{ArcCosh}\left(\frac{S+m_2^2-m_1^2}{2 \, \sqrt{S \, 
(p_T^2+m_2^2)}}\right).
\eea

\subsection{Supersymmetry Breaking Models}
\label{susybreak}

The physical gluino and gaugino masses that we use, as well as the gaugino 
mixing matrices, are based on four popular SUSY breaking models plus a fifth  
scenario in which the gluino mass is relatively light.

For our {\it default} minimal SUGRA scenario \cite{sugra}, we select 
the common scalar and fermion masses at the GUT scale to be $m_0 = 100$ GeV
and $m_{1/2} = 150$ GeV. The trilinear coupling $A_0 = 300$ GeV, and the
ratio of the Higgs vacuum expectation values, $\tan\beta = 4$. The absolute value
of the Higgs mass parameter $\mu$ is fixed by electroweak symmetry breaking,
and we choose $\mu > 0$.  (Our sign convention for $A_0$ is opposite to 
that in the ISASUGRA code \cite{isacode}.)  For this scenario, the 
neutralino masses $m_{\neutralino_{1-4}}$ are 55, 104, 283, and 309 GeV with
$m_{\neutralino_3} < 0$ inside a polarization sum. The chargino masses
$m_{\chargino_{1,2}}$ are 101 and 308 GeV and therefore almost degenerate
with the masses of $\neutralino_{2,4}$.  The gluino mass $m_{\tilde g}$ is  
410 GeV, and the squark mass is 359 GeV.  All of these masses are above 
the exclusion limits established from LEP and Tevatron 
collider data~\cite{Abel:2000,lepdata}.  Since the gluino and gaugino masses vary 
principally with $m_{1/2}$, we freeze the values of the other four parameters, 
and we vary $m_{1/2}$ over the range 100 to 400 GeV.  The squark, gluino, and 
gaugino masses all increase as $m_{1/2}$ increases.  

In considering gauge-mediated SUSY breaking (GMSB), we adopt the parameters of 
model I studied for the SUSY/Higgs Run II workshop \cite{lowscaleGMSB}, with 
$\tan\beta = 2.5$, $\mu > 0$, one messenger SU(5) generation, and a messenger 
scale $M = 2 \Lambda$.  Parameter $\Lambda$ is the scale of SUSY breaking. 
We examine six cases in which $\Lambda$ varies from 40 to 150 TeV.  
GMSB does not favor associated production at Tevatron energies because it results 
in a pattern of gaugino masses in which $M_3(M)/M_2(M) = \alpha_3(M)/\alpha_2(M)$, 
where $M_3$ and $M_2$ are the masses of the gluino and weak gaugino, and 
$\alpha_3$ and $\alpha_2$ are the SU(3) and SU(2) gauge couplings.  
Since $M$ is a low scale, $\alpha_3$ is still quite strong.  The gluino 
is generally heavy compared to the other gauginos. Selecting $\Lambda = 40$ TeV, 
$M_1, ~M_2, ~\mu =$ 56.47, 112.8, 241.7 GeV, and we obtain 
neutralino masses $m_{\neutralino_{1-4}} =$ 45, 88, 245, and 281 GeV with
$m_{\neutralino_3} < 0$ inside a polarization sum.  The chargino masses
$m_{\chargino_{1,2}}$ are 82 and 277 GeV and again almost degenerate
with the masses of $\neutralino_{2,4}$.  The gluino mass $m_{\tilde g} = 
367$ GeV, and the squark mass is 471 GeV. The spectrum of masses is similar 
to that of our default SUGRA 
scenario, and the masses grow as we increase $\Lambda$.  For comparable 
gluino and gaugino masses, we find that LO 
cross sections are roughly a factor of 5 (3) smaller at Tevatron (LHC) energies 
than in the SUGRA model, related to the larger squark mass. 

Our anomaly mediated (AMSB) scenario is based on the work in Ref.~\cite{amsb}. It 
is less well-defined in the sense that scalar masses are not understood, and thus 
the value of $\mu$ is not determined through radiative electroweak symmetry 
breaking.  However, $M_1, M_2$,~and~$M_3$ are well specified.  We fix the 
squark masses to be 350 GeV and  
choose $\tan\beta = 2$.  
The gaugino masses are controlled by the gravitino mass.  
The gluino tends to be 
heavy in this scenario, disfavoring associated production.  
However, the gluino 
mass has phase $\pi$ relative to $M_1$ and $M_2$, resulting in constructive 
interference at LO in the production of $\neutralino_1$, $\neutralino_2$, 
and $\neutralino_4$, and negative interference in production of
$\neutralino_3$, in contrast to the SUGRA scenario.  In AMSB the
lightest  neutralino is always a $\tilde{W}$ and has a large 
coupling to (s)quarks,
in contrast to the $\tilde{B}$-like lightest neutralino of the SUGRA model.
We vary the gravitino mass parameter $m_{3/2}$ from 30 to 60 TeV.  
For $M_1, M_2, \rm{and} ~
\mu =$ 272, 80, and -300 GeV, we obtain 
neutralino masses $m_{\neutralino_{1-4}} =$ 91, 269, 309, and 371 GeV with
$m_{\neutralino_4} < 0$ inside a polarization sum.  The chargino masses are 
$m_{\chargino_{1,2}} =$ 91 and 318 GeV. The gluino mass $m_{\tilde g} = 
- 672$ GeV.  The masses grow as we increase $m_{3/2}$.  For the 
$\tilde g \neutralino_1$ 
channel the LO cross section is roughly a factor of 15 larger at  Tevatron energies 
than in the SUGRA model, for comparable masses.  However, the fact that the 
combination of the gluino and neutralino masses exceeds 750 GeV makes this model an 
unlikely candidate for discovery at the Tevatron.  

Gaugino dominated boundary conditions~\cite{gaugino1} offer 
another interesting possibility, exemplified by gaugino-mediated SUSY 
breaking (\~gMSB)~\cite{gaugino2,gaugino3}.  In this class of models, 
gauge fields propagate freely in the five-dimensional bulk, whereas
fermions are confined to one or more four-dimensional hyper-surfaces.
Supersymmetry is
broken at a distant point in the extra dimension, giving mass to the gauginos,
and the gauge
interactions communicate this breaking to the scalar fermions as well.
We present results based on the model of Ref.~\cite{gaugino3}, which combines
this simple mechanism of SUSY breaking with a model of quark masses and
mixings.
This model has two input parameters, $m_{1/2}$ and the mass of the down-type
Higgs at
the GUT scale, $m_{H_d}$.  Fixing, for example, $m_{H_d} = 200$ GeV
and $m_{1/2} = 150$ GeV (which determines $\tan \beta \sim$ 15
\cite{gaugino3}),
we find a spectrum with gluino mass 379 GeV,
neutralino masses 57, 103, -224, and 249 GeV, chargino masses 101 and 251 GeV,
and squark masses of about 330 GeV.  This spectrum is similar to the default
SUGRA
scenario, with the principal differences that the Higgsino masses are somewhat
lighter
because the non-universal boundary conditions at the GUT scale result (through
the
requirement of radiative EWSB) in a non-SUGRA $\mu$-term at the weak scale.
For the light gaugino-like states the cross sections are virtually
unchanged with
respect to the SUGRA scenario, whereas the LO cross sections for
$\widetilde{\chi}^0_3$,
$\widetilde{\chi}^0_4$, and $\widetilde{\chi}^\pm_2$ are 7, 5, and 5 times
larger than those in the default SUGRA scenario, because of the increased phase 
space for these states.

An intriguing scenario is that of a light gluino LSP \cite{raby}.  Gluino masses 
in the range of 25 to 35 GeV may still be allowed \cite{raby2}.  Since the work 
of Ref.~\cite{raby2} treats only the strong SUSY sector, we make some assumptions 
about the weak parameters, respecting LEP limits~\cite{lepdata} 
on neutralino and chargino masses. 
We choose $m_{\gluino} =$ 30 GeV and $m_{\tilde q} =$ 450 GeV.  For the weak sector, 
we adopt masses typical of SUGRA models, discussed above.  Since the gluino mass is 
light, there is much more phase space available, and cross sections for associated 
production are substantial at Tevatron energies, reaching $\sim 1 {\rm pb}^{-1}$ 
for a wide range of values of $m_{1/2}$.  

Because the GMSB, $\tilde{g}$MSB and AMSB cross sections at LO are not too 
dissimilar from those of the SUGRA case at Tevatron energies, we focus our 
NLO work on the SUGRA and light gluino models.  

\section{Next-to-Leading Order Contributions}
\label{nlocorr}

The next-to-leading order contributions to the associated production of
gluinos and gauginos can be separated into virtual corrections that 
contain internal loops of colored particles, and 2-to-3 parton real 
emission contributions in which a light gluon or quark is emitted.  The
kinematics of the virtual contributions are identical to the Born case
described in Sec.~\ref{leading} whereas the presence of an
additional out-going particle in the emission contributions requires
integration over a three-body (rather than two-body) phase space.
It is useful to separate the real emission contributions into parts 
in which the additional parton's energy approaches zero (and thus the three 
body final state effectively becomes a two-body one) and parts in which 
the additional parton is hard (energetic).  We refer to these two parts as 
soft emission and hard emission contributions, respectively.

All of these NLO contributions contain singularities.  The virtual corrections 
contain ultraviolet (UV) singularities that may be absorbed into the definitions of
the couplings and operators in the usual renormalization procedure.  Both virtual 
and emission contributions contain infrared (IR) singularities when the
energy of the produced or exchanged particle approaches zero.  These
singularities cancel when the virtual and emission contributions are combined.  
Finally, there are collinear singularities when the produced particle is
emitted collinearly with another massless colored object.  These singularities
are absorbed into the (universal) NLO definition of the parton distribution 
functions.

\subsection{Virtual Corrections}
\label{virtcor}

\begin{figure}[h]
 \begin{center}
  \epsfig{file=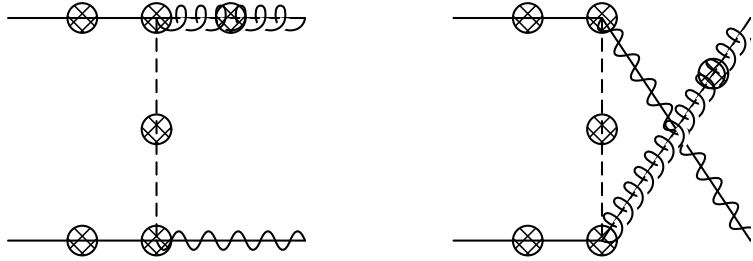,bbllx=70pt,bblly=360pt,bburx=253pt,bbury=432pt,%
          width=10cm,clip=}
 \end{center}
 \caption{Generic virtual one-loop diagrams for the associated production
 of a gluino and a gaugino. The crossed regions denote self-energy and vertex
 corrections that are present only one at a time at next-to-leading order.}
 \label{gn_virt}
\end{figure}

\begin{figure}[h]
 \begin{center}
  \epsfig{file=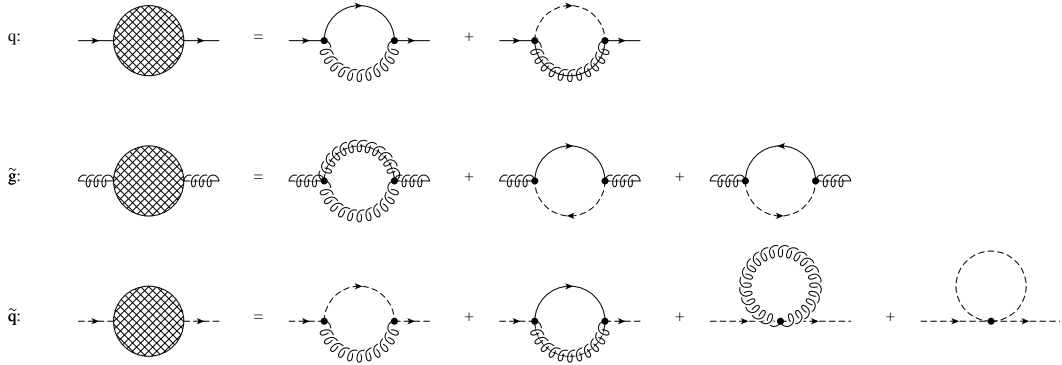,bbllx=35pt,bblly=216pt,bburx=577pt,bbury=432pt,%
          width=14cm,clip=}
 \end{center}
 \caption{Self-energy diagrams for external quarks ($q$) and gluinos
 ($\tilde{g}$) and internal squarks ($\tilde{q}$) including the full
 supersymmetric QCD particle spectrum. Contributions from the tadpole graphs 
 vanish.}
 \label{gn_self}
 \end{figure}

\begin{figure}[h]
 \begin{center}
  \epsfig{file=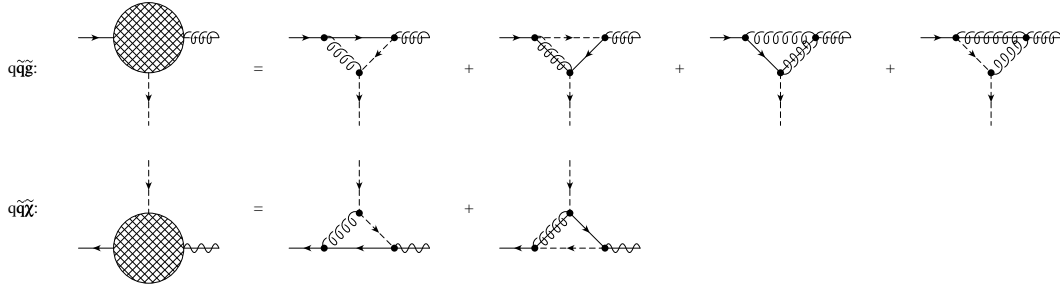,bbllx=35pt,bblly=288pt,bburx=577pt,bbury=435pt,%
          width=14cm,clip=}
 \end{center}
 \caption{Vertex corrections for the quark-squark-gluino
 ($q\tilde{q}\tilde{g}$) and quark-squark-gaugino ($q\tilde{q}\tilde{\chi}$)
 vertices including the full supersymmetric QCD particle spectrum.}
 \label{gn_vert}
\end{figure}

In this subsection we present the virtual corrections to the associated
production of gauginos and gluinos in hadron collisions.
They arise from the interference of the Born matrix elements presented
in Sec.~\ref{leading} with the one-loop amplitudes shown generically
in Fig.~\ref{gn_virt}.
In these diagrams, the crossed regions indicate contributions from self-energy
corrections (Fig.~\ref{gn_self}) and vertex corrections (Fig.~\ref{gn_vert})
that are present one at a time at next-to-leading order.
Additional contributions arise from the box diagrams in Fig.~\ref{gn_box}. 
We include the full supersymmetric spectrum of strongly interacting
particles in the virtual loops, i.e.\ squarks and gluinos as well as
quarks and gluons.

Since these virtual loop contributions contain ultraviolet and infrared
singularities, we regularize the cross sections by computing the phase
space and matrix elements in $n=4-2\epsilon$ dimensions.
We then obtain the virtual differential cross section from
\begin{eqnarray}
   \frac{d^2\hat{\sigma}^V_{ij}}{dt_2\,du_2} &=& \frac{\pi 
    S_{\epsilon}}{s^2\,\Gamma (1 - \epsilon)} \left[\frac{ t_2\,u_2 - m^{2}_{2}
    s}{\mu^2\,s} \right]^{- \epsilon} \Theta(t_2\,u_2 - m^{2}_{2}\,s)
    \Theta (s - (m_1 + m_2)^2)\,
    \nonumber \\[1mm]
 & & 
   \times \delta (s + t + u - m^{2}_1 - m^{2}_2)\,
   \overline{({\cal M}^B{\cal M}^{V\ast}+{\cal M}^V{\cal M}^{B\ast})}
 . \label{virtph}
\end{eqnarray}
As in the Born case, the matrix elements are summed (averaged) over the
colors and spins of the outgoing (incoming) particles.

We calculate the traces of Dirac matrices with the help of the computer
algebra program FORM \cite{form} using the so-called ``naive'' $\gamma_5$
scheme.  In this scheme, $\gamma_5$ anticommutes with all other 
$\gamma_{\mu}$ matrices, which is justified for anomaly-free one-loop 
amplitudes~\cite{naive}.
The $\gamma_5$ matrix enters the calculation through both the
quark-squark-gluino and quark-squark-gaugino Yukawa couplings.
The integration over the internal loop momenta is simplified by
reducing all tensorial integration kernels to expressions that are only
scalar functions of the loop momentum \cite{pasvelt}.
The resulting one-, two-, three-, and some four-point functions were 
computed in the context of other physical processes \cite{squarkgluino}.
However, two previously unknown divergent four-point functions are 
computed here for the first time due to the fact that the final state
particles, i.e.\ the gluino and the gaugino, have different
masses in general.  
The absorptive parts are obtained with Cutkosky cutting rules and the real 
parts with dispersion techniques. The results are collected in 
Appendix~\ref{scalarint}.

The virtual one-loop corrections contain ultraviolet divergences that
appear as poles in $1/\epsilon$ in the one- and two-point functions.
They are removed by renormalization of the coupling constants in the
modified-minimal-subtraction scheme ($\overline{\rm MS}$) scheme at the 
renormalization scale $\mu$~\cite{msbar}, and of the masses
of the heavy particles (squarks and gluinos) in the on-shell scheme.
A difficulty arises from the fact that gluons have $n-2$ possible
polarizations, whereas gluinos have 2, leading to violation of
supersymmetry in the $\overline{\rm MS}$ scheme.
The simplest procedure to restore supersymmetry, which we adopt here, is through 
a finite shift in the quark-squark-gluino Yukawa coupling:
\begin{equation}
 \hat{g}_s = g_s \le 1+\frac{\alpha_s}{4\pi} \lr \frac{2}{3}
 N-\frac{1}{2}C_F \rr \re = g_s \le 1+\frac{\alpha_s}{3\pi} \re .
\end{equation}
This shift was discussed first in Ref.~\cite{mismatch}.  

The virtual corrections can be classified into a $C_F$ and a $N_C$ color
class depending on the color flow or the Abelian or non-Abelian nature of the
correction vertices.
In addition to UV singularities they have collinear and infrared
singularities that appear as $1/\epsilon$ or $1/\epsilon^2$ poles in the
derivatives of the two-point- and in the three- and four-point functions.
The generally divergent scalar integrals are always multiplied by finite
coefficient functions proportional to parts of the Born matrix elements.
The full result is given in Appendix~\ref{virtapp}.

\subsection{Real Emission Contributions}
\label{realcorr}

At NLO, the production of gluinos and gauginos receives contributions
from real emission of gluons or massless quarks and anti-quarks.  In
the following sub-sections both of these types of two-to-three
partonic contributions are dealt with separately.

Following the notation developed in Ref.~\cite{hquark}, we express our
results in terms of the following sets of invariants,
\begin{alignat}{2}
  s &= (p_a + p_b)^2 \qquad\quad & s_5 &= (p_1 +p_2)^2 
  \label{reinvar}\\
  s_3 &= (p_3 +p_2)^2 - m_1^2 \qquad\quad & s_4 &
  = (p_3 +p_1)^2 -m_1^2 \nonumber\\
  t & = (p_b -p_2)^2 \qquad\quad &  t' &= (p_b -p_3)^2 \nonumber \\
  u & = (p_a -p_2)^2 \qquad\quad & u' &= (p_a -p_3)^2 \nonumber\\
  u_6 &= (p_b -p_1)^2 - m_1^2 &\qquad\quad u_7 &
  = (p_a -p_1)^2 - m_1^2, \nonumber
\end{alignat}
where $p_a$ and $p_b$ are the four-momenta of the incoming (massless)
partons, $p_1$ and $p_2$ are the $\gluino$ and $\gaugino$ momenta,
and $p_3$ is the momentum of the additional massless parton.  We
also find it useful to define the following derived quantities:
\begin{alignat}{2}
  t_1 &= t - m_1^2 \qquad\quad & t_2 &= t - m_2^2 
  \label{reauxinvar}\\
  u_1 &= u - m_1^2 \qquad\quad & u_2 & = u -m_2^2 \nonumber\\
  \delu &= m_1^2 - \msqu{u}^2\qquad\quad & 
  \delt &= m_1^2 - \msqu{t}^2 \nonumber\\
  \u6d & = u_6 + \delu \qquad\quad & \ud7 &= u_7 + \delt \nonumber \\
  \s4d & = s_4 + \delu \qquad\quad & \sd3 &= s_3 + \delt
  \,.\nonumber
\end{alignat}
Energy and momentum conservation provide relations among these
quantities:
\begin{alignat}{2}
  s_4 &= s + t_2 + u_1 \qquad\qquad\qquad & s_3 &= s + u_6 + u_7 
  \label{relations}\\
  s_5 &= s + t' + u'   \qquad\qquad\qquad & u_6 &= -s - t_2 - t'
  \nonumber  \\
  u_7 &= -s - u_2 -u' \,. \nonumber
\end{alignat}

The $n$-dimensional three-body phase space may be derived conveniently
if we integrate the general fully differential cross 
section in the 1-3 rest frame \cite{hquark}.  In this frame
the $4$-dimensional components of the
$n$-dimensional momenta are expressed as:
\bea
  p_a &=& (\omega_a,0,\omega_a \sin \psi, 
  \omega_a \cos \psi)\\
  p_b &=& (\omega_b,,0,0,\omega_b)\nonumber\\
  p_1 &=& (E_1,-\omega_3\sin\theta_1\sin\theta_2,
  -\omega_3\sin\theta_1\cos\theta_2,-\omega_3\cos\theta_1)\nonumber  \\
  p_2 &=& (E_2,\omega_a \sin \psi, 
  \omega_a \cos \psi + \omega_b)\nonumber\\
  p_3 &=& (\omega_3,\omega_3\sin\theta_1\sin\theta_2,
  \omega_3\sin\theta_1\cos\theta_2,\omega_3\cos\theta_1)\nonumber ,
\eea
with
\begin{alignat}{2}
  \omega_a &= \frac{s+u_2}{2\sqrt{s_4+ m_1^2}}& \qquad
  \omega_b &= \frac{s+t_2}{2\sqrt{s_4+ m_1^2}} \\[0.2cm]
  \omega_3 &= \frac{s_4}{2\sqrt{s_4+ m_1^2}} & \qquad
  E_1 &= \frac{s_4 +2 \, m_1^2}{2\sqrt{s_4+ m_1^2}} \nonumber \\[0.3cm]
  E_2 &= {}-\frac{t_2 +u_2 +2 \, m_1^2}{2\sqrt{s_4+ m_1^2}}& \qquad
  \cos\psi &= \frac{-s (s_4 + m_1^2 + m_2^2) + t_2 u_2}
  {(s+t_2)(s+u_2)} \, . \nonumber 
\end{alignat}
Using this parameterization, we may express the invariants defined in 
Eq.~(\ref{reinvar}) in terms of $\theta_1$, $\theta_2$, and the 
$\theta$-independent variables $\omega_{(a,b,3)}$, $E_{(1,2)}$,
and $\psi$.  For the real emission contributions, it is
sometimes convenient to parameterize these momenta with the $\hat{z}$
axis aligned along $p_a$ or $p_2$.  As these alternate frames are
related by a simple spatial rotation, the expressions for
$E_{(1,2)}$, and $\omega_{(a,b,3)}$ remain unchanged.
The general three-body cross section may be expressed in this frame as
\bea
  \label{ph3}
  \frac{d^3\hat{\sigma}^R_{ij}}{ds_4\,du_2 \, dt_2} &=& \frac{ {S_\epsilon}^2
    \mu^{2 \epsilon}}{2 \, s^2 \, \Gamma(1-2\epsilon)}
    \left[\frac{t_2 \, u_2 - s \, m_2^2}{s \, \mu^2}\right]^{-\epsilon}
    \Theta(t_2 \, u_2 - s \, m_2^2) \, \Theta (s_4) \\[.1cm]
   & \times & \Theta(s - (m_1 + m_2)^2)
   \, \frac{s_4^{1-2\epsilon}}
   {(s_4+m_1^2)^{1-\epsilon}} \, \delta (s + t_2 + u_1 - s_4)
   \int d\Omega_n  \overline{|{\cal M}^R|}^2 , \nonumber
\eea
in which $\overline{|{\cal M}^R|}^2$ is the real emission matrix
element squared, summed over final spins and colors and averaged over
initial spins and colors,
and the $n$-dimensional angular integration is
$d\Omega_n = \sin^{1-2\epsilon}(\theta_1)\, d\theta_1\, 
\sin^{-2\epsilon}(\theta_2) \,d\theta_2$.

\begin{figure}
\centerline{\hbox{
\psfig{figure=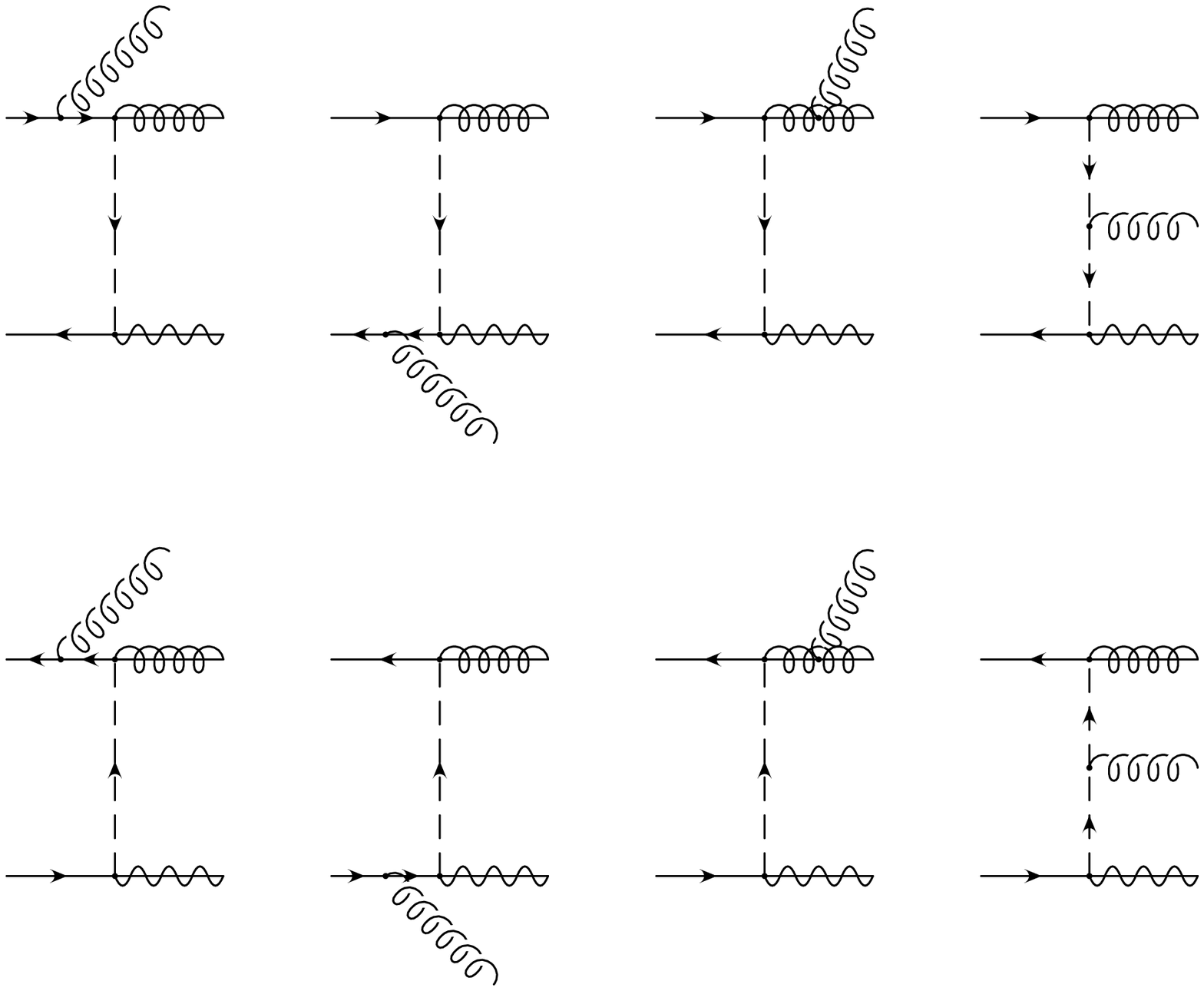,height=4.0in}}}
\vskip 0.1 in
\caption{Feynman diagrams for the gluon emission contribution,
$q \bar{q} \ra g \, \gluino \, \gaugino$.}
\label{realglu}
\end{figure}

\begin{figure}
\centerline{\hbox{
\psfig{figure=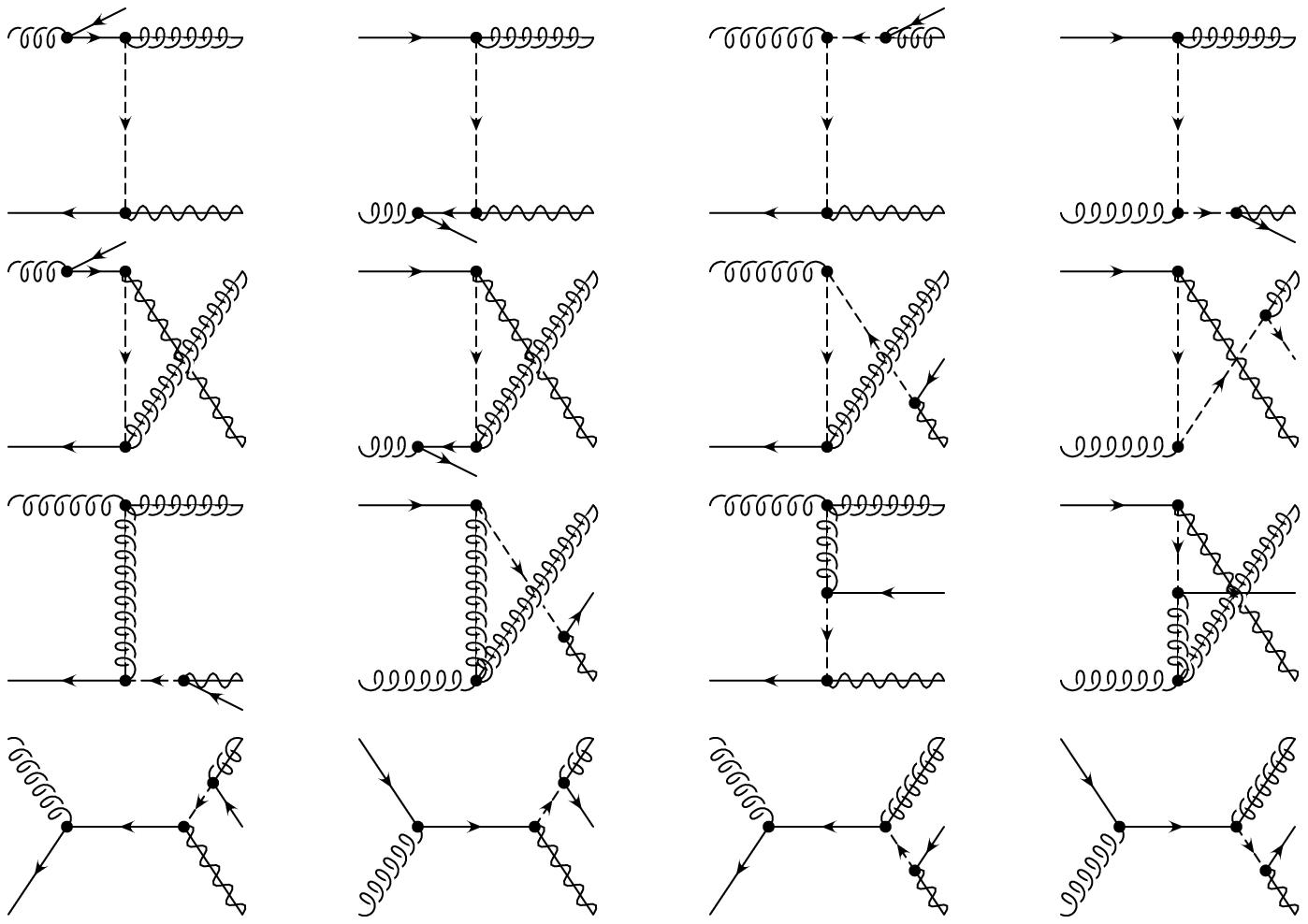,height=4.1in}}}
\vskip 0.1 in
\caption{Feynman diagrams for the emission of a light quark,
$q + g \ra q + \gluino + \gaugino$.}
\label{realq}
\end{figure}

In evaluating the integration over the angular variables in 
Eq.~(\ref{ph3}), we follow the procedure 
outlined in Ref.~\cite{hquark}, in
which we use the relations among the invariants, 
Eq.~(\ref{relations}), to reduce all of the angular integrals to the form,
\bea
  I_n^{(k,l)} &=& \int_0^\pi \sin^{1-2\epsilon}(\theta_1)\, d\theta_1\,
  \int_0^\pi \sin^{-2\epsilon}(\theta_2) \,d\theta_2 \\
  & & \times (a + b \cos\theta_1)^{-k}
  ( A + B\cos\theta_1 + C\sin\theta_1\cos\theta_2)^{-l} \nonumber,
\eea
the necessary expressions for which may be found in Ref.~\cite{hquark}.
The angular integrations involving negative powers of $t'$ and $u'$
produce poles in $\epsilon$ which correspond to collinear
singularities in which particle 3 is collinear with particle $a$ or
$b$ (c.f. Figs.~(\ref{realglu}) and (\ref{realq})).
Because these singularities follow a universal structure,
they may be removed from the cross section and absorbed into the
parton distribution functions according to the usual mass 
factorization procedure \cite{facto}.  The non-zero mass of the gluino 
kinematically forbids collinear emission, and thus the gluino has no 
associated collinear singularities.

The collinear singular pieces have the factorized form 
\bea
  \label{masterfac}
  \frac{d^2\hat{\sigma}^R_{ij}(\, s,t_2,u_2,\mu^2)}{dt_2\, du_2} &=&
  \int_0^1 dx_1 \, x_1 \int_0^1 dx_2 \, x_2 \, \sum_{k,l} \,
  \Gamma_{ki}(\, x_1,\mu_F^2,\mu^2,\epsilon) \\
  & & \times \,\Gamma_{lj}(\, x_2,\mu_F^2,\mu^2,\epsilon)
  \, \frac{d^2\hat{\sigma}^R_{kl}( \,
  \hat{s},\hat{t}_2,\hat{u}_2,\mu_F^2)}{ d\hat{t}_2\, d\hat{u}_2} ,
  \nonumber
\eea
in which $\hat{s} = x_1 \, x_2 \, s$, $\hat{u}_2 = x_1 \, u_2$, 
$\hat{t}_2 = x_2 \, t_2$.  The
universal splitting functions $\Gamma_{ij}$ contain the
collinear divergences associated with incoming parton $j$ splitting into
parton $i$, and the hard scattering cross section, 
$d^2\hat{\sigma}^R_{kl} / d\hat{t}_2\, d\hat{u}_2$, is free from
singularities.  The splitting functions may be redefined by an arbitrary
finite term, and thus one must choose a factorization scheme.  In order
to use recent sets of parton distributions extracted from
data we adopt the \MS scheme, in which
the splitting functions at ${\cal O}(\alphas)$ are 
\bea
  \Gamma_{ij}(x,\mu_F^2,\mu^2,\epsilon) = \delta_{ij}\,\delta(\, x - 1)
  +\frac{\alphas}{2\pi}\left[-\frac{1}{\epsilon} 
  + \gamma_E - \log (4 \pi)
  + \log\left(\frac{\mu_F^2}{\mu^2}\right)\right] P_{ij}(x).
\eea
In the above expression, the $P_{ij}(x)$ are the 
Altarelli-Parisi evolution kernels \cite{ap},
\bea
  P_{q q}(x_i) = P_{\bar{q} \bar{q}}(x_i) &=& 
  C_F \left[ \frac{1 + x_i^2}{1 - x_i} \Theta (1 - \delta_i - x_i)
  + \left( 2 \log \delta_i + \frac{3}{2} \right) \delta (1 - x_i)
  \right], \\
  P_{gq}(x_i) = P_{g \bar{q}}(x_i) &=& C_F\,\frac{1 + (1-x_i)^2}{x_i}  
  \nonumber , \\
  P_{qg}(x_i) = P_{\bar{q} g}(x_i) &=& T_F\left[x_i^2 + (1-x_i)^2\right]
  \nonumber , \\
  P_{gg}(x_i) &=&  2 N_C \left[\frac{1}{x_i (1-x_i)}
   + x_i(1-x_i) -2\right] \Theta(1-x_i-\delta_i) \nonumber \\
  & &{} +\left[2 N_C \log\delta_i + \frac{1}{2}\beta_0^L\right] 
  \delta(1-x_i) \nonumber ,
\eea
with $C_F = 4 / 3$ and $T_F = 1 / 2$ for $N_C = 3$ colors of quarks; 
$\beta_0^L = 11N_C/6 - 2n_F T_F/3$.  The quantities $\delta_i$ express 
the slicing of $s_4$ into hard and soft regimes in terms of the $x_a$
and $x_b$ variables.  They can be related to the $\Delta$ 
of Section \ref{gemiss} by
$\delta_a = \Delta / (s + u_2)$ and $\delta_b = \Delta / (s + t_2)$.

We set the renormalization
and factorization scales equal to each other, $\mu_F = \mu$,
and expand Eq.~(\ref{masterfac}) to ${\cal O}(\alphas)$ to 
derive the expression for the reduced cross section,
\begin{eqnarray}
  \label{subtracter}
  \frac{d^2\hat{\sigma}^R_{ij}(s,t_2,u_2,\mu^2)}{dt_2\,du_2} &=&
  \frac{d^2{\hat{\sigma}}^R_{ij}(s,t_2,u_2,\mu^2)}{dt_2\,du_2}\\
  & & + \, \frac{\alphas}{2\pi} \, \frac{1}{\bar\epsilon} \,
  \int_0^1 dx_1 \, x_1 \, P_{li}(x_1)\,
  \frac{d^2{\hat{\sigma}}^0_{lj}(x_1 s,t_2,x_1 u_2,\mu^2)}{dt_2\,d\hat{u}_2}  
  \nonumber\\
  & & + \, \frac{\alphas}{2\pi} \, \frac{1}{\bar\epsilon} \,
  \int_0^1 dx_2 \, x_2 \, P_{kj}(x_2)\,
  \frac{d^2{\hat{\sigma}}^0_{ik}(x_2 s,x_2 t_2,u_2,\mu^2)
    }{d \hat{t}_2 \, du_2}.  \nonumber
\end{eqnarray}
We employ the compact notation 
$\bar{\epsilon}^{-1} = \epsilon^{-1} - \gamma_E + \log(4 \pi)$; 
$d^2{\hat{\sigma}}^0_{ik} / dt_2 \, du_2$ is the leading order cross
section for $i \, k \ra \gluino \, \gaugino$.
The resulting hard scattering cross section is free from
collinear singularities, as the implicit $\epsilon$-dependence of
$d^2{\hat{\sigma}}_{ij} / dt_2\,du_2$ cancels with the explicit 
$\epsilon$-dependence of the second and third terms.

\subsubsection{Gluon Emission}
\label{gemiss}

The NLO real contributions with an additional gluon in the final state,
\bea
q \, \bar{q} \, \ra \, g \, \gluino \, \gaugino,
\eea
proceed from the Feynman diagrams shown in Fig.~\ref{realglu}.
As was the case for the leading order cross section for
production of neutralinos with gluinos, the set of graphs
in which a right-handed squark is exchanged cannot interfere with the
graphs in which a left-handed squark is exchanged because the incoming
quark and anti-quark must have definite helicity.  Production of
charginos in association with gluinos involves only left-handed squarks.


In addition to the collinear singularities described above,
this set of corrections also has infrared singularities that 
arise when the energy of the emitted gluon approaches zero.  These
singularities appear as poles in $s_4$ in the reduced cross section,
and must also be extracted so that they can be combined with
corresponding terms in the virtual corrections and shown to cancel.

To make this cancellation conveniently, we slice the gluon emission
corrections into hard and soft pieces,
\begin{equation}
  \label{softhard}
  \frac{d^2\hat{\sigma}^R_{ij}}{dt_2 \, du_2} =
  \int_0^{\Delta} ds_4 \,\frac{d^3\hat{\sigma}^S_{ij}}{dt_2\, du_2 \, ds_4} +
  \int_\Delta^{{s_4}^{max}} ds_4 \,\frac{d^3\hat{\sigma}^H_{ij}}
  {dt_2\, du_2 \, ds_4},
\end{equation}
where $\Delta$ is an arbitrary cut-off between what we call soft gluon
radiation and hard gluon radiation.  When the cut-off is much
smaller than the other invariants, $\Delta \ll s, \, t, \, u, \, m_i^2$,
the $s_4$ integration for the soft term becomes very simple and can be
evaluated analytically.  This operation results in singular terms 
\bea
  \label{cclass}
  \frac{d^2\hat{\sigma}^B_{ij}}{dt_2 \, du_2} \;& & \left[
  \left( \frac{C_F \, \alpha_S}{\pi} \right)
  \left\{ \frac{1}{\bar \epsilon^2}\,
  + \, \left( \frac{3}{2} \, + \, \log \frac{\mu^2}{s} \right)
    \frac{1}{\bar \epsilon} \; \right\} \right. \\[0.2cm]
  & & \: - \left.
  \left( \frac{N_C \, \alpha_S}{2 \pi} \right)
  \left\{ \, \log \left( \frac{(s+t_2)(s+u_2)}{s \, m^2_1}  \right) - 1
  \; \right\} \frac{1}{\bar \epsilon}
  \right] . \nonumber
\eea
In Eq.~(\ref{cclass}) we distinguish the contributions from the $N_C$ 
and $C_F$ color classes. 

The singular expression may then be combined with the
virtual corrections discussed in Sec.~\ref{virtcor} to yield the
combined ``soft and virtual" 
contribution free from infrared singularities.  The residual finite 
soft contributions are presented in Appendix~\ref{softgapp}.

In the hard gluon regime, there are collinear singularities,
but no IR singularities, and thus the most singular terms are
proportional to $\epsilon^{-1}$.  After the mass subtraction 
described above is performed, the results are
singularity-free, and they can be presented as minimally subtracted, 
singularity-free integrals,
\bea
  \hat{I}\left( f(\theta_1, \theta_2) \right) &=& 
  \int_0^{\pi} d\theta_1\, \int_0^{\pi} \, d\theta_2
  \sin^{1-2\epsilon}(\theta_1) \, \sin^{-2\epsilon}(\theta_2) \, 
  f(\theta_1, \theta_2) \label{finiteint} \\
  & & - \, \frac{1}{\epsilon} \, \lim_{\epsilon \ra 0} 
  \left( \epsilon
  \int_0^{\pi} d\theta_1\, \int_0^{\pi} \, d\theta_2
  \sin^{1-2\epsilon}(\theta_1) \, \sin^{-2\epsilon}(\theta_2) \,
    f(\theta_1, \theta_2) \right) , \nonumber
\eea
which contain only the finite terms with the
$\epsilon^{-1}$ poles subtracted.  The resulting expression
consists of a simple power series in $\epsilon$, which may then be
evaluated in 4 dimensions by setting $\epsilon \ra 0$.  
Note that the function $f(\theta_1, \theta_2)$ can
involve coefficients for angular expressions that have mass dimension,
and thus the mass dimension of 
$\hat{I}(f(\theta_1, \theta_2))$ will depend on 
$f(\theta_1, \theta_2)$.  The gluon emission matrix elements are 
presented in Appendix~\ref{hardgapp}.

\begin{figure}
\centerline{\hbox{
\psfig{figure=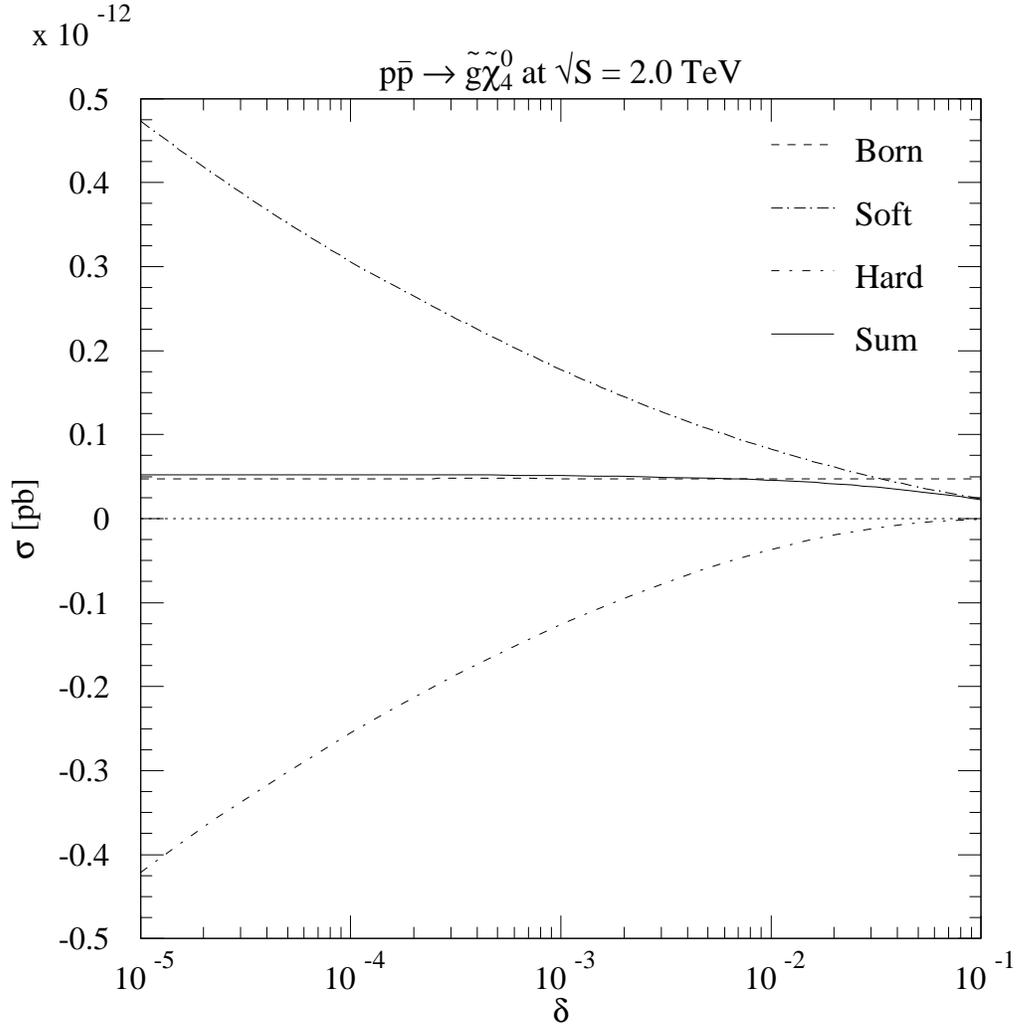,height=6.0in}}}
\vskip 0.1 in
\caption{Dependence of the total hadronic cross section on the cutoff 
parameter $\delta$ for the $\gluino \neutralino_4$ channel in the SUGRA 
model at the Tevatron, with $m_{\gluino} = 1012$ GeV and 
$m_{\neutralino_4} = 679 $ GeV.}
\label{cutoff}
\end{figure}

The cutoff on the $s_4$ integration
introduces an implicit logarithmic dependence on $\Delta$ that is
matched by the explicit logarithms of $\Delta$ which appear in the 
combined soft and virtual term.  The total correction is
independent of the value of $\Delta$.  
Choosing for illustration $m_{1/2} = 400$ GeV, we display in 
Fig.~\ref{cutoff} the dependence of various contributions on our 
cutoff $\delta$.  The Born contribution is obviously independent of 
$\delta$, but its contribution helps to show the relative magnitude 
of different terms.  The combined soft and virtual contribution is 
positive but falls as an explicit analytic function of 
$\rm{log}\delta$.  The hard part of the gluon emission contribution 
is negative, but its numerical value grows more positive as a 
implicit function of $\rm{log}\delta$.  The figure shows that two 
contributions balance each other well, and the combined 
soft/virtual plus hard contribution is independent of the cutoff
for $\delta < 2~10^{-3}$.  The figure also shows that the net small 
next-to-leading order contribution is obtained after large cancellations 
take place.  The case chosen for display in 
Fig.~\ref{cutoff} is a worst case.  With $m_{1/2} = 400$ GeV, 
$m_{\neutralino_4} = 679$ GeV, and $m_{\gluino} = 1012$ GeV.  The energy 
chosen is that of the Tevatron, $\sqrt S = 2$ TeV, so phase space 
limitations are relatively severe for this set of masses.  For all other 
cases, the cross 
sections are independent of $\delta$ for a larger range of $\delta$.  

\subsubsection{Light Quark (Anti-Quark) Emission}
\label{qemiss}

A second set of real emission corrections involves 
an additional light quark
(or anti-quark) in the final state, through the partonic reactions,
\bea
q \, g \ra q \, \gluino \, \gaugino &,& 
\bar{q} \, g \ra \bar{q} \, \gluino \, \gaugino .
\eea
The set of Feynman diagrams
contributing to emission of a quark is shown in Fig.~\ref{realq}.
They include diagrams in which an incoming gluon splits into a 
$q \bar{q}$ pair as well as diagrams in which an intermediate
squark splits into a quark and either a gluino or gaugino.
This set of corrections does not have an IR divergence, and thus it
is not necessary to slice it into hard and soft regimes.  
However, after all of the initial state collinear singularities are removed 
by the mass factorization procedure described above, the matrix elements may 
still contain integrable singularities if the mass of the
squark is larger than the mass of the gluino or gaugino.  In these cases, 
the intermediate squark state can be on its mass-shell, and 
the variables $\s4d$ and $\sd3$ go to zero inside the
region of integration.  This problem was encountered previously 
\cite{squarkgluino}, and we follow the same procedure.
These singularities represent the LO production of a squark
and a gluino or gaugino, followed by the LO decay of the squark.
They may be removed if one includes the full Breit-Wigner form for 
the squark propagator, which regulates the squark resonance by the 
squark width.  This procedure amounts to the replacements,
\bea
  \label{bwprop}
  \frac{1}{\s4d} &\ra& \frac{1}
  {\s4d + i \, \msqu{u} \Gamma_{\tilde{q}_u} } 
  \ra {\cal P}\left( \frac{1}{\s4d} \right) 
  - i \, \pi \, \delta ( \s4d )  \\
  \frac{1}{\sd3} &\ra& \frac{1}
  {\sd3 + i \, \msqu{t} \Gamma_{\tilde{q}_t} } 
  \ra {\cal P}\left( \frac{1}{\sd3} \right) 
  - i \, \pi \, \delta ( \sd3 ) , \nonumber
\eea
where ${\cal P}$ indicates the principal value function, and the final
distribution identity holds in the limit of small squark widths,
$\Gamma_{\tilde{q}} \ll \msqu{}$.  The replacement removes the singularities,
and when both $\s4d$ and $\sd3$ are zero it generates an additional
real term from the product of $\delta$ functions.

There is a further subtlety associated with
the requirement that we not double-count the region of phase space
in which the squark is on-shell.  Properly, the kinematic configuration
with an on-shell squark is included in the LO production of
a squark and a gluino or a squark and a gaugino, and thus should not be
considered as a genuine higher order correction to the production of gluinos
with gauginos.  To avoid double counting, we thus subtract the on-shell squark
contribution by defining the total cross section (for illustration, we
deal with the $\s4d$ singular case),
\bea
  \hat{\sigma} = \int_0^{s_{4}^{max}} ds_{4}
  \int_{t_2^-(s_{4})}^{t_2^+(s_{4})} dt_2\,
  \frac{d^2\hat{\sigma}}{dt_2\,ds_{4}}
  = \int_0^{s_{4}^{max}} ds_{4}\, \frac{f(\s4d)}
  {\s4d^2 + \msqu{ }^2 \, \Gamma^2_{\tilde{q}} }.
\eea
The on-shell contribution then corresponds to 
$f(0) \, / \: (\s4d^2 + \msqu{ }^2 \, \Gamma^2_{\tilde{q}})$, with 
\bea
   f(0) = {\hat{\sigma}}^B_{\gaugino \tilde{q}} \,
   \frac{ \msqu{ } \Gamma_{\tilde{q}} }{ \pi } \,
   \frac{\Gamma^B_{\tilde{q} \ra q \, \gluino} }{\Gamma_{\tilde{q}}}
   \ra {\hat{\sigma}}^B_{\gaugino \tilde{q}} \: 
   BR( \tilde{q} \, \ra \, q \, \gluino ) \:
   (\s4d^2 + \msqu{ }^2 \, \Gamma^2_{\tilde{q}}) \:
   \delta( \s4d ) .
\eea
It can be subtracted leaving a genuine NLO contribution 
\bea
   {\hat{\sigma}}^{\rm NLO} &=& \int_0^{s_4^{max}} ds_4
   \frac{ f(\s4d) - f(0) }
   { \s4d^2 + \msqu{ }^2 \, \Gamma^2_{\tilde{q}} }
\eea
which may once again be expressed as a principal value function, since
$\s4d \, / \, (\s4d + \msqu{ }^2 \, \Gamma^2_{\tilde{q}}) 
\ra {\cal P}( 1 / \s4d)$
in the limit of small squark width.  The $\sd3$ singular terms may be 
treated in a similar way, with the added complication that 
the integration over $\sd3$
is hidden in the angular integrations \cite{squarkgluino}.

The quark emission matrix elements are presented in Appendix~\ref{qapp}.

For the parameters of the SUGRA model that we adopt, the gluino mass 
remains greater than the squark mass for all values of $m_{1/2}$, and 
there is never an intermediate squark-to-gluino-plus-quark final state 
singularity. However, the two chargino masses and all four neutralino 
masses are always less than the squark mass, and the final-state 
on-shell squark-to-gaugino-plus-quark singularity comes into play in 
all cases.  In the light gluino model, with $m_{\tilde{q}} = 450$~GeV, 
the gluino mass of 30 GeV is light enough that on-shell intermediate 
squark decay into a gluino is always active. In this light gluino model, 
as $m_{1/2}$ is varied from 100 to 400 GeV, the masses of the lighter 
gauginos ($\neutralino_1$, $\neutralino_2$, and $\chargino_1$) remain 
less than the squark mass so that the on-shell intermediate squark decay 
into a gaugino is active over the whole range of $m_{1/2}$ for these 
light channels.  The situation changes for the heavier gauginos 
($\neutralino_3$, $\neutralino_4$, and $\chargino_2$).  For small 
$m_{1/2}$ their masses are below the squark mass, and the on-shell 
decay is active.  However, above roughly $m_{1/2} =250$~GeV, the masses 
of the heavier gauginos exceed the squark mass and the on-shell 
possibility closes.  


\section{Quantitative Results}
\label{results}

In this section we collect our main results on total 
and differential cross sections for the associated production of 
gauginos and gluinos at Tevatron and LHC energies.  

\subsection{Scaling Functions}
\label{scafns}

We begin with the cross section at the parton level expressed as
\bea
 \hat{\sigma}_{ij} &=& \frac{\alpha\alphas (\mu)}{m^2} \left\{
 f_{ij}^B(\eta)+4\pi\alphas (\mu)\left[ f_{ij}^{V+S}(\eta,\mu)+f_{ij}^H
 (\eta,\mu) \right] \right\}.
 \label{partonic}
\eea
It has been integrated over the Mandelstam invariants $t$ and $s_4$ and
depends on the partonic center-of-mass energy $s$ through the scaling
variable
\bea
 \eta &=& \frac{s}{4m^2}-1, 
\eea
where $m$ is the average mass of the produced sparticles, 
\bea
 m &=& \frac{m_1+m_2}{2}.
\eea
It also depends on the produced masses $m_1$ and $m_2$ and on the squark 
mass $\ms$ (through the internal squark propagator).  The common renormalization 
and factorization scale is denoted by $\mu$.  The partonic initial state is 
labeled $i,j=g,q,\bar{q}$.

Equation~(\ref{partonic}) defines the dimensionless scaling functions 
$f_{ij}$, studied in Ref.~\cite{nason}. These functions are 
independent of the coupling constants $\alpha$ and $\alphas$,
of parton densities, and of the collider type and energy.
They permit precise checks of individual contributions and of the threshold, 
resonance, and high energy behaviors of the production process.

The Born $f_{ij}^B$ and the summed virtual and soft scaling functions 
$f_{ij}^{V+S}$ receive contributions only from $q\bar{q}$ initial states,
where $q=u~{\rm or}~d$, with the possible emission of a soft and/or collinear gluon. 
The hard scaling function $f_{ij}^H$ has contributions from $qg$ initial 
states when an additional quark or antiquark is emitted together with the gluino
and the gaugino. We eliminate the explicit dependence of the soft
contributions on the technical cut-off $\delta=\Delta/m^2$ by subtracting the
$\log^{(1,2)}\delta$ terms. These terms are then added to the hard
contribution such that this contribution is also independent of $\delta$.
In Sec.~\ref{nlocorr} we show that our results are independent of $\delta$ at 
least in the range $\delta \in [10^{-5};10^{-3}]$, and we use the value 
$\delta = 10^{-4}$ in the following.

\begin{figure}
\centerline{\hbox{
\psfig{figure=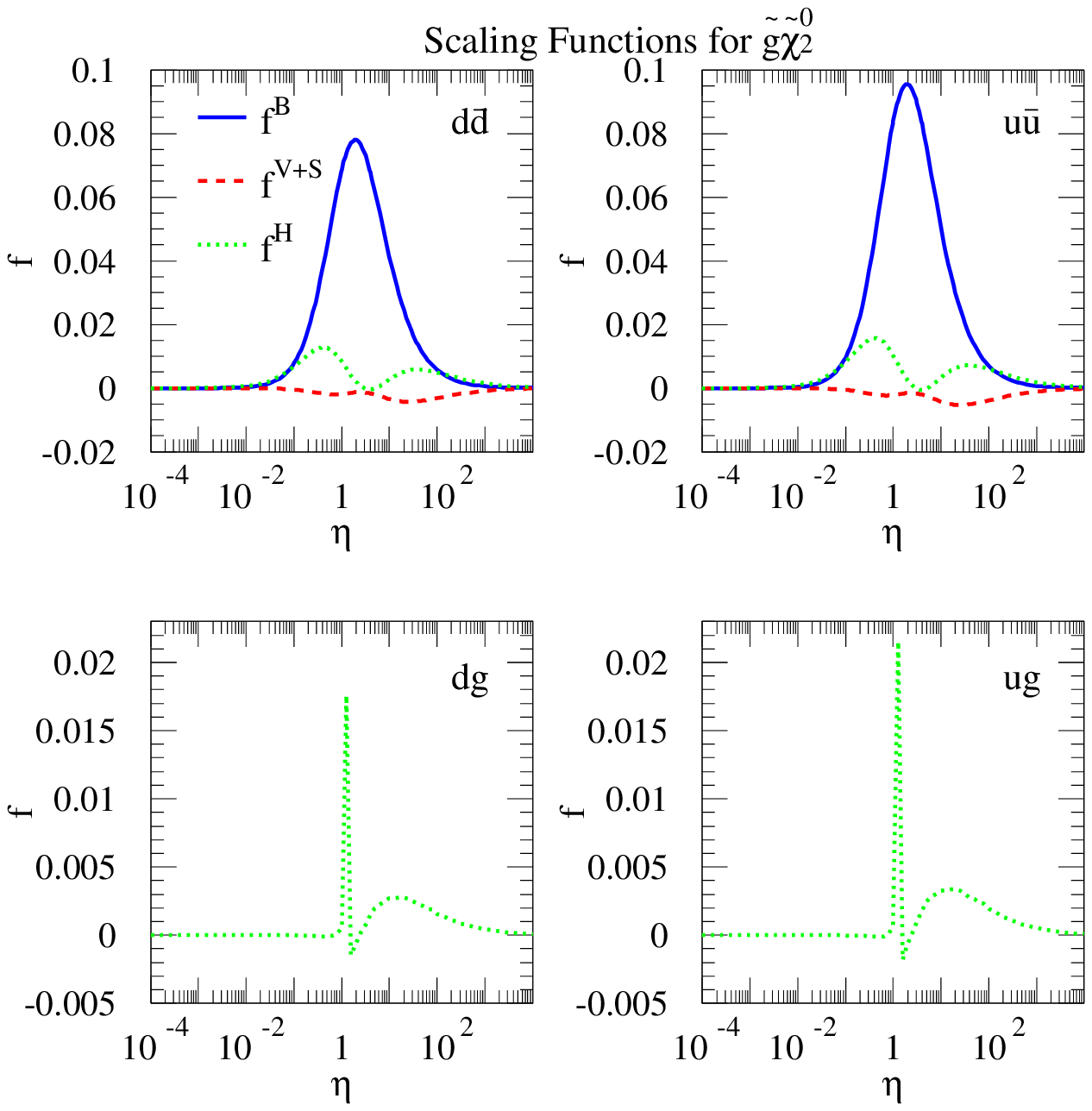,height=6.0in}}}
\vskip 0.1 in
\caption{Scaling functions for the SUGRA $\gluino \neutralino_2$ channel, 
with $m_{\gluino} = 410$ GeV and 
$m_{\neutralino_2} = 104 $ GeV.}
\label{scafun0}
\end{figure}

\begin{figure}
\centerline{\hbox{
\psfig{figure=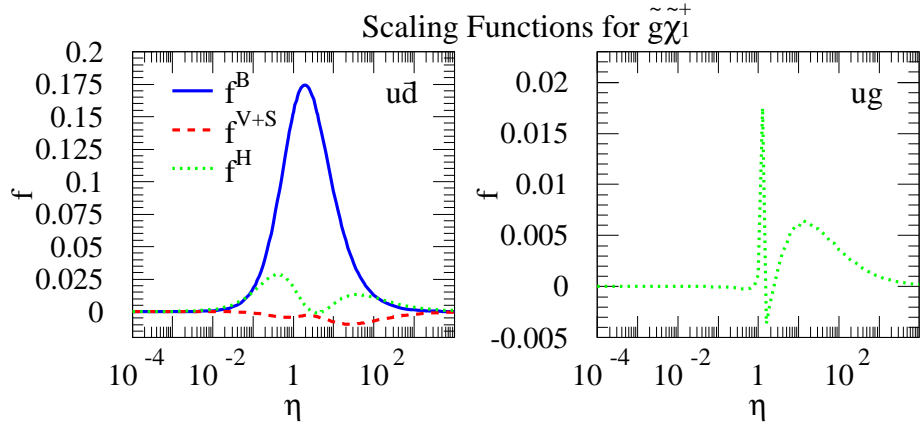,height=3.0in}}}
\vskip 0.1 in
\caption{Scaling functions for the $\gluino \tilde{\chi}^{+}_1$ 
channel in the SUGRA model with $m_{\gluino} = 410 $ GeV and 
$m_{\tilde{\chi}^{+}_1} = 101 $ GeV.}
\label{scafun1}
\end{figure}

The scaling functions for the production of a $\gluino$ and a $\neutralino_2$
are presented in Fig.~\ref{scafun0} and those for a $\gluino$ and a $\chargino_1$ 
in Fig.~\ref{scafun1}.  Here we set the scale $\mu$ equal to the average 
particle mass $m$.  The masses are those of our default SUGRA scenario.  
As discussed in Sec.~\ref{qemiss}, the emission of an additional quark or antiquark 
can lead to intermediate on-shell squarks and therefore to a singular squark 
propagator in Feynman diagrams. After the LO two-body 
$q + g \rightarrow \tilde{g} + \tilde{q}$ contribution is removed, the remaining 
integrable singularities can be identified as 
spikes in the $gu$ and $gd$ scaling functions in the two figures.  They yield 
finite contributions after integration over the momentum fractions 
$x_a$ and $x_b$ of the incoming partons or, equivalently, over the partonic 
center-of-mass energy $s=x_ax_bS$.  

Evident from Figs.~\ref{scafun0} and~\ref{scafun1} is that next-to-leading order 
contributions do not alter either the threshold or high energy asymptotic behaviors 
in $\eta$, unlike, e.g., the situation for pair production of heavy 
quarks~\cite{nason,jsmith}.  The combined virtual and soft scaling functions 
$f_{ij}^{V+S}$ contribute negatively but are small in magnitude when compared with 
the hard scaling functions $f_{ij}^H$.  The figures show that one should expect 
only modest enhancements in predicted rates when the $u {\bar u}$, $d {\bar d}$, 
and $u {\bar d}$ 
production channels are dominant, as is true at Tevatron energies, for which the 
quark/antiquark parton luminosity is large and the range in $\eta$ is limited (
$\eta < 15 $ for these two channels in our default scenario).  The contribution of 
the $qg$ channel can become important if phase space is open to large values of 
$\eta$.  At LHC energies, $\eta$ extends to nearly 800.  This range in $\eta$, along 
with the large $qg$ luminosity, suggests that the $qg$ channel will supply 
significant enhancements in the predicted rates at LHC energies, as is 
demonstrated below.  
   
Scaling functions for the $\gluino \neutralino_1$, $\gluino \neutralino_4$, 
and $\gluino \chargino_2$ channels show behavior similar to that seen in 
Figs.~\ref{scafun0} and~\ref{scafun1}, with the notable exception that the 
positive excursion at large $\eta$ in the $qg$ channels is relatively more 
prominent for $\gluino \neutralino_4$ and $\gluino \chargino_2$ than for 
$\gluino \neutralino_2$ and $\gluino \chargino_1$.  The $\gluino \neutralino_3$ 
channel is distinguished primarily by the fact that the peak in the $\eta$ 
distribution at the Born level occurs near $\eta = 0.4$ whereas the peaks 
occur at larger $\eta$, in the range of $\eta =$ 1 to 2, in all other cases.  
There is also a noticeable difference in threshold behaviors of the Born and 
NLO hard-gluon emission contributions for this channel.  We relate 
these differences to the fact that only the $\gluino \neutralino_3$ channel 
exhibits positive interference at the Born level between the t- and u-channel 
contributions, c.f., Eq.~(\ref{bornme}).  Because $m_{\neutralino_3}$ is negative 
in the SUGRA model, the term proportional to $X_{tu}$ in Eq.~(\ref{bornme}) is 
positive for $\gluino \neutralino_3$ production but negative in all other cases.  

\subsection{Hadronic Total Cross Sections}
\label{totcross}

The hadronic total cross section is obtained from the
partonic cross section through
\bea
 \sigma^{h_1h_2}(S,\mu) &=& \sum_{i,j=g,q,\overline{q}}
 \int_{\tau}^1{\rm d}x_a\int_{\tau/x_a}^1{\rm d}x_b
 f_i^{h_1}(x_a,\mu) f_j^{h_2}(x_b,\mu)
 \hat{\sigma}_{ij}(x_ax_bS,\mu), \nonumber \\
\eea
where
\bea
 \tau &=& \frac{4m^2}{S} ,
\eea
and $\sqrt S$ is the hadronic center-of-mass energy (2 TeV for Run II at the
Fermilab $p\bar{p}$ collider Tevatron and 14 TeV at the CERN $pp$ collider
LHC). Our NLO predictions are calculated in the \MS scheme with the
CTEQ5M parametrization \cite{cteq5} for the parton densities $f(x,\mu)$ in
the proton and antiproton and a two-loop approximation for the strong
coupling constant $\alphas$ with $\Lambda^{(5)}=226$ MeV.  This 
value of $\Lambda^{(5)}$ is used also in the renormalization group
evolution equations for our SUSY scenarios.  In obtaining LO 
cross sections, we use the CTEQ5L LO parton densities and the one-loop 
approximation for $\alphas$, with $\Lambda^{(5)}=146$ MeV.


\subsubsection{SUGRA model results}
\label{sugracross}

\begin{figure}
\centerline{\hbox{
\psfig{figure=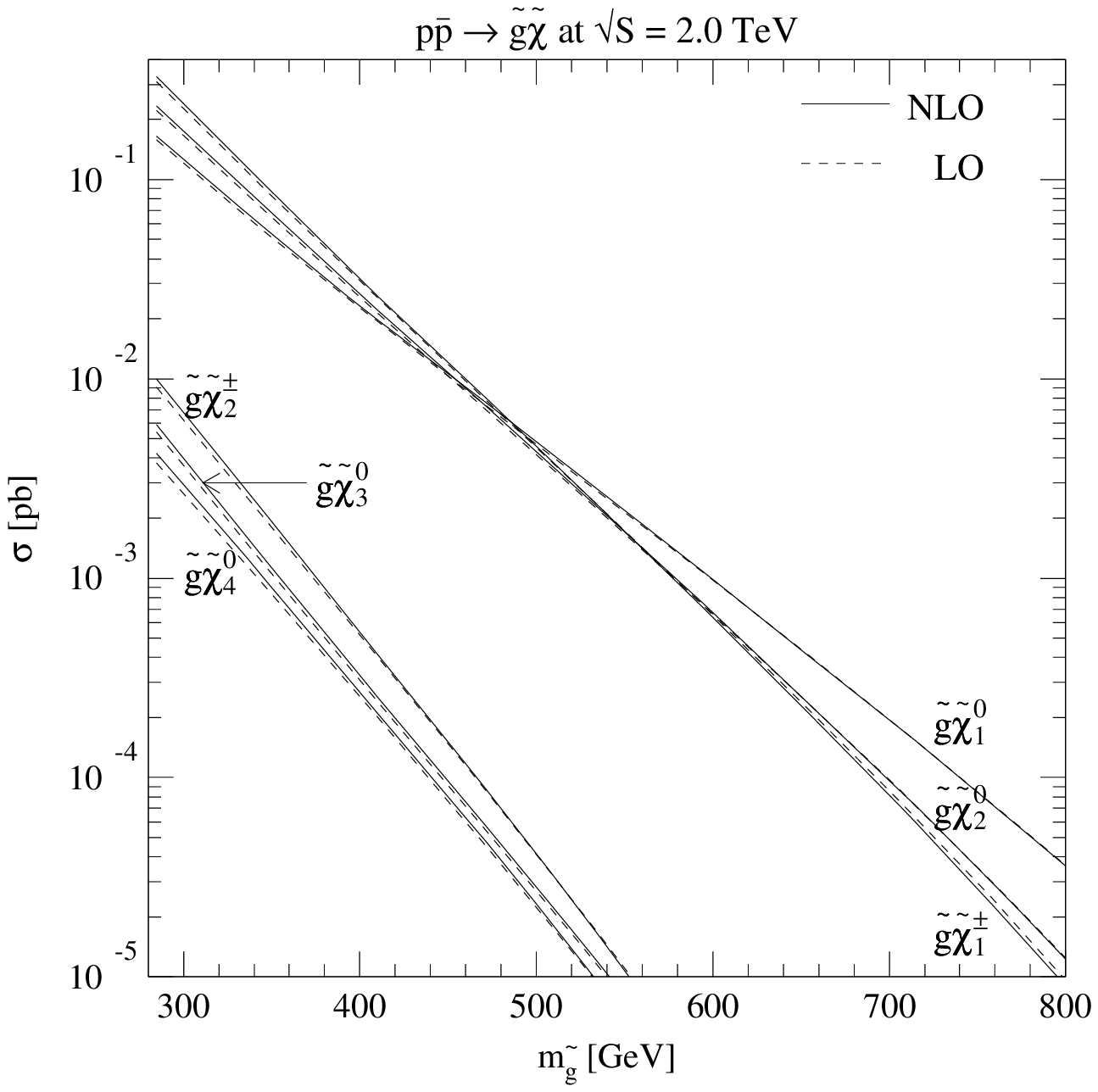,height=6.0in}}}
\vskip 0.1 in
\caption{Predicted total cross sections at the Tevatron for all six  
 $\gluino \gaugino$ channels in the SUGRA model as functions of the mass of 
 the gluino. See also Table I.}
\label{xsectev}
\end{figure}

\begin{figure}
\centerline{\hbox{
\psfig{figure=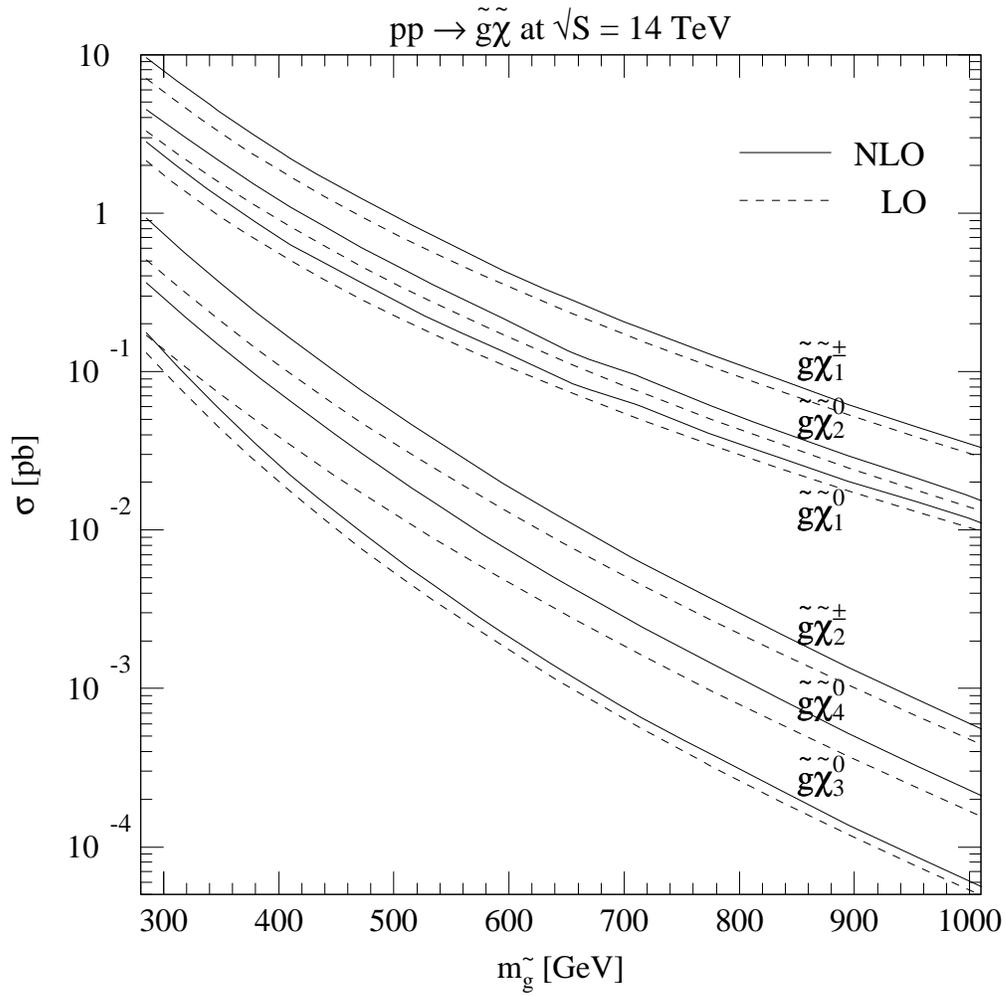,height=6.0in}}}
\vskip 0.1 in
\caption{Predicted total cross sections at LHC energies for all six  
 $\gluino \gaugino$ channels in the SUGRA model as functions of the mass of 
 the gluino. See also Table I.}
\label{xseclhc}
\end{figure}

For the SUGRA scenario, we present the total hadronic cross sections for the 
associated production of gluinos and gauginos at Run II of the Tevatron 
in Fig.~\ref{xsectev} and for the LHC in Fig.~\ref{xseclhc}.  We vary the 
SUGRA parameter $m_{1/2}$ from 100 to 400 GeV and keep the other SUGRA 
parameters fixed at the values described in Sec.~\ref{susybreak}. 
The squark mass runs from 250 GeV to 890 GeV in this region. The cross 
sections are presented as a function of the physical gluino mass $\mg$. 
The corresponding gaugino mass ranges are 31 to 163 GeV for $\neutralino_1$, 
62 to 317 GeV for $\neutralino_2$ and $\chargino_1$, 211 to 666 GeV for 
$\neutralino_3$, and 240 to 679 GeV for $\neutralino_4$ and $\chargino_2$. 
The chargino cross sections are summed over positive and negative charges. 
The renormalization and factorization scale $\mu$ is set equal to the 
average particle mass $m$. We truncate Fig.~\ref{xsectev} at a cross section 
of $10^{-5} {\rm pb}$ since the anticipated integrated luminosity at Run II 
is at most 30 ${\rm fb}^{-1}$.  
For the convenience of the reader, we provide numerical values of the cross 
sections in Table~I for a few selected points.

For small $\mg$ one might expect the largest cross section for the lightest 
gaugino, $\neutralino_1$.  However, its coupling is dominantly
of type $\tilde B$ and therefore smaller than the $\tilde{W}_3$-type coupling 
of $\neutralino_2$ which, in turn, has a larger cross section despite its larger
mass. The heavier gauginos $\neutralino_{3,4}$ and $\chargino_2$ are dominantly
Higgsino and are therefore suppressed by several orders of magnitude
with respect to the lighter gauginos
because of the light quark Yukawa couplings.  
At the LHC, the $\chargino_1$ cross 
section is dominant.  At small values of $\mg$, the LHC cross sections are 
a factor of about 30 greater than at the Tevatron, and at large $\mg$, the 
factor is about $10^4$.  

\begin{figure}
\centerline{\hbox{
\psfig{figure=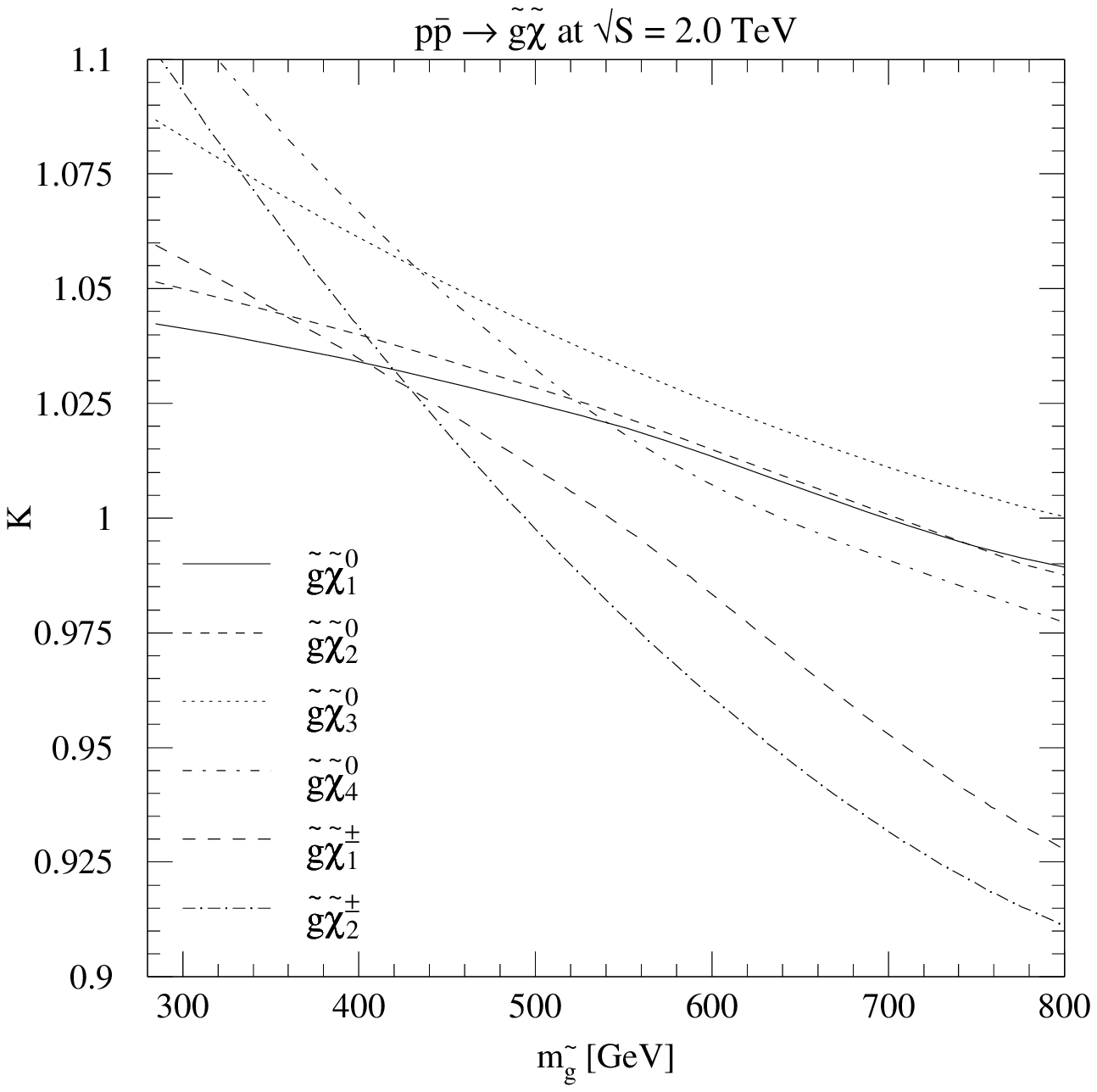,height=6.0in}}}
\vskip 0.1 in
 \caption{Calculated enhancements of the cross sections from NLO contributions 
at the Tevatron for all six SUGRA
$\gluino \gaugino$ channels as functions of the 
mass of the gluino.}
 \label{kfactev}
\end{figure}

\begin{figure}
\centerline{\hbox{
\psfig{figure=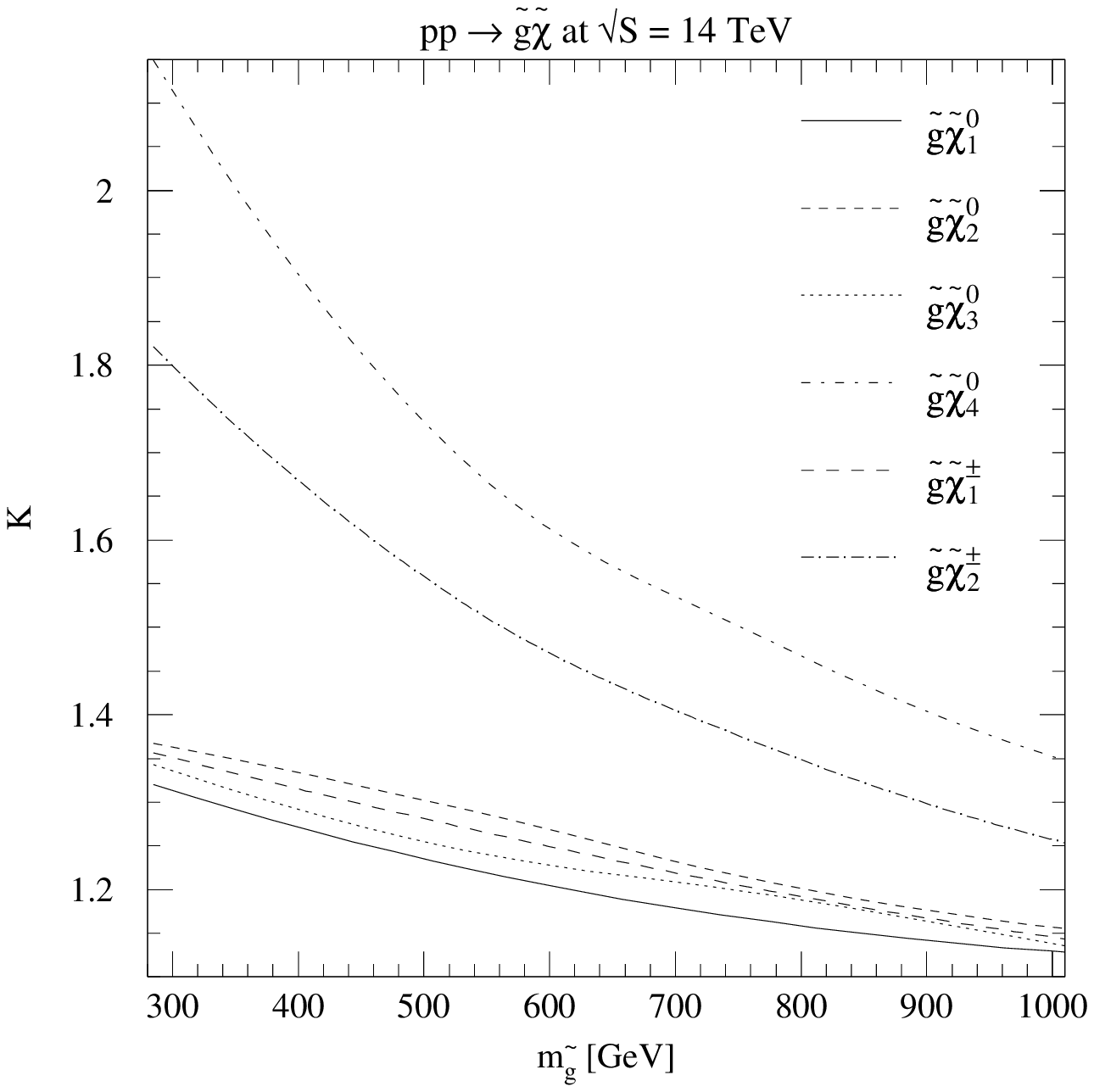,height=6.0in}}}
\vskip 0.1 in
 \caption{Calculated enhancements of the cross sections from NLO contributions 
at the LHC for all six SUGRA
$\gluino \gaugino$ channels as functions of the mass 
of the gluino.}
 \label{kfaclhc}
\end{figure}

Comparing the NLO predictions in Figs.~\ref{xsectev} and~\ref{xseclhc} 
(solid curves) with the LO predictions (dashed curves) we observe that the 
NLO corrections are all positive and substantially larger at the LHC than
at the Tevatron. At the Tevatron, some of the NLO predictions fall below the 
LO predictions at large mass, a point to which we return below.  The NLO 
enhancements are more evident in the ratio of the NLO over the
LO cross section ($K$ factors) shown in Figs.~\ref{kfactev} and~\ref{kfaclhc}. 
The $K$ factors are computed at the scale $\mu = m$. 
At the Tevatron, the NLO corrections amount to at most a 10\% increase in cross 
section.  At the LHC, they appear generally in the range of 20 to 40\% but can 
amount to a factor of 2 for $\neutralino_4$ and $\chargino_2$. 

The very modest size of the NLO enhancement at the Tevatron is somewhat  
expected from the behavior of the scaling functions, but it is also 
attributable partly to differences in the NLO and LO parton densities.  
Recalculating the $K$ factors with the CTEQ4 parametrization \cite{cteq4}, 
we find increases in $K$ by as much as 0.1 at the Tevatron energy.  The 
change from CTEQ4 to CTEQ5 is interesting.  The $u$ quark density at NLO 
decreased by 1 to 5\% and the $d$ quark density at NLO increased by up to 
10\% over the range 0.1 $< x <$ 0.6.  On the other hand, the LO $u$ quark 
density increased by about 1\% and the LO $d$ quark density by up to 20\%.
These purely parton density effects result in a net increase in the LO cross 
sections and decrease in the NLO cross sections, a drop in the calculated 
$K$ factors from CTEQ4 to CTEQ5.  

\begin{figure}
\centerline{\hbox{
\psfig{figure=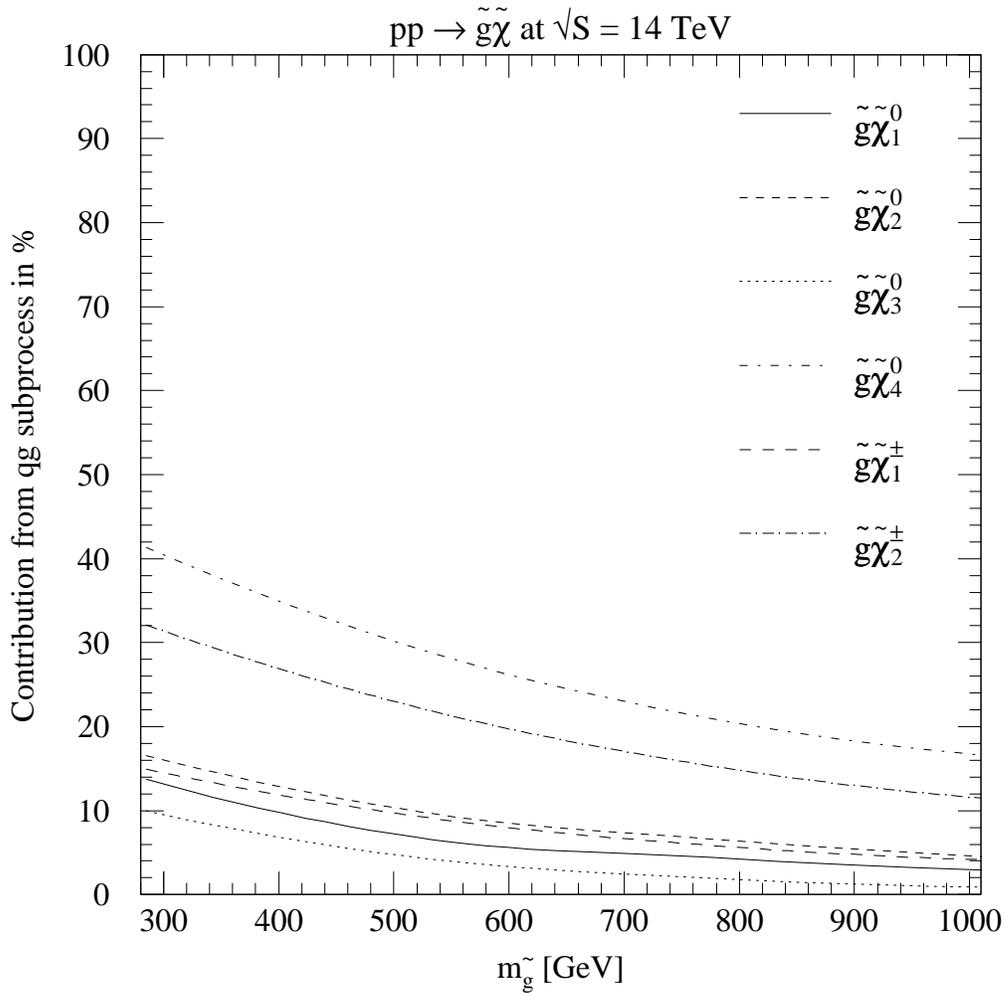,height=6.0in}}}
\vskip 0.1 in
 \caption{Fraction of the NLO cross sections at the LHC from the 
 $qg$ initial state for all six SUGRA $\gluino \gaugino$ channels as functions 
 of the mass of the gluino.}
 \label{qgfractlhc}
\end{figure}

The contribution from the $qg$ initial state at the energy of the Tevatron 
is insignificant (less than $10^{-3}$ of the total) for all six gaugino channels, 
but it is considerable at the energy of the LHC.  In Fig.~\ref{qgfractlhc} 
we show the fraction of the NLO cross section at the LHC attributed to the 
$qg$ initial state.  At the energy of the LHC, the contribution from the 
$q {\bar q}$ channel is less dominant, and the $q g$ contribution becomes 
significant owing to the large gluon density. A similar effect is seen in 
single top quark production~\cite{smithwillen} for comparable values of the 
average produced mass $m$.  

The large $K$ factor for $\neutralino_4$ and $\chargino_2$ production at 
the LHC is associated with a large contribution from the $gq$ channel at this 
energy. For these two channels, there is strong interference 
at LO between the t- and u-channel exchange diagrams that does not occur 
in the NLO quark emission graphs.  

\begin{figure}
\centerline{\hbox{
\psfig{figure=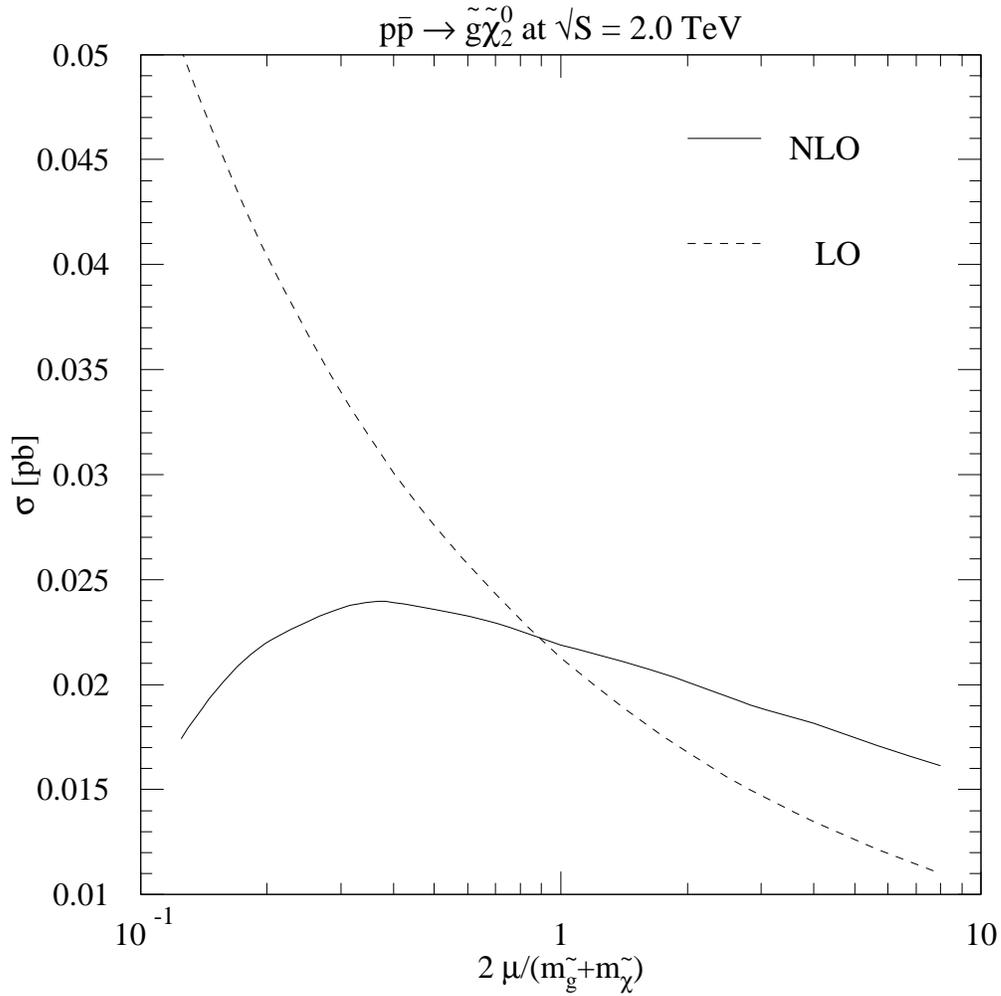,height=6.0in}}}
\vskip 0.1 in
 \caption{Dependence of the predicted NLO and LO total cross sections at the 
  Tevatron on the renormalization and factorization scale.  We show the case 
  $\gluino \neutralino_2$ production in the SUGRA model, 
  with $m_{\gluino} = 410 $ GeV and $m_{\neutralino_2} = 104 $ GeV.}
\label{mutev}
\end{figure}

\begin{figure}
\centerline{\hbox{
\psfig{figure=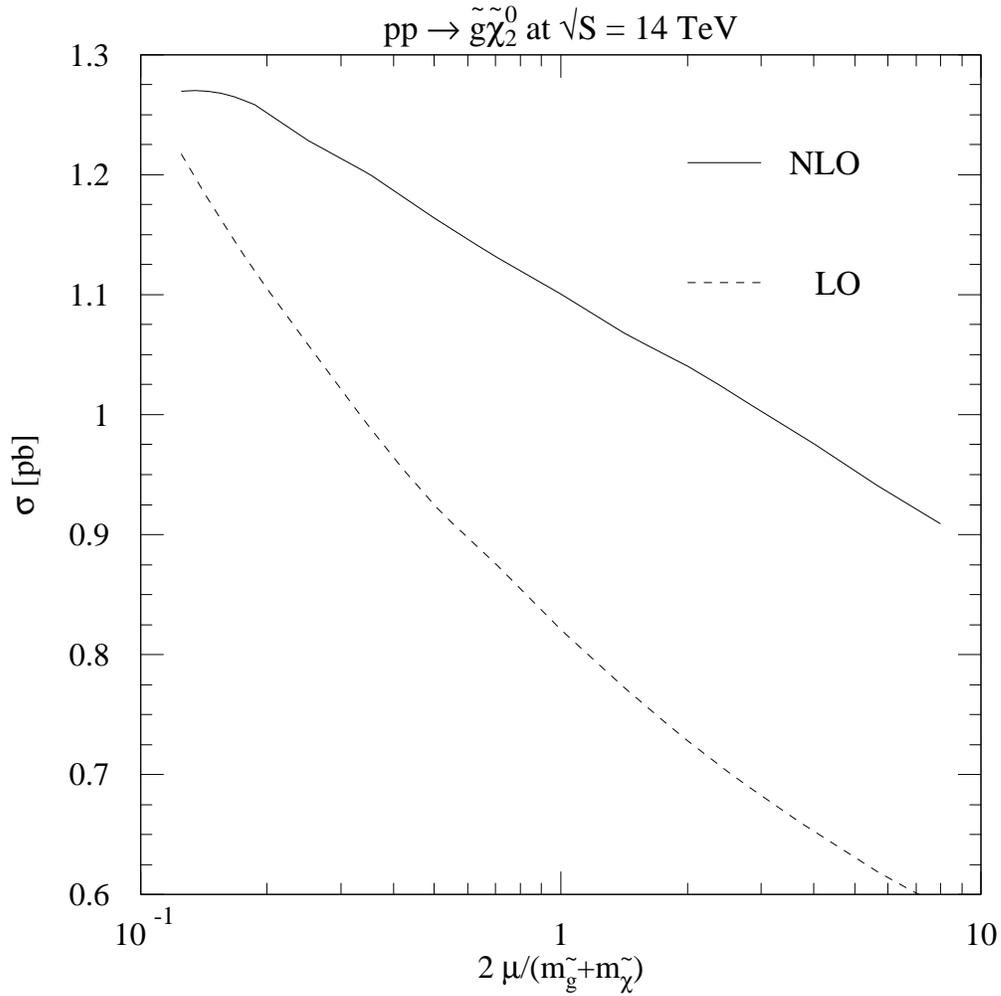,height=6.0in}}}
\vskip 0.1 in
 \caption{Dependence of the predicted NLO and LO total cross sections at the 
  LHC on the renormalization and factorization scale. We select 
  $\gluino \neutralino_2$ production in the SUGRA model, 
  with $m_{\gluino} = 410 $ GeV and $m_{\neutralino_2} = 104 $ GeV.}
\label{mulhc}
\end{figure}

The dependence of the predicted cross sections on the renormalization and 
factorization scale is reduced considerably at next-to-leading order.  As 
an example, in Fig.~\ref{mutev} we show the scale dependences for the 
$\neutralino_2$ channel at the Tevatron.  All six channels show similar 
behavior at the Tevatron.  Cross sections vary by $\pm 23\%$ at LO but 
only by $\pm 8\%$ at NLO as the scale ratio 
$\mu/m$ is varied over the range 0.5 to 2.0, a substantial improvement in 
reliability.  Here $m$ is the average mass for the default scenario.  At the 
Tevatron, the NLO and LO cross sections intersect at scale ratio near unity.  
In Fig.~\ref{mulhc} we show the $\mu$ dependences for the 
$\neutralino_2$ channel at the LHC.  Again, all six channels show similar 
behavior.  Cross sections vary by $\pm 12\%$ at LO but 
only by $\pm 4.5\%$ at NLO in the region $0.5 < \mu/m < 2.0$. However, unlike 
the situation at the Tevatron, owing to the important contribution of the $qg$ 
channel at the energy of the LHC, the NLO and LO cross sections do not 
intersect at scale ratio near unity.  Instead, if they intersect at all, the 
crossing point is at a very low scale.  

Uncertainties in the cross section from parton density variation may be 
estimated roughly if we compare NLO results obtained with CTEQ4M and CTEQ5M.  
For $\neutralino_2$ production at the Tevatron in our default SUGRA scenario, 
we compute cross sections of 0.0244 pb (CTEQ4M) and 0.0219 pb (CTEQ5M), 
a $12\%$ difference.  At the LHC, the cross sections are 1.138 pb (CTEQ4M), 
and 1.095 pb (CTEQ5M), a $4\%$ difference.  

\subsubsection{Light gluino model}
\label{lightcross}

\begin{figure}
\centerline{\hbox{
\psfig{figure=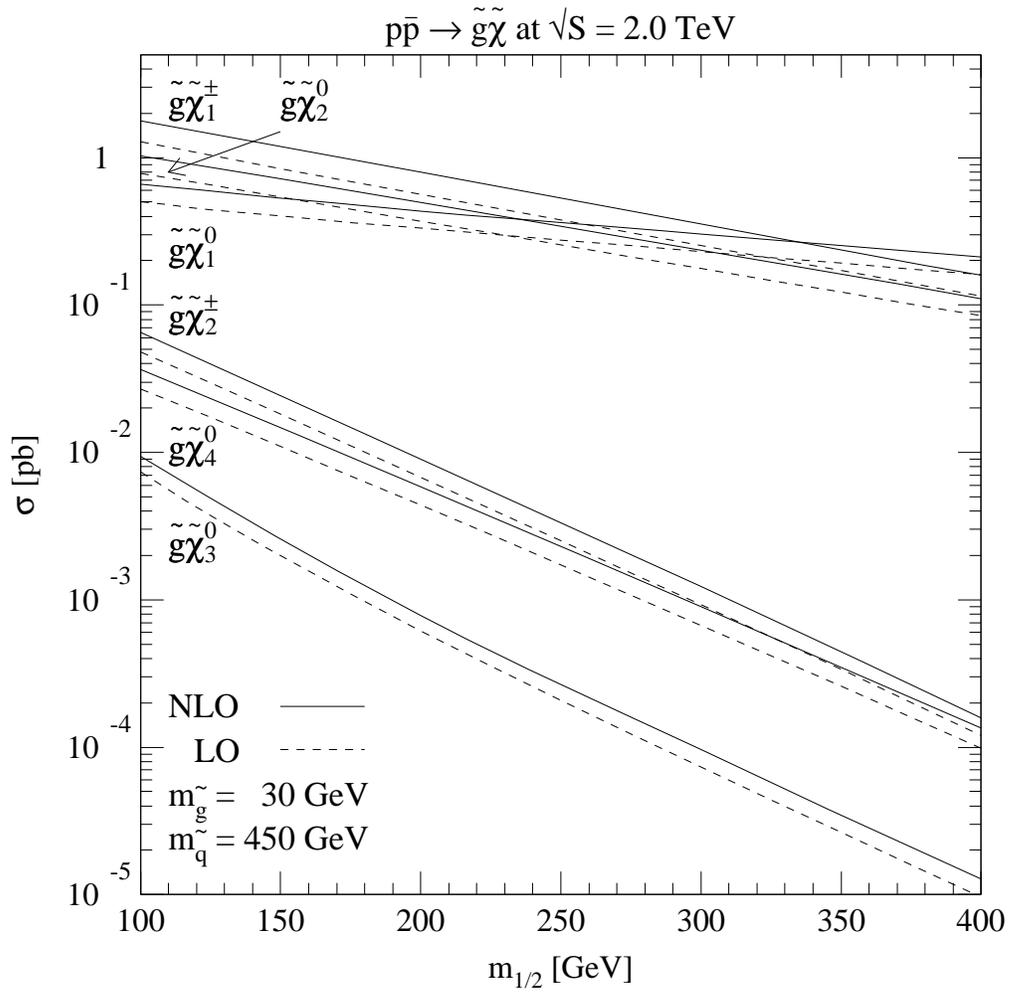,height=6.0in}}}
\vskip 0.1 in
\caption{Predicted total cross sections at the Tevatron for all six  
 $\gluino \gaugino$ channels for a gluino with mass 30 GeV as functions of 
 $m_{1/2}$.}
\label{xsectevlight}
\end{figure}

\begin{figure}
\centerline{\hbox{
\psfig{figure=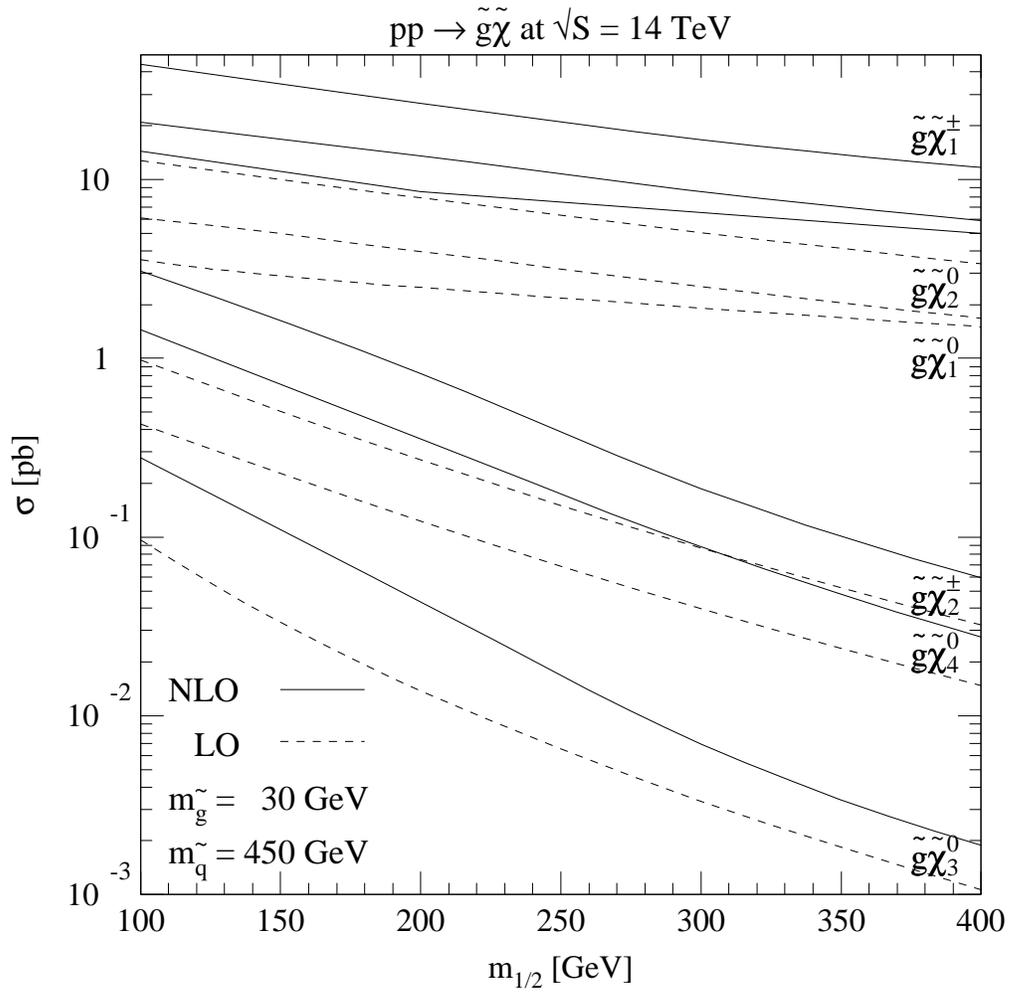,height=6.0in}}}
\vskip 0.1 in
\caption{Predicted total cross sections at LHC energies for all six  
 $\gluino \gaugino$ channels for a gluino with mass 30 GeV as functions of 
 $m_{1/2}$.}
\label{xseclhclight}
\end{figure}

For the light gluino model, we present the total hadronic cross sections 
for the associated production of gluinos and gauginos at Run II of the Tevatron 
in Fig.~\ref{xsectevlight} and for the LHC in Fig.~\ref{xseclhclight}.  As 
mentioned in Sec.~II~A, we fix $m_{\gluino} =$ 30 GeV and $m_{\tilde q} =$ 
450 GeV.  We display the cross sections as functions of the mass of the 
common GUT-scale fermion mass $m_{1/2}$.  As we vary $m_{1/2}$ from 100 to 400 
GeV, the gaugino masses range from 31 to 163 GeV for $\neutralino_1$, 
62 to 317 GeV for $\neutralino_2$ and $\chargino_1$, 211 to 666 GeV for 
$\neutralino_3$, and 240 to 679 GeV for $\neutralino_4$ and $\chargino_2$.  
It is worth noting that the coupling strengths also vary with $m_{1/2}$.  
The chargino cross sections are summed over positive and negative charges. 
The renormalization and factorization scale $\mu$ is set equal to the average 
of the masses of the gaugino and gluino in each of the channels.  

Evident in Figs.~\ref{xsectevlight} and~\ref{xseclhclight} is that the 
cross sections do not depend strongly on the gaugino masses.  The 
values of $m_{1/2}$ in these figures extend below the value $\simeq 150$ GeV 
believed excluded since LEP data~\cite{lepdata} set a lower bound on the mass 
of $\chargino_1$ of about 100 GeV.  Nevertheless, even above $m_{1/2} = 
150$~GeV, the Tevatron cross sections for the three lighter gluino 
channels are predicted to be 
in the range of 0.1 to 0.5 $\rm{pb}$.  The cross sections would be increased 
if $m_{\tilde q}$ were reduced from the value 450 GeV that we use.  

The relatively large cross sections suggest that associated production is a good 
channel for discovery of a light gluino at the Tevatron, for closing the window 
on this possibility, and/or for setting limits on light gaugino masses.  We 
remark in this connection that the usual searches for a light gluino LSP 
begin with the assumption of pair production of gluinos.  In this situation, 
the dominant background is QCD pair production of hadronic jets.  Hard cuts 
on transverse momentum must be made to reduce this background to tolerable 
levels.  The cuts, in turn, mitigate against gluinos of modest mass.  By 
contrast, if light gluinos are produced in association with gauginos, one 
can search for light gluino monojets accompanied by leptons and/or missing 
transverse energy from gaugino decays.    

\begin{figure}
\centerline{\hbox{
\psfig{figure=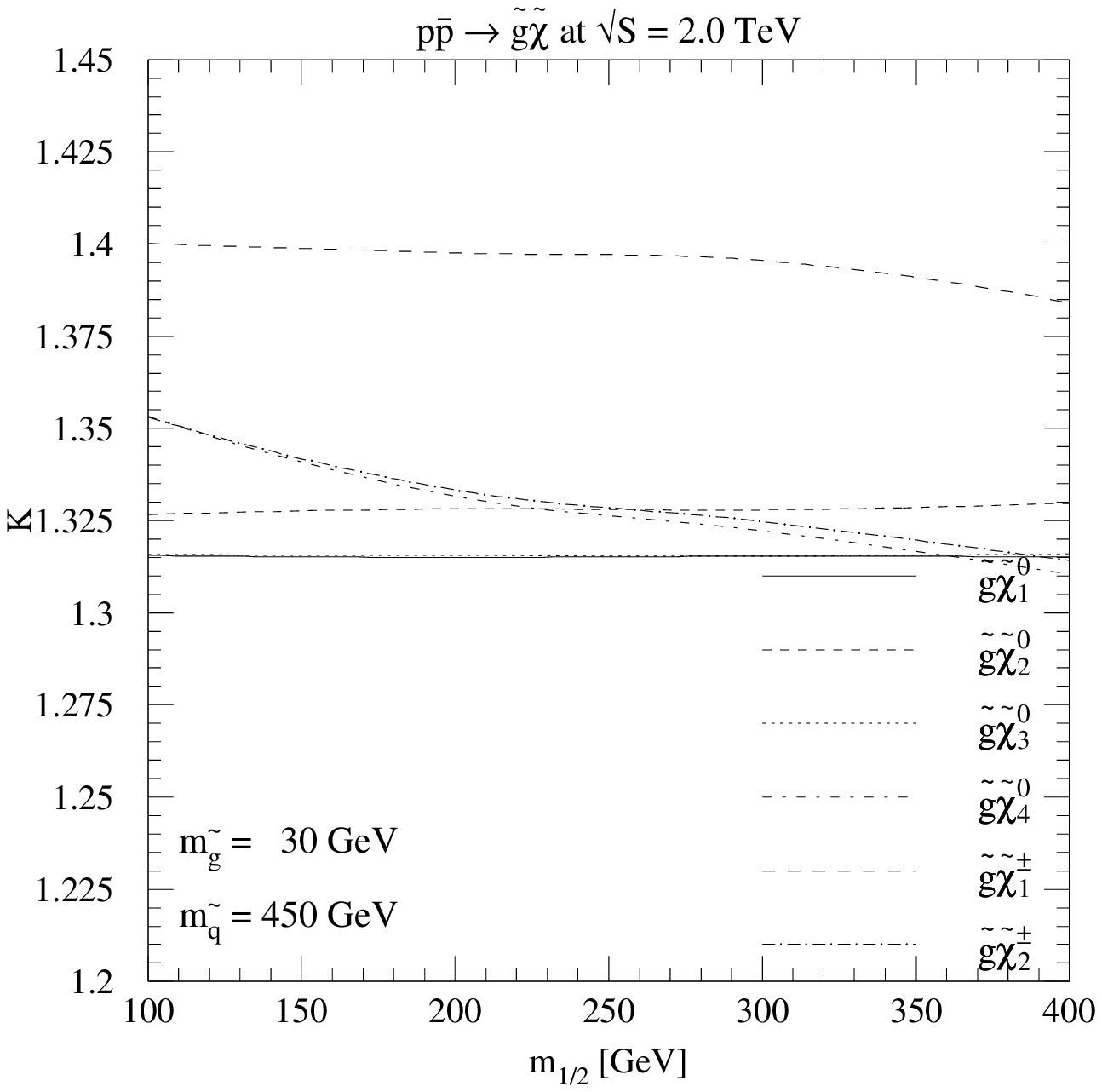,height=6.0in}}}
\vskip 0.1 in
 \caption{Calculated enhancements of the cross sections from NLO contributions 
at the Tevatron for all six $\gluino \gaugino$ channels with a light
gluino of mass 30 GeV as functions of $m_{1/2}$.}
 \label{kfactevlite}
\end{figure}

\begin{figure}
\centerline{\hbox{
\psfig{figure=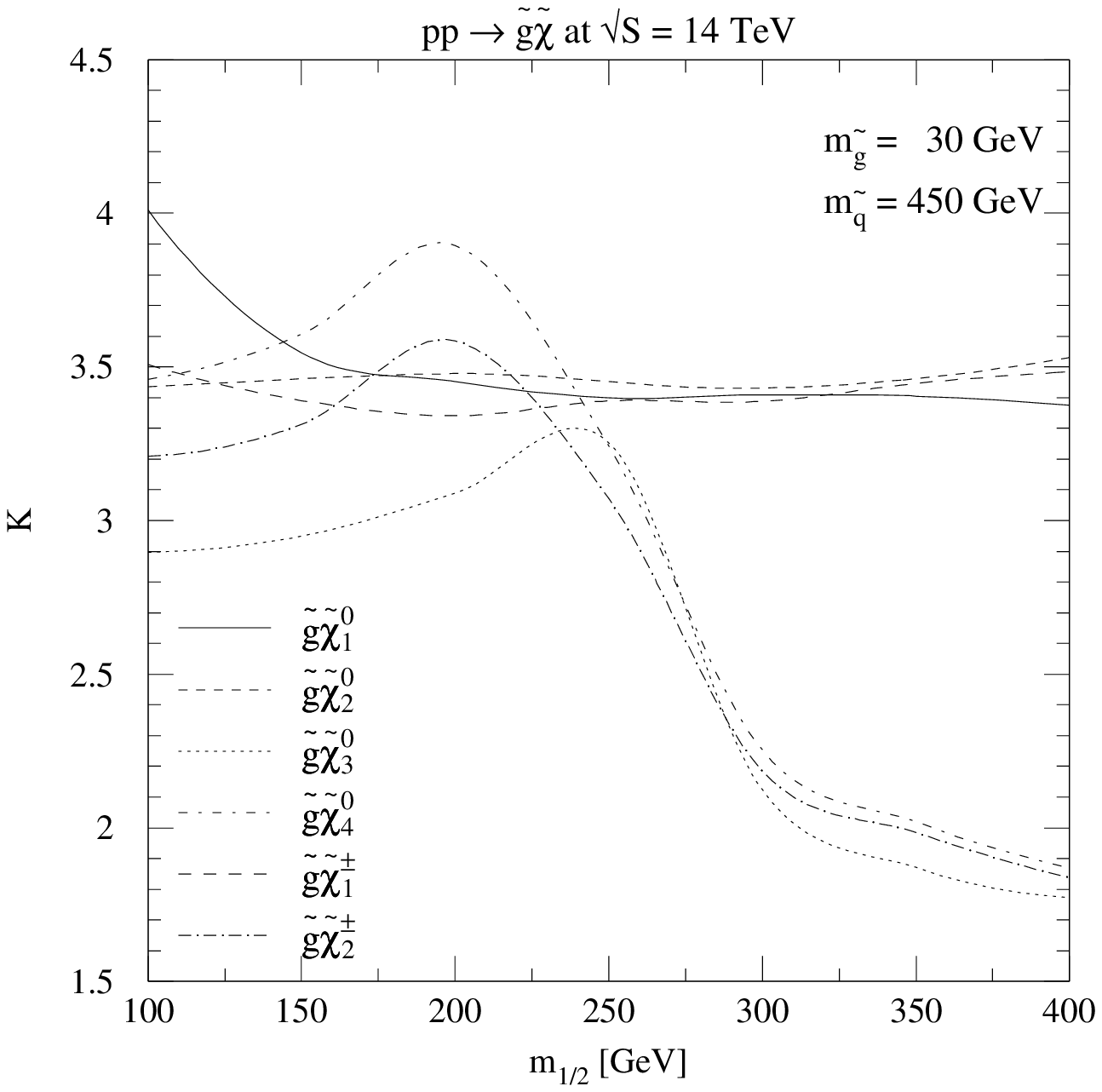,height=6.0in}}}
\vskip 0.1 in
 \caption{Calculated enhancements of the cross sections from NLO contributions 
at the LHC for all six $\gluino \gaugino$ channels with a light
gluino of mass 30 GeV as functions of $m_{1/2}$.}
 \label{kfaclhclite}
\end{figure}

The $K$ factors for the light gluino case are shown in Figs.~\ref{kfactevlite} 
and~\ref{kfaclhclite}.  At the Tevatron, the NLO corrections amount to a 
$30\%$ to $40\%$ increase in cross section.  At the LHC, they are generally 
in the range of a factor of 2 to 3.5.  The large $K$ factors owe their origins 
to the important role of the $gq$ channel.  The gluon parton density is very 
large at small values of $x$.  Contributions from the $gq$ production channel 
are more intense in the light gluino case than in the SUGRA case where average 
produced masses and, thus, typical values of $x$ are larger.  At the energy of 
the Tevatron, the $gq$ channel accounts for more than 20\% of the NLO cross 
section in the $\neutralino_1$ and $\chargino_1$ channels at small values of 
$m_{1/2}$ and more than 10\% at large $m_{1/2}$.  At the energy of the LHC, 
the fraction exceeds 60\% in the $\neutralino_1$, $\neutralino_2$, and 
$\chargino_1$ channels for the entire range of $m_{1/2}$.  It hovers above 
50\% in the $\neutralino_3$, $\neutralino_4$, and $\chargino_2$ channels at 
small values of $m_{1/2}$ and near 20\% at large $m_{1/2}$.  

Motivated by the curious behavior of the $K$ factors for the three heavier 
gaugino channels at the LHC energy, we examined the change in the scaling 
functions for the $\neutralino_4$ case as $m_{1/2}$ is varied.  In the 
$q {\bar q}$ channel, the net (virtual plus soft plus hard) NLO contribution 
is positive for all $\eta$, and, relative to the Born contribution, its 
magnitude grows gradually with $m_{1/2}$.  The $q {\bar q}$ channel 
accounts for a slightly increasing component of $K$, hovering about 1.5.  
On the other hand, the contribution from the $gq$ channel changes markedly 
as $m_{1/2}$ increases.  Below about $m_{1/2} = 200$~GeV, its scaling 
function is large, with significant support below $\eta = 1$, where the 
gluon parton density is large.  As $m_{1/2}$ increases above 200 GeV, 
the $gq$ scaling function decreases in magnitude.  The $gq$ channel supplies 
a component of $K$ that increases slightly from about 2 to 2.5 as 
$m_{1/2}$ increases to 200 GeV and then falls gradually to below 0.5 
at $m_{1/2} =400$~GeV.  

We attribute the sharp decrease of the $K$ factor for the three 
heavier gaugino channnels to the role of 
on-shell intermediate squark decay into a gaugino, discussed in 
Sec.~\ref{qemiss}.  In the light gluino model, as $m_{1/2}$ is varied from 
100 to 400 GeV, the masses of the lighter gauginos ($\neutralino_1$, 
$\neutralino_2$, and $\chargino_1$) remain less than the squark mass so 
that the on-shell intermediate squark decay into a gaugino is active over 
the whole range of $m_{1/2}$ for these light channels.  The situation changes 
for the heavier gauginos ($\neutralino_3$, $\neutralino_4$, and 
$\chargino_2$).  For small $m_{1/2}$ their masses are below the squark mass, 
and the on-shell decay is active.  However, above roughly 
$m_{1/2} =250$~GeV, the masses of the heavier gauginos exceed the squark 
mass and the on-shell possibility closes.  


\subsection{Differential Cross Sections}
\label{difcross}
\begin{figure}
\centerline{\hbox{
\psfig{figure=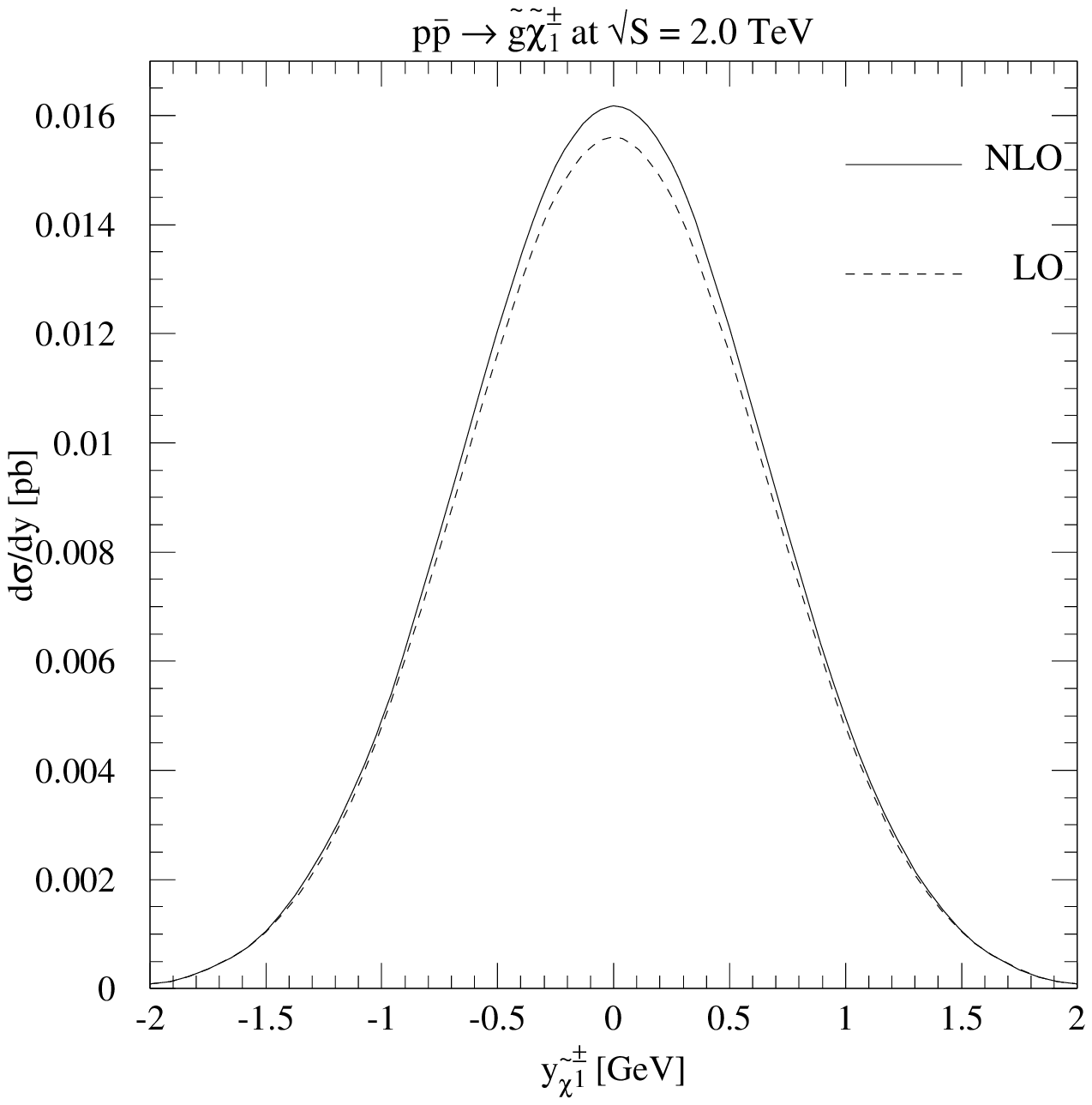,height=6.0in}}}
\vskip 0.1 in
 \caption{Differential cross section in rapidity $d\sigma/dy$ for the 
  production of $\chargino_1$ with mass 101 GeV in association with a 
  $\gluino$ of mass 410 GeV at the Tevatron.}
\label{ytev}
\end{figure}

\begin{figure}
\centerline{\hbox{
\psfig{figure=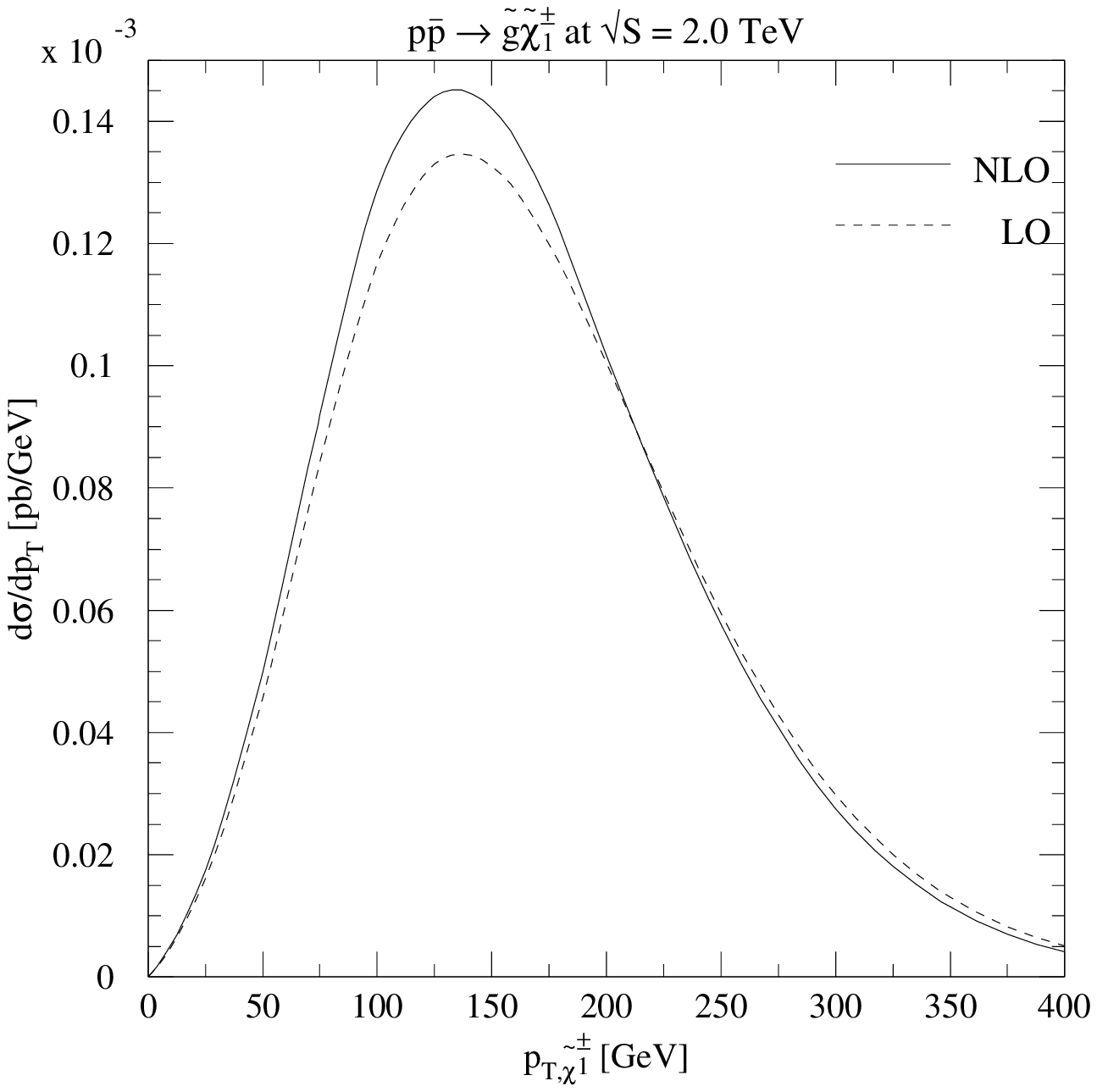,height=6.0in}}}
\vskip 0.1 in
 \caption{Differential cross section in transverse momentum $d\sigma/dp_T$ 
  for the production of $\chargino_1$ with mass 101 GeV in association with 
  a $\gluino$ of mass 410 GeV at the Tevatron.}
\label{pttev}
\end{figure}

In Figs.~\ref{ytev} and~\ref{pttev}, we display the differential cross sections 
in the rapidity $y$ and in the transverse momentum $p_T$ of $\chargino_1$ at the 
energy of the Tevatron collider.  Here the $\chargino_1$ and $\tilde g$ masses 
are set at their default values in the SUGRA scenario, 101 and 410 GeV, 
respectively.  The rapidity distribution is obtained after 
integration over all $p_T$, and the $p_T$ distribution after integration over 
all $y$. More restrictive selections could be made, but until experimental 
conditions are better known, any such restrictions (cuts) would be unmotivated.  
The NLO (solid) curve in Fig.~\ref{ytev} shows a modest enhancement in the 
rapidity distribution in the central region with respect to the LO (dashed) 
curve, but the shape of the distribution is unchanged qualitatively.  
The $p_T$ distributions in Fig.~\ref{pttev} show that the NLO contribution tends 
to shift the distribution to somewhat smaller values of $p_T$.  Since the 
contribution of the $gq$ initial state is very small at the energy of 
the Tevatron, the shift in the $p_T$ distribution is associated with 
next-to-leading order corrections in the dominant $q {\bar q}$ initial state.  

The features of the $y$ distributions are qualitatively similar 
for all gauginos except for the expected and systematic narrowing of the $y$
distribution with increasing gaugino mass.  We show one example representative 
of the full set. The $p_T$ distributions for the different gauginos are also 
qualitatively similar except that the maximum in the distribution 
moves to larger $p_T$ as the gaugino mass is increased.  The location of the 
peak is specified roughly by $m/2$, where $m$ is the average 
mass of the produced sparticles.  The one exception is $\neutralino_3$ 
production. In this case, the peak occurs at a smaller value (about 100 GeV 
for the default masses), an effect correlated with the fact that the location 
of the peak in the scaling function occurs at a smaller value of $\eta$.  
Interference effects enhance the cross section at small $p_T$ for 
$\neutralino_3$ production.  

\begin{figure}
\centerline{\hbox{
\psfig{figure=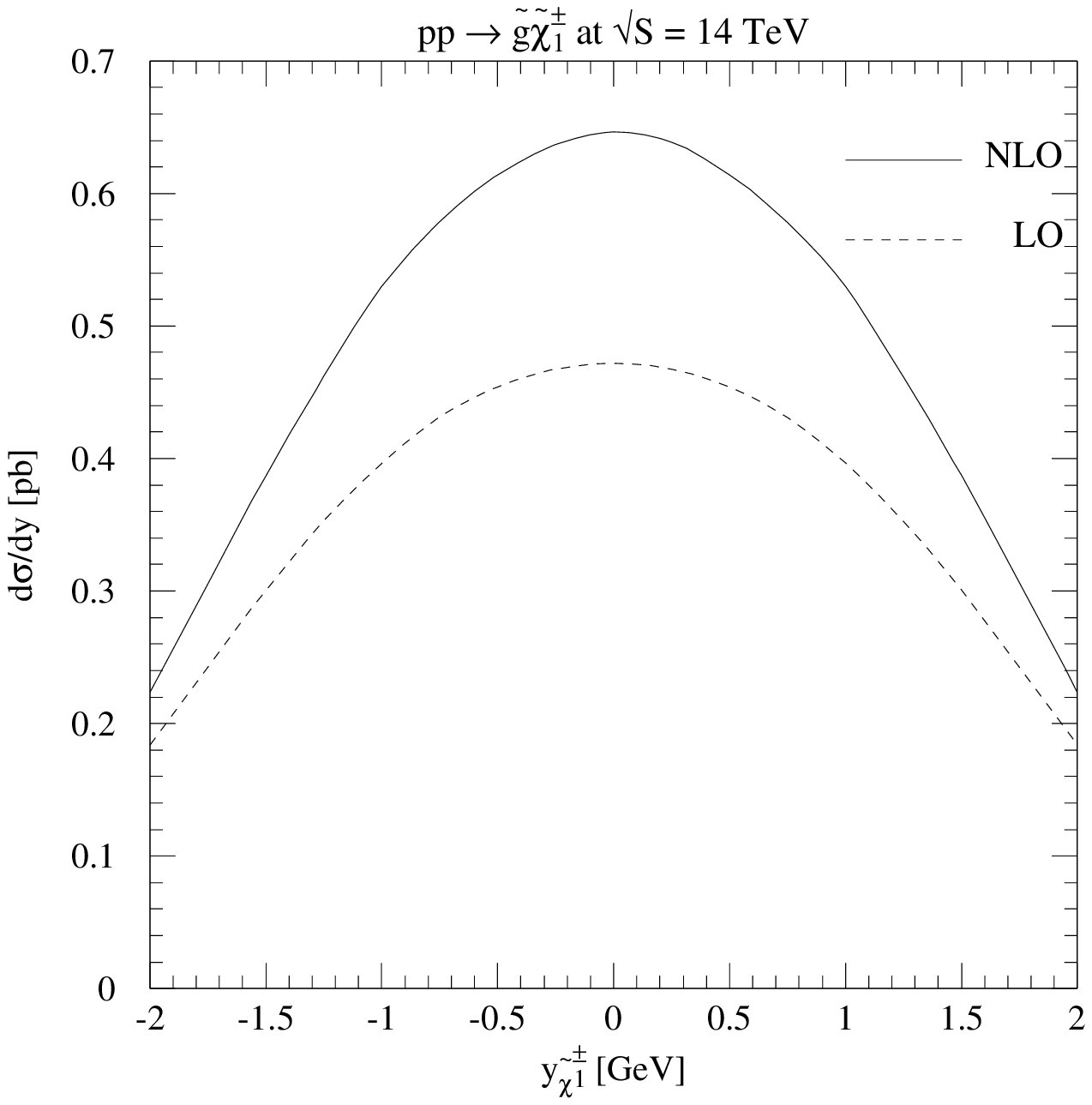,height=6.0in}}}
\vskip 0.1 in
 \caption{Differential cross section in rapidity $d\sigma/dy$ for the 
  production of $\chargino_1$ with mass 101 GeV in association with a 
  $\gluino$ of mass 410 at the LHC.}
\label{ylhc}
\end{figure}

\begin{figure}
\centerline{\hbox{
\psfig{figure=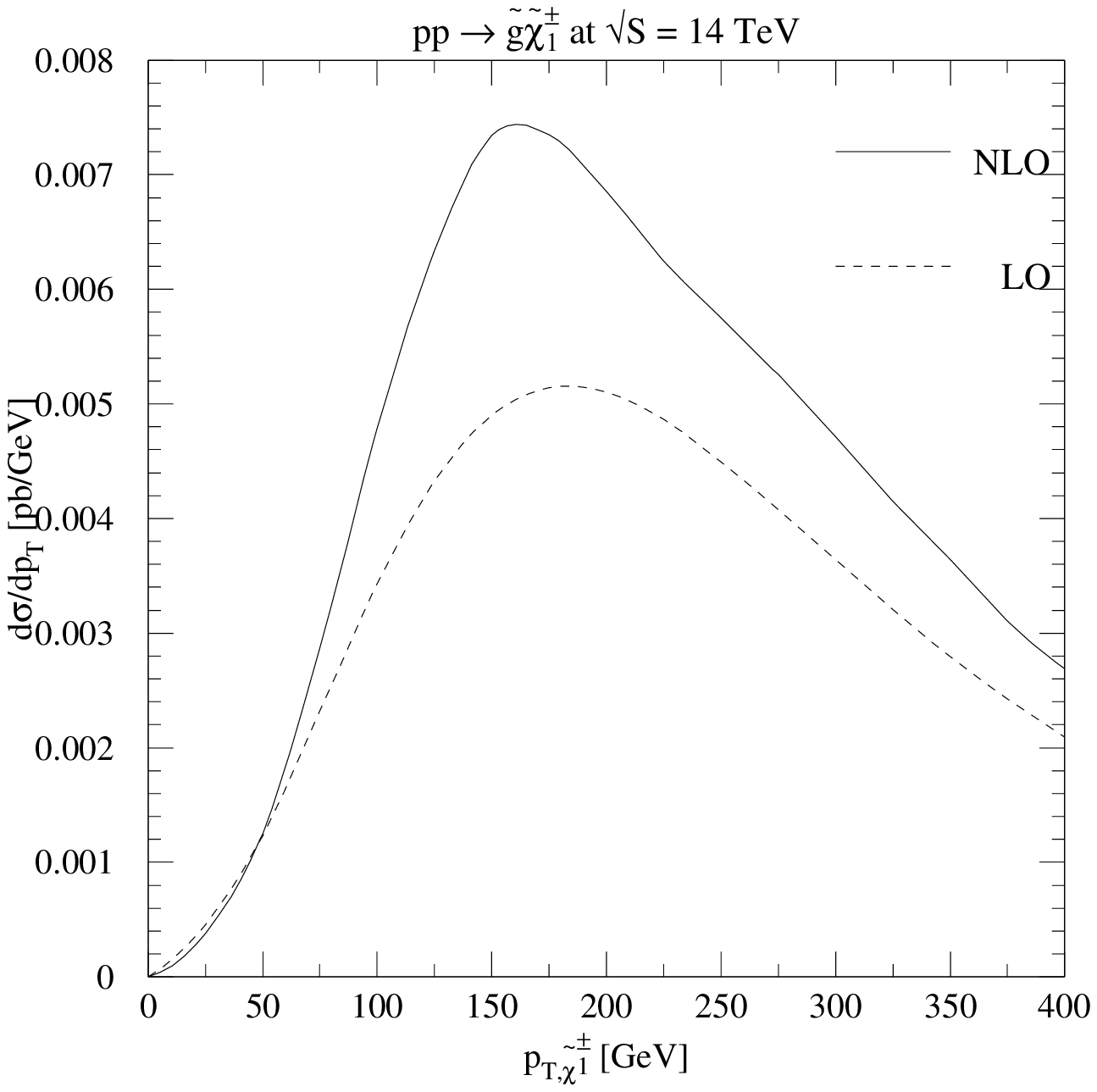,height=6.0in}}}
\vskip 0.1 in
 \caption{Differential cross section in transverse momentum $d\sigma/dp_T$ 
  for the production of $\chargino_1$ with mass 101 GeV in association with 
  a $\gluino$ of mass 410 GeV at the LHC.}
\label{ptlhc}
\end{figure}

\begin{figure}
\centerline{\hbox{
\psfig{figure=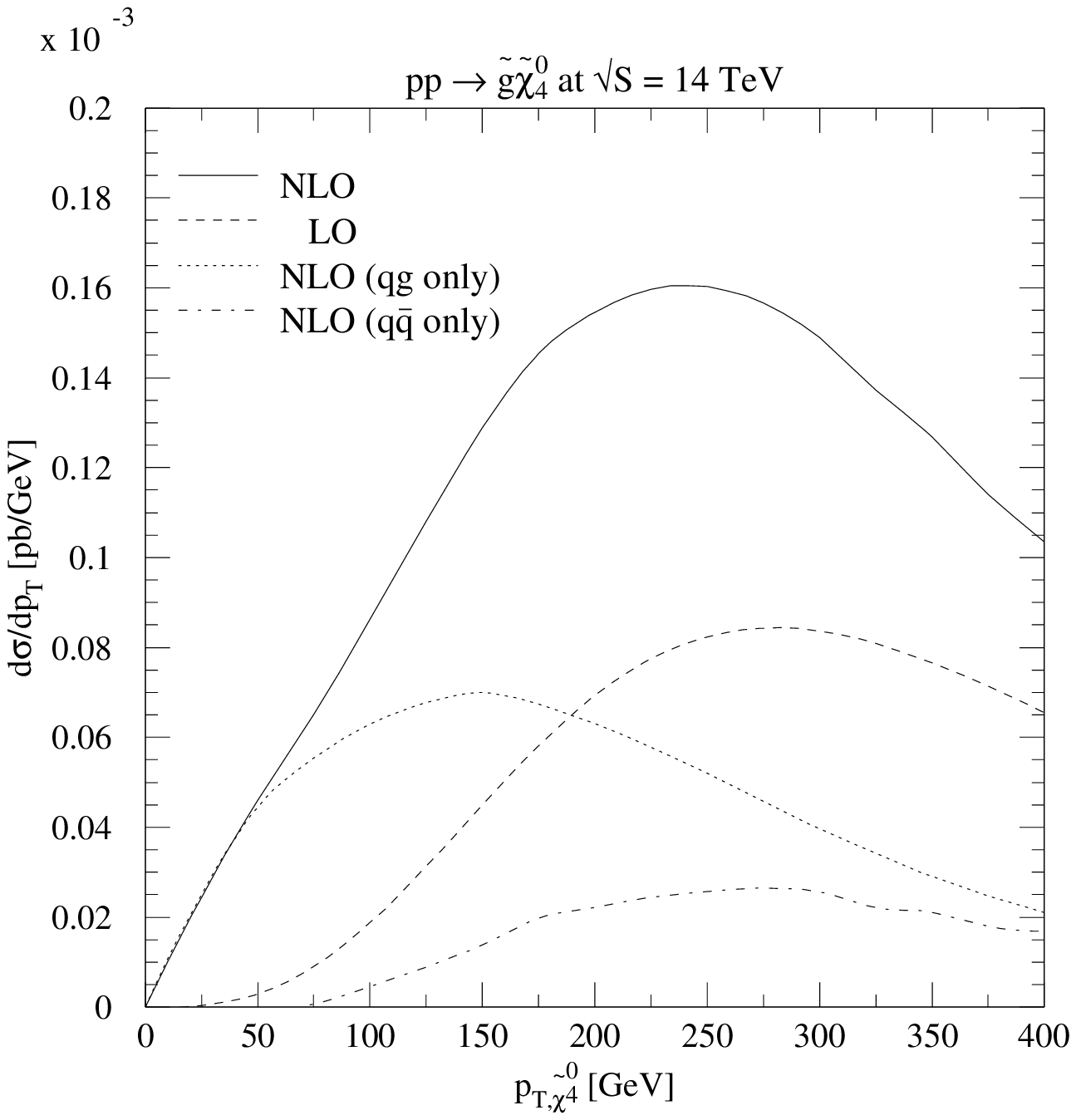,height=6.0in}}}
\vskip 0.1 in
 \caption{Differential cross section in transverse momentum $d\sigma/dp_T$ 
  for the production of $\neutralino_4$ with mass 309 GeV in association 
  with a $\gluino$ of mass 410 GeV at the LHC.  The solid line is the full 
  NLO cross section.  The dashed curve shows the LO Born cross section.  The 
  dotted line shows the contribution of the $qg$ initial state subprocess.  
  The dot-dashed curve is the difference between the full NLO result and the 
  sum of the Born and $qg$ contributions.}
\label{ptlhc2}
\end{figure}

We show differential cross sections in $y$ and $p_T$ for $\chargino_1$ at the 
energy of the LHC collider in Figs.~\ref{ylhc} and~\ref{ptlhc}, and in 
Fig.~\ref{ptlhc2} we present the $p_T$ distribution for $\neutralino_4$.  
The rapidity and transverse momentum distributions are much broader than at 
the Tevatron.  As at the energy of the Tevatron, the features of the 
$y$ distributions are qualitatively similar for all gauginos except for the 
expected and systematic narrowing of the $y$ distribution with increasing 
gaugino mass. The NLO contributions enhance the $y$ distributions at all $y$.  

At the energy of the LHC, the $p_T$ spectra are qualitatively similar for 
the relatively light $\neutralino_1$, $\neutralino_2$, and $\chargino_1$ 
states, illustrated by the $\chargino_1$ case in Fig.~\ref{ptlhc}. The 
enhancement factor $K$ is near unity at small $p_T$ but becomes sizeable 
at larger $p_T$.  For $\neutralino_4$ and $\chargino_2$, the 
$p_T$ spectra are altered significantly by NLO contributions.  As 
illustrated in Fig.~\ref{ptlhc2}, the NLO contribution associated with 
the important $qg$ channel fills in the distribution at small $p_T$   
and softens the overall $p_T$ distribution in these two cases.  
At LHC energies, it is therefore not a good approximation to assume that the 
enhancement factor $K$ is roughly independent of $p_T$.


\section{Conclusions}
\label{conclusions}

In this paper we report a complete next-to-leading order analysis of
the associated production of gauginos and gluinos at hadron colliders.
If supersymmetry exists at the electroweak scale, the cross section for
this process is expected to be observable at the Fermilab Tevatron
and/or the CERN LHC. It is enhanced by the large color charge of the gluino
and the relatively small mass of the light gauginos in many SUSY models.
Associated production represents a chance to study in detail the
parameters of the soft SUSY-breaking Lagrangian.  The rates are proportional 
to the phases of the gaugino and gluino masses, and to the mixings in the 
squark and chargino/neutralino sectors.  Thus, in combination with other 
channels, associated production could allow one to measure some or all of 
these quantities.

The physical gluino and gaugino masses that we use, as well as the gaugino 
mixing matrices, are based on four popular SUSY breaking models plus a fifth  
scenario in which the gluino mass is relatively light.
Because the LO cross sections in gauge-mediated, gaugino-mediated, and 
anomaly-mediated supersymmetry breaking models are not too 
dissimilar from those of the SUGRA case at Tevatron energies, we focus our 
NLO work on the SUGRA model and on a model with a light gluino LSP, with 
$m_{\gluino} =$ 30 GeV.  

In the SUGRA model, the largest cross sections at the 
Fermilab Tevatron energy are those 
for $\neutralino_2$, enhanced by its $\tilde{W}_3$-like coupling with respect 
to the $\tilde B$-like $\neutralino_1$, and the $\chargino_1$, which is about 
equal in mass with the $\neutralino_2$. The NLO corrections to associated production 
are generally positive, but they can be modest in size, ranging in the SUGRA 
model from a few percent at the energy of the Tevatron to 100\% at 
the energy of the LHC, depending on the sparticle masses.  In the light-gluino 
case, NLO contributions increase the cross section by factors of 1.3 to 1.4 at 
the energy of the Tevatron and by factors of 2 to 3.5 at the energy of the 
LHC.  The large $K$ factors owe their origins to the important role of the $gq$ 
channel that enters first at NLO.  

Owing to the NLO enhancements, collider searches for signatures of associated 
production will generally discover or exclude sparticles with 
masses larger than one would estimate based on LO production rates alone.  
More significant from the viewpoint of reliability, the renormalization and 
factorization scale dependence of the cross sections is reduced by a 
factor of more than two when NLO contributions are included. 

At Run II of the Fermilab Tevatron, for an integrated luminosity of 
2 $\rm{fb}^{-1}$, we expect that 10 or more events could be produced in 
each of the lighter gaugino channels of the SUGRA model, 
$\gluino \neutralino_1$, 
$\gluino \neutralino_2$, and $\gluino \chargino_1$, provided that the 
gluino mass $m_{\gluino}$ is less than 450 GeV.  The cross sections 
for the three heavier gaugino channels, $\gluino \neutralino_3$, 
$\gluino \neutralino_4$, and $\gluino \chargino_2$, are smaller by 
an order of magnitude or more than those of the lighter gaugino 
channels.  In the light gluino model, more than 100
events could be produced in the three lighter gaugino channels provided 
that the common GUT-scale fermion mass $m_{1/2}$ is less than 400 GeV, 
and as many as 10 events in the three heavier gaugino channels as long 
as $m_{1/2}$ is less than 200 GeV.  At the higher energy and luminosity 
of the LHC, at least a few events should be produced in every channel in 
the SUGRA model and many more in the light gluino model.    

The shapes of the rapidity distributions of the gauginos are not altered 
appreciably by NLO contributions, but the locations of the maximum cross 
section in transverse momentum ($p_T$) are shifted to smaller values 
by NLO contributions.  At LHC energies where the contribution of the 
$qg$ initial state is important, modifications of the $p_T$ spectra 
can be pronounced.  

The relatively large cross sections suggest that associated production is a 
good channel for discovery of a light gluino at the Tevatron, for closing the 
window on this possibility, and/or for setting limits on light gaugino masses.  
The usual searches for a light gluino LSP are based on the assumption 
that gluinos are produced in pairs.  In this situation, the dominant background 
is QCD production of hadronic jets.  Hard cuts 
on transverse momentum must be made to reduce this background to tolerable 
levels.  The cuts, in turn, mitigate against gluinos of modest mass.  By 
contrast, if light gluinos are produced in association with gauginos, one 
can search for light gluino monojets accompanied by leptons and/or missing 
transverse energy from gaugino decays.    


\section*{Acknowledgments}
Work in the High Energy Physics Division at Argonne National Laboratory
is supported by the U.S. Department of Energy, Division of High Energy
Physics, under Contract W-31-109-ENG-38.  M. Klasen is supported by the 
Bundesministerium f\"ur Bildung und Forschung under Contract 05 HT9GUA 3, 
by the Deutsche Forschungsgemeinschaft under Contract KL 1266/1-1, and by 
the European Commission under Contract ERBFMRXCT980194.  The authors are 
grateful for correspondence with W.~Beenakker and T.~Plehn and 
conversations with S.~Mrenna.  T.~Tait benefitted from discussions with 
D.~Kaplan,~G.~Kribs, and C.--P.~Yuan.

\begin{appendix}

\newpage
\section{Virtual Loop Contributions}
\label{virtapp}

\begin{figure}[h]
 \begin{center}
  \epsfig{file=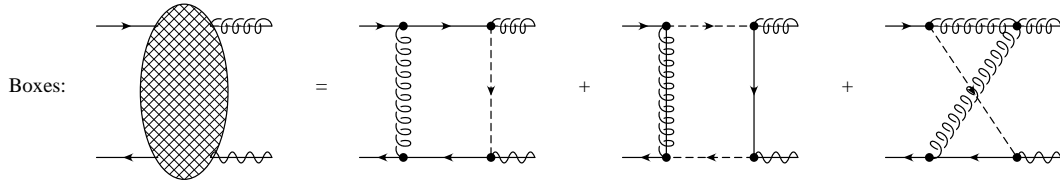,bbllx=35pt,bblly=360pt,bburx=469pt,bbury=435pt,%
          width=14cm,clip=}
 \end{center}
 \caption{Box diagrams for associated production of a gluino and a
 gaugino. There are $u$-channel box diagrams in addition to the diagrams shown.}
\label{gn_box}
\end{figure}

At the one-loop level in SUSY-QCD, virtual corrections contribute to the
hadroproduction of supersymmetric particles through the interference of
self-energy corrections, vertex corrections, and box diagrams with the
tree-level diagrams. For the associated production of gluinos and gauginos
one has to calculate the self-energy corrections in 
Fig.~\ref{gn_self}, the vertex corrections in Fig.~\ref{gn_vert}, and 
the box diagrams in Fig.~\ref{gn_box}.

The self-energy corrections for the external quark and antiquark factorize
the complete Born matrix element and are independent of the
underlying scattering process. Since the quark and antiquark are treated as
massless, only the squark-gluino loop contributes to
\bea
 \overline{|{\cal M}^q|}^2&=&i[B_0(0;\mg,\ms)+B_0'(0;\mg,\ms)(\mg^2-\ms^2)]
 2 C_F \hat{g}_s^2 \overline{|{\cal M}^B|}^2,
\eea
where the factor of two accounts for the sum of the quark and antiquark
contributions. Functions $B_0(p^2;m_1,m_2)$ and $B_0'(p^2;m_1,m_2)$ stand 
for the scalar two-point integral and its derivative.  They are defined in 
Appendix~\ref{scalarint}. 
The quark-gluon loop contains an infrared singularity
\bea
 \overline{|{\cal M}^q_{\rm IR}|}^2&=&C_{\epsilon}\frac{1}{\epsilon}
 2 C_F g_s^2 \overline{|{\cal M}^B|}^2 ,
\eea
where $C_{\epsilon} = (4\pi)^{\epsilon}/(16\pi^2)e^{-\epsilon\gamma_E}$.
This infrared singularity is not shown in the result above since it is
canceled by an ultraviolet singularity when infrared and ultraviolet
singularities are not distinguished in dimensional regularization. 

The external gluino self-energy also factorizes the complete Born matrix
element and is independent of the underlying scattering process:
\bea
 \overline{|{\cal M}^{\gluino}|}^2&=&i\left[\frac{A_0(\mg)}{\mg^2}(1-\epsilon)
 -4 B_0'(\mg^2;0,\mg)\mg^2\right] N_C g_s^2 \overline{|{\cal M}^B|}^2 \nonu
 &+& i\left[                 -\frac{A_0(\ms)}{2\mg^2}+B_0(\mg^2;\ms,0  )
     \frac{\mg^2      +\ms^2}{2\mg^2}
     +B_0'(\mg^2;\ms,0  )(\mg^2      -\ms^2)\right]n_f
     4 \frac{1}{2} \hat{g}_s^2 \overline{|{\cal M}^B|}^2 \nonu
 &+& i\left[\frac{A_0(m_t)}{2\mg^2}-\frac{A_0(\ms)}{2\mg^2}+B_0(\mg^2;\ms,m_t)
     \frac{\mg^2-m_t^2+\ms^2}{2\mg^2} +B_0'(\mg^2;\ms,m_t)\right. \nonu
 & & \left.\times(\mg^2+m_t^2-\ms^2)\right]
     4 \frac{1}{2} \hat{g}_s^2 \overline{|{\cal M}^B|}^2.
\eea
The scalar one-point integral $A_0(m)$ is defined in 
Appendix~\ref{scalarint}.
Except for the gluino mass and the $N_C$ color factor, the gluon-gluino loop
is identical to the heavy-quark self-energy. It contains an infrared
singularity in the derivative of the scalar two-point integral:
\bea
 \overline{|{\cal M}^{\gluino}_{\rm IR}|}^2
 &=&-C_{\epsilon} \frac{1}{\epsilon} 4
     \frac{N_C}{2} g_s^2 \overline{|{\cal M}^B|}^2 .
\eea
The quark-squark loop with a color factor of 1/2 contributes through two
different fermion number flows due to the Majorana nature of gluinos. We take
into account $n_f=5$ light (s)quark flavors and a heavy top (s)quark. We do
not include mixing in the top squark sector and take the top squark 
mass equal to the light squark masses.  We set $m_t=175$ GeV.

The gaugino couples only electroweakly to the quarks and squarks and thus
does not give rise to strong self-energy corrections.  All external particle
self-energies have been renormalized on-shell and multiplied by a factor
of 1/2 for proper wave function renormalization.

The self-energy correction of the internal squark propagator depends on the
off-shell squark four-momentum squared. Therefore, it factorizes only the
corresponding $t$- or $u$-channel interference piece of the Born matrix
element:
\bea
 \overline{|{\cal M}^{\tilde{q_t}}|}^2  
 &=& i\left[B_0(t    ;\mg,0)\frac{t    -\mg^2}{t-\ms^2}
           -B_0(\ms^2;\mg,0)\frac{\ms^2-\mg^2}{t-\ms^2}\right]
     4 C_F \hat{g}_s^2 [\overline{|{\cal M}^{t}{\cal M}^{t\ast}|}
                       +\overline{|{\cal M}^{t}{\cal M}^{u\ast}|}] \nonu
 &-& i\left[B_0(t    ;\ms,0)\frac{t+\ms^2}{t-\ms^2}
           -B_0(\ms^2;\ms,0)\frac{2\ms^2  }{t-\ms^2}\right]
     4 C_F      g_s^2  [\overline{|{\cal M}^{t}{\cal M}^{t\ast}|}
                       +\overline{|{\cal M}^{t}{\cal M}^{u\ast}|}] .
\eea
The ultraviolet divergences cancel between the quark-gluino loop contribution
and its supersymmetric counterpart, the squark-gluon loop contribution. Since
the gluon tadpole contribution is quadratic in the loop momentum it vanishes
in dimensional regularization. The squark tadpole contribution vanishes
after renormalization. The $u$-channel result can be obtained from the
$t$-channel result given above through the exchange $t\leftrightarrow u$.

Like the self-energy correction of the squark propagator, the corrections to
the quark-squark-gluino and quark-squark-gaugino vertices depend on the
four-momentum squared of the squark and factorize only the $t$- or $u$-channel
interference pieces of the Born matrix element. The quark-squark-gaugino vertex
receives corrections through a gluon and a gluino exchange between the initial
state quark or antiquark and the squark that are proportional to the gauge
and Yukawa coupling, respectively. For the $t$-channel we find
\bea
 \overline{|{\cal M}^{q\tilde{q}\gaugino}|}^2
 &=& i\left[B_0(t;\mg,0)-B_0(\mc^2;\ms,0)+C_0(\mc^2,0,t;0,\ms,\mg)
           (t+\ms^2-\mg^2-\mc^2)\right] \nonu
 & & \times \frac{\mg\mc}{t-\mc^2} 4 C_F \hat{g}_s^2
     \frac{X_t^{q'\tilde{q}\gaugino\ast}}{X_t^{q\tilde{q}'\gaugino}}
     [\overline{|{\cal M}^{t}{\cal M}^{t\ast}|}
     +\overline{|{\cal M}^{t}{\cal M}^{u\ast}|}] \nonu
 &+& i\left[B_0(t;\ms,0) \frac{t+\mc^2}{2(t-\mc^2)}
	   -B_0(\mc^2;\ms,0) \frac{\mc^2}{t-\mc^2}
	   +C_0(\mc^2,0,t;\ms,0,0) \right. \nonu
 & & \left.\times (\mc^2-\ms^2) \right] 4 C_F g_s^2
     [\overline{|{\cal M}^{t}{\cal M}^{t\ast}|}
     +\overline{|{\cal M}^{t}{\cal M}^{u\ast}|}] .
\eea
Function $C_0(p_1^2,p_2^2,(p_1+p_2)^2;m_1,m_2,m_3)$ stands 
for the scalar three-point
integral defined in Appendix~\ref{scalarint}. The ratio of quark-squark-gaugino
couplings $X_t^{q'\tilde{q}\gaugino\ast}/X_t^{q\tilde{q}'\gaugino}$ accounts
for reversed flavor flow in the vertex correction with respect to the
underlying Born matrix element in the case of the exchange of an additional
gluino. For neutralinos with real couplings it reduces to unity, whereas for
charginos with real couplings it is given by a ratio of chargino mixing matrix
elements. The infrared singularities
\bea
 \overline{|{\cal M}^{q\tilde{q}\gaugino}_{\rm IR}|}^2
 &=&-C_{\epsilon}\left[\frac{1}{\epsilon}+\frac{1}{\epsilon}\log\left(
     \frac{\ms^2-t}{\ms^2-\mc^2}\right)\frac{\ms^2-\mc^2}{t-\mc^2}\right]
     4 C_F g_s^2 [\overline{|{\cal M}^{t}{\cal M}^{t\ast}|}
                 +\overline{|{\cal M}^{t}{\cal M}^{u\ast}|}]
\eea
arise from the gluon exchange correction. In dimensional regularization, the
gauge bosons have $(n-2)$ degrees of freedom whereas their supersymmetric
counterparts, the gauginos, have two. This difference 
leads to a mismatch between the
quark-quark-gauge boson gauge couplings and the quark-squark-gaugino Yukawa
couplings through finite next-to-leading order terms. The (super-)symmetry
between the gauge and Yukawa couplings can be restored through a finite
renormalization contribution
\bea
 \overline{|{\cal M}^{q\tilde{q}\gaugino}_{\rm finite}|}^2
 &=&-C_{\epsilon} C_F g_s^2 [\overline{|{\cal M}^{t}{\cal M}^{t\ast}|}
                              +\overline{|{\cal M}^{t}{\cal M}^{u\ast}|}],
\eea
that can be found by comparing the quark-squark-gaugino vertex correction
given above with the corresponding quark-quark-gauge boson vertex correction
in exact supersymmetry. All $u$-channel results can again be obtained through
the exchange $t \leftrightarrow u$.

The quark-squark-gluino vertex correction can be obtained from the
quark-squark-gaugino vertex correction if $\mc$ is replaced with $\mg$ and the
color factor $C_F$ with $C_F-N_C/2$. There are, however, two additional
contributions with a color factor of $N_C/2$, due to the non-Abelian gauge
coupling of the gluino to the gluon, when a gluon is exchanged between the
final state gluino and the initial state quark or antiquark and the squark.
For the total quark-squark-gluino vertex correction in the $t$-channel we 
find
\bea
 \overline{|{\cal M}^{q\tilde{q}\gluino}|}^2
 &=& \overline{|{\cal M}^{q\tilde{q}\gaugino}|}^2 (\mc\rightarrow\mg,
     C_F\rightarrow C_F-N_C/2) \nonu
 &+& i\left[B_0(t;\mg,0)\left(\frac{\mg^2}{t-\mg^2}+\epsilon\right)
           -B_0(\mg^2;\mg,0)\frac{t}{t-\mg^2}
           +C_0(\mg^2,0,t;\mg,0,0)(t-\mg^2)\right. \nonu
 &+&  \left.B_0(t;0,\ms) \frac{t+\mg^2}{2(t-\mg^2)}
           -B_0(\mg^2;0,\mg) \frac{t}{t-\mg^2}
           +C_0(\mg^2,0,t;0,\mg,\ms)\frac{\mg^4-t\ms^2}{t-\mg^2}
     \right] \nonu
 & & \times 4 \frac{N_C}{2} g_s^2
     [\overline{|{\cal M}^{t}{\cal M}^{t\ast}|}
     +\overline{|{\cal M}^{t}{\cal M}^{u\ast}|}] .
\eea
It contains the following infrared singularities:
\bea
 \overline{|{\cal M}^{q\tilde{q}\gluino}_{\rm IR}|}^2
 &=& \overline{|{\cal M}^{q\tilde{q}\gaugino}_{\rm IR}|}^2 (\mc\rightarrow\mg,
     C_F\rightarrow C_F-N_C/2) \nonu
 &-& C_{\epsilon}\left[\frac{1}{2\epsilon^2}+\frac{1}{\epsilon}+\frac{1}
     {2\epsilon}\log\left(\frac{Q^2}{\mg^2}\right)-\frac{1}
     {\epsilon}\log\left(\frac{\mg^2-t}{\mg^2}\right)\right] \nonu
 & & \times 4 \frac{N_C}{2} g_s^2 [\overline{|{\cal M}^{t}{\cal M}^{t\ast}|}
                                  +\overline{|{\cal M}^{t}{\cal M}^{u\ast}|}].
\eea
The finite renormalization contribution for the gluino vertex correction is
\bea
 \overline{|{\cal M}^{q\tilde{q}\gluino}_{\rm finite}|}^2
 &=& C_{\epsilon} \left(\frac{4}{3}N_C-C_F\right) g_s^2
     [\overline{|{\cal M}^{t}{\cal M}^{t\ast}|}
     +\overline{|{\cal M}^{t}{\cal M}^{u\ast}|}].
\eea
For the $u$-channel contribution, $t$ and $u$ have to be exchanged as before.

Turning to the box diagrams that contribute to the associated production of
gluinos and gauginos, we notice that they depend naturallly on the
four-momentum squared of the exchanged squark. Therefore, we do not expect the
full Born matrix element to factorize. In addition, the traces of the
Dirac matrices project out terms that depend on the final state masses
separately from those that do not, so that the squared $t$-channel and
$u$-channel diagrams and the interference term can only be factorized
individually. 

For the first box diagram in Fig.~\ref{gn_box} we find
\bea
 \overline{|{\cal M}^{\rm Box 1}|}^2 &=&
           i B_0(t;\ms,0) (t-\ms^2) g_s^2 (C_F-N_C/2) 4 \overline{|{\cal M}^{t}{\cal M}^{u\ast}|} /s
      \nonumber \\ &+& i B_0(t;\ms,0) (t-\ms^2) g_s^2 (C_F-N_C/2) 2 \overline{|{\cal M}^{t}{\cal M}^{t\ast}|}
\nonumber \\ &&   (  - 2 
       \mg^2  /(t-\mc^2) /(t-\mg^2)  - 2 /(t-\mc^2) )
      \nonumber \\ &+& i B_0(s;0,0) (t-\ms^2) /K(s,\mc^2,\mg^2) g_s^2 (C_F-N_C/2) 
      2 \overline{|{\cal M}^{t}{\cal M}^{u\ast}|}\nonumber \\ &&   ( 2 s  + 4 (t-\mg^2) - 2 \mc^2  + 2 \mg^2 )
      \nonumber \\ &+& i B_0(s;0,0) (t-\ms^2) /K(s,\mc^2,\mg^2) g_s^2 (C_F-N_C/2) 
      2 \overline{|{\cal M}^{t}{\cal M}^{t\ast}|}\nonumber \\ &&   ( 8 s \mg^4/(t-\mc^2) /(t-\mg^2)   
+ 2 s  \mc^2/(t-\mc^2)  + 8 s  \mg^2 /(t-\mc^2)
        \nonumber \\ &&  - 6 s  \mg^2  /(t-\mg^2)
 + 2  \mc^2 \mg^2 /(t-\mc^2) - 2  \mc^4 /(t-\mc^2)  + 2 \mc^2  \mg^2 /(t-\mg^2)
     \nonumber \\ &&    - 2 \mg^4 /(t-\mg^2)  )
      \nonumber \\ &+& i B_0(s;0,0) (t-\ms^2) g_s^2 (C_F-N_C/2) 2 \overline{|{\cal M}^{t}{\cal M}^{t\ast}|}
\nonumber \\ &&   ( 2 \mg^2 /(t-\mc^2) 
         /(t-\mg^2)  + 2 /(t-\mc^2) )
      \nonumber \\ &+& i B_0(\mg^2;\ms,0) (t-\ms^2) /K(s,\mc^2,\mg^2) g_s^2 (C_F-N_C/2) 2 
\overline{|{\cal M}^{t}{\cal M}^{u\ast}|}\nonumber \\ &&
         (  - 2 (t-\mg^2)\mc^2/s  + 2 (t-\mg^2)\mg^2/s  - 2 s - 2 (t-\mg^2)  + 2 \mc^2  + 2 \mg^2 )
      \nonumber \\ &+& i B_0(\mg^2;\ms,0) (t-\ms^2) /K(s,\mc^2,\mg^2) g_s^2 (C_F-N_C/2) 
      2 \overline{|{\cal M}^{t}{\cal M}^{t\ast}|}\nonumber \\ &&   (  - 4 s  \mg^4 /(t-\mc^2) /(t-\mg^2) 
- 2 s \mc^2 /(t-\mc^2) - 4 s \mg^2 /(t-\mc^2) 
           \nonumber \\ && + 4 s \mg^2/(t-\mg^2)  
 - 2  \mc^2 \mg^2 /(t-\mc^2)  + 2 \mc^4 /(t-\mc^2)  )
      \nonumber \\ &+& i B_0(\mc^2;\ms,0) (t-\ms^2) /K(s,\mc^2,\mg^2) g_s^2 (C_F-N_C/2) 2 
\overline{|{\cal M}^{t}{\cal M}^{u\ast}|}\nonumber \\ &&
         ( 2 (t-\mg^2)\mc^2/s - 2 (t-\mg^2)/s \mg^2 - 2 (t-\mg^2) - 4 \mg^2 )
      \nonumber \\ &+& i B_0(\mc^2;\ms,0) (t-\ms^2) /K(s,\mc^2,\mg^2) g_s^2 (C_F-N_C/2) 
      2 \overline{|{\cal M}^{t}{\cal M}^{t\ast}|}\nonumber \\ &&   (  - 4 s /(t-\mc^2) /(t-\mg^2) \mg^4 - 4 s 
/(t-\mc^2) \mg^2  + 2 s /(t-\mg^2) 
         \mg^2 \nonumber \\ && - 2 /(t-\mg^2) \mc^2 \mg^2  + 2 /(t-\mg^2) \mg^4 )
      \nonumber \\ &-& i B_0(\mc^2;\ms,0) (t-\ms^2) g_s^2 (C_F-N_C/2) 4 \overline{|{\cal M}^{t}{\cal M}^{u\ast}|}
 /s
      \nonumber \\ &+& i C_0(\mg^2,\mc^2,s;0,\ms,0) (t-\ms^2) /K(s,\mc^2,\mg^2) g_s^2 (C_F-N_C/2) 
      2 \overline{|{\cal M}^{t}{\cal M}^{u\ast}|}\nonumber \\ &&   ( 2 \ms^2 s  + 4 \ms^2 (t-\mg^2) - 2 \ms^2 
\mc^2  + 2 \ms^2 \mg^2  + 2 s (t-\mg^2)
           + 2 s \mg^2 \nonumber \\ && - 2 (t-\mg^2) \mc^2  - 2 (t-\mg^2) \mg^2  + 2 \mc^2 \mg^2 - 2 \mg^4 )
      \nonumber 
\eea

\bea
  &+& i C_0(\mg^2,\mc^2,s;0,\ms,0) (t-\ms^2) /K(s,\mc^2,\mg^2) g_s^2 (C_F-N_C/2) 
      2 \overline{|{\cal M}^{t}{\cal M}^{t\ast}|}\nonumber \\ &&   ( 8 \ms^2 s /(t-\mc^2) /(t-\mg^2) \mg^4  
+ 2 \ms^2 s /(t-\mc^2) \mc^2  + 8 \ms^2
          s /(t-\mc^2) \mg^2 \nonumber \\ && - 6 \ms^2 s /(t-\mg^2) \mg^2  + 2 \ms^2 /(t-\mc^2) \mc^2 \mg^2 
- 2 \ms^2
          /(t-\mc^2) \mc^4  \nonumber \\ && + 2 \ms^2 /(t-\mg^2) \mc^2 \mg^2 - 2 \ms^2 /(t-\mg^2) \mg^4  
+ 8 s /(t-\mc^2) 
         /(t-\mg^2) \mg^6  \nonumber \\ && + 6 s /(t-\mc^2) \mc^2 \mg^2  + 8 s /(t-\mc^2) \mg^4 
- 2 s /(t-\mg^2) \mc^2 
         \mg^2 - 8 s /(t-\mg^2) \mg^4  \nonumber \\ && + 2 /(t-\mc^2) \mc^2 \mg^4 - 2 /(t-\mc^2) \mc^4 \mg^2 - 2 
         /(t-\mg^2) \mc^2 \mg^4  + 2 /(t-\mg^2) \mc^4 \mg^2 ) \nonumber \\
   &+& i C_0(\mg^2,\mc^2,s;0,\ms,0) (t-\ms^2) g_s^2 (C_F-N_C/2) 8 \overline{|{\cal M}^{t}{\cal M}^{u\ast}|}
      \nonumber \\
   &+& i C_0(\mg^2,\mc^2,s;0,\ms,0) (t-\ms^2) g_s^2 (C_F-N_C/2) 2 \overline{|{\cal M}^{t}{\cal M}^{t\ast}|}
\nonumber \\ &&   (  - \ms^2 s
          /(t-\mc^2) /(t-\mg^2)  + 2 \ms^2 /(t-\mc^2) /(t-\mg^2) \mg^2  + \ms^2 /(t-\mc^2) \nonumber \\ && - 
\ms^2 /(t-\mg^2)  + s 
         /(t-\mc^2) /(t-\mg^2) \mg^2 - s /(t-\mg^2)  \nonumber \\ && + 2 /(t-\mc^2) /(t-\mg^2) \mg^4   
+ 3 /(t-\mc^2) \mg^2  + 
         \mc^2 /(t-\mg^2)  - 2 /(t-\mg^2) \mg^2 )
      \nonumber \\ &-& i C_0(\mg^2,0,t;\ms,0,0) (t-\ms^2) g_s^2 (C_F-N_C/2) 4 
\overline{|{\cal M}^{t}{\cal M}^{u\ast}|}
      \nonumber \\ &+& i C_0(\mg^2,0,t;\ms,0,0) (t-\ms^2) g_s^2 (C_F-N_C/2) 2 
\overline{|{\cal M}^{t}{\cal M}^{t\ast}|}\nonumber \\ &&   (  - 1  + 
         \ms^2 /(t-\mg^2) - \mc^2 /(t-\mg^2)  )
      \nonumber \\ &-& i C_0(\mc^2,0,t;\ms,0,0) (t-\ms^2) g_s^2 (C_F-N_C/2) 4 
\overline{|{\cal M}^{t}{\cal M}^{u\ast}|}
      \nonumber \\ &+& i C_0(\mc^2,0,t;\ms,0,0) (t-\ms^2) g_s^2 (C_F-N_C/2) 2 
\overline{|{\cal M}^{t}{\cal M}^{t\ast}|}\nonumber \\ &&   (  - 1  + 
         \ms^2 /(t-\mc^2) - \mg^2 /(t-\mc^2)  )
      \nonumber \\ &+& i C_0(0,0,s;0,0,0) (t-\ms^2) g_s^2 (C_F-N_C/2) 2 \overline{|{\cal M}^{t}{\cal M}^{t\ast}|}
\nonumber \\ &&   ( \ms^2 s 
         /(t-\mc^2) /(t-\mg^2) - s /(t-\mc^2) /(t-\mg^2) \mg^2  + s /(t-\mg^2) )
      \nonumber \\ &+& i D_0(\mc^2,\mg^2,0,0;0,\ms,0,0) (t-\ms^2) g_s^2 (C_F-N_C/2) 4 
\overline{|{\cal M}^{t}{\cal M}^{u\ast}|}   s
      \nonumber \\ &+& i D_0(\mc^2,\mg^2,0,0;0,\ms,0,0) (t-\ms^2) g_s^2 (C_F-N_C/2) 4 
\overline{|{\cal M}^{t}{\cal M}^{t\ast}|}   s
      \nonumber \\ &+& i D_0(\mc^2,\mg^2,0,0;0,\ms,0,0) (t-\ms^2)^2 g_s^2 (C_F-N_C/2) 2 
\overline{|{\cal M}^{t}{\cal M}^{t\ast}|}\nonumber \\ &&   ( 
          - s /(t-\mc^2) - s /(t-\mg^2) )
      \nonumber \\ &+& i D_0(\mc^2,\mg^2,0,0;0,\ms,0,0) (t-\ms^2)^3 g_s^2 (C_F-N_C/2) 2 
\overline{|{\cal M}^{t}{\cal M}^{t\ast}|}\nonumber \\ &&   ( s
          /(t-\mc^2) /(t-\mg^2) ).
\eea
Function $K(s,\mc^2,\mg^2) = s^2 - 2 s (\mc^2+\mg^2) + (\mc^2-\mg^2)^2 $ 
is the triangle (K\"all\'en) function 
of the partonic center-of-mass energy and the masses of the produced particles. 
The corresponding $u$-channel contribution is obtained by exchanging $t$ and $u$.

For the second box diagram in Fig.~\ref{gn_box} we find
\bea
 \overline{|{\cal M}^{\rm Box 2}|}^2 &=&
                     i B_0(t;\mg,0) (u-\ms^2) g_s^2 (C_F-N_C/2) 4 \overline{|{\cal M}^{t}{\cal M}^{u\ast}|} /s
      \nonumber \\ &+& i B_0(t;\mg,0) (u-\ms^2) g_s^2 (C_F-N_C/2) 2 \overline{|{\cal M}^{u}{\cal M}^{u\ast}|}
\nonumber \\ &&   ( 2 /(t-\mc^2) 
         /(u-\mc^2) \mc^2  + 2 /(t-\mg^2) /(u-\mg^2) \mg^2 \nonumber \\ && + 2 /(u-\mc^2) /(u-\mg^2) \mg^2  
+ 2 /(u-\mc^2) )
      \nonumber \\ &+& i B_0(s;\ms,\ms) (u-\ms^2) /K(s,\mc^2,\mg^2) g_s^2 (C_F-N_C/2) 
      2 \overline{|{\cal M}^{t}{\cal M}^{u\ast}|}\nonumber \\ &&   (  - 2 s - 4 (u-\mg^2)  + 2 \mc^2 - 2 \mg^2 )
      \nonumber \\ &+& i B_0(s;\ms,\ms) (u-\ms^2) /K(s,\mc^2,\mg^2) g_s^2 (C_F-N_C/2) 
      2 \overline{|{\cal M}^{u}{\cal M}^{u\ast}|}\nonumber \\ &&   (  - 8 s /(u-\mc^2) /(u-\mg^2) \mg^4 - 2 s 
/(u-\mc^2) \mc^2 - 8 s /(u-\mc^2) 
         \mg^2  \nonumber \\ && + 6 s /(u-\mg^2) \mg^2 - 2 /(u-\mc^2) \mc^2 \mg^2  + 2 /(u-\mc^2) \mc^4 
- 2 /(u-\mg^2) 
         \mc^2 \mg^2  \nonumber \\ && + 2 /(u-\mg^2) \mg^4 )
      \nonumber \\ &+& i B_0(s;\ms,\ms) (u-\ms^2) g_s^2 (C_F-N_C/2) 2 \overline{|{\cal M}^{u}{\cal M}^{u\ast}|}
\nonumber \\ &&   (  - 2 
         /(u-\mc^2) /(u-\mg^2) \mg^2 - 2 /(u-\mc^2) )
      \nonumber \\ &+& i B_0(\mg^2;\ms,0) (u-\ms^2) /K(s,\mc^2,\mg^2) g_s^2 (C_F-N_C/2) 2 
\overline{|{\cal M}^{t}{\cal M}^{u\ast}|}\nonumber \\ &&
         ( 2 (u-\mg^2)\mc^2/s - 2 (u-\mg^2)/s \mg^2  + 2 s  + 2 (u-\mg^2) - 2 \mc^2 - 2 \mg^2 )
      \nonumber \\ &+& i B_0(\mg^2;\ms,0) (u-\ms^2) /K(s,\mc^2,\mg^2) g_s^2 (C_F-N_C/2) 
      2 \overline{|{\cal M}^{u}{\cal M}^{u\ast}|}\nonumber \\ &&   ( 4 s /(u-\mc^2) /(u-\mg^2) \mg^4  + 
2 s /(u-\mc^2) \mc^2  + 4 s /(u-\mc^2) \mg^2
       \nonumber \\ &&   - 4 s /(u-\mg^2) \mg^2  + 2 /(u-\mc^2) \mc^2 \mg^2 - 2 /(u-\mc^2) \mc^4 )
      \nonumber \\ &-& i B_0(\mg^2;\ms,0) (u-\ms^2) g_s^2 (C_F-N_C/2) 4 \overline{|{\cal M}^{t}{\cal M}^{u\ast}|}
 /s \nonumber \\ &+& i B_0(\mg^2;\ms,0) (u-\ms^2) g_s^2 (C_F-N_C/2) 2 \overline{|{\cal M}^{u}{\cal M}^{u\ast}|}
\nonumber  \\ &&   (  - 2 /(t-\mc^2)    /(u-\mc^2) \mc^2 ) \nonumber 
\eea

\bea
      &+& i B_0(\mc^2;\ms,0) (u-\ms^2) /K(s,\mc^2,\mg^2) g_s^2 (C_F-N_C/2) 2 
\overline{|{\cal M}^{t}{\cal M}^{u\ast}|}\nonumber \\ &&
         (  - 2 (u-\mg^2)\mc^2/s  + 2 (u-\mg^2)/s \mg^2  + 2 (u-\mg^2)  + 4 \mg^2 )
      \nonumber \\ &+& i B_0(\mc^2;\ms,0) (u-\ms^2) /K(s,\mc^2,\mg^2) g_s^2 (C_F-N_C/2) 
      2 \overline{|{\cal M}^{u}{\cal M}^{u\ast}|}\nonumber \\ &&   ( 4 s /(u-\mc^2) /(u-\mg^2) \mg^4  + 
4 s /(u-\mc^2) \mg^2 - 2 s /(u-\mg^2) \mg^2
           \nonumber \\ && + 2 /(u-\mg^2) \mc^2 \mg^2 - 2 /(u-\mg^2) \mg^4 )
      \nonumber \\ &+& i B_0(\mc^2;\ms,0) (u-\ms^2) g_s^2 (C_F-N_C/2) 2 \overline{|{\cal M}^{u}{\cal M}^{u\ast}|}
\nonumber \\ &&   (  - 2 /(t-\mg^2) 
         /(u-\mg^2) \mg^2 ) \nonumber \\
  &+& i C_0(\mg^2,\mc^2,s;\ms,0,\ms) (u-\ms^2) /K(s,\mc^2,\mg^2) g_s^2 (C_F-N_C/2) 
      2 \overline{|{\cal M}^{t}{\cal M}^{u\ast}|}\nonumber \\ &&   ( 2 \ms^2 s  + 4 \ms^2 (u-\mg^2) - 2 \ms^2\mc^2
+ 2 \ms^2 \mg^2 - 2 s (u-\mg^2)
          - 2 s \mg^2  \nonumber \\ &&+ 2 (u-\mg^2) \mc^2   + 2 (u-\mg^2) \mg^2 - 2 \mc^2 \mg^2  + 2 \mg^4 )
      \nonumber \\ &+& i C_0(\mg^2,\mc^2,s;\ms,0,\ms) (u-\ms^2) /K(s,\mc^2,\mg^2) g_s^2 (C_F-N_C/2) 
      2 \overline{|{\cal M}^{u}{\cal M}^{u\ast}|}\nonumber \\ &&   ( 8 \ms^2 s /(u-\mc^2) /(u-\mg^2) \mg^4  
+ 2 \ms^2 s /(u-\mc^2) \mc^2  + 8 \ms^2
          s /(u-\mc^2) \mg^2 \nonumber \\ && - 6 \ms^2 s /(u-\mg^2) \mg^2  + 2 \ms^2 /(u-\mc^2) \mc^2 \mg^2 
- 2 \ms^2
          /(u-\mc^2) \mc^4  \nonumber \\ && + 2 \ms^2 /(u-\mg^2) \mc^2 \mg^2  - 2 \ms^2 /(u-\mg^2) \mg^4 - 
8 s /(u-\mc^2) 
         /(u-\mg^2) \mg^6 \nonumber \\ && - 6 s /(u-\mc^2) \mc^2 \mg^2 - 8 s /(u-\mc^2) \mg^4  + 2 s /(u-\mg^2) 
\mc^2 
         \mg^2  + 8 s /(u-\mg^2) \mg^4 \nonumber \\ && - 2 /(u-\mc^2) \mc^2 \mg^4  + 2 /(u-\mc^2) \mc^4 \mg^2 +2 
         /(u-\mg^2) \mc^2 \mg^4 - 2 /(u-\mg^2) \mc^4 \mg^2 )
      \nonumber \\ &+& i C_0(\mg^2,\mc^2,s;\ms,0,\ms) (u-\ms^2) g_s^2 (C_F-N_C/2) 2 
\overline{|{\cal M}^{t}{\cal M}^{u\ast}|} \nonumber \\ &&   ( 1 - 2 \ms^2 
         /s  + 2 (u-\mg^2)/s -  \mc^2/s  + 3 \mg^2/s )
      \nonumber \\ &+& i C_0(\mg^2,\mc^2,s;\ms,0,\ms) (u-\ms^2) g_s^2 (C_F-N_C/2) 2
\overline{|{\cal M}^{u}{\cal M}^{u\ast}|} \nonumber \\ &&   (  - 2 
         \ms^2 s /(u-\mc^2) /(u-\mg^2)  + 4 \ms^2 /(u-\mc^2) /(u-\mg^2) \mg^2  + 2 \ms^2 /(u-\mc^2) \nonumber \\ 
&& - 2 \ms^2 
         /(u-\mg^2) - s /(u-\mc^2) /(u-\mg^2) \mg^2 - s /(u-\mc^2)  + s /(u-\mg^2)  \nonumber \\ && + 
s^2 /(u-\mc^2) /(u-\mg^2)  - 4 
         /(u-\mc^2) /(u-\mg^2) \mg^4 - 3 /(u-\mc^2) \mg^2 \nonumber \\ && + 3 /(u-\mg^2) \mg^2 )
      \nonumber \\ &+& i C_0(\mg^2,0,t;0,\ms,\mg) (u-\ms^2) g_s^2 (C_F-N_C/2) 2 
\overline{|{\cal M}^{t}{\cal M}^{u\ast}|}\nonumber  \\ &&   (  - 1  + 2 
         \ms^2 /s - (u-\mg^2)/s - 2 \mg^2/s ) \nonumber 
\eea

\bea
      &+& i C_0(\mg^2,0,t;0,\ms,\mg) (u-\ms^2) g_s^2 (C_F-N_C/2) 2 
\overline{|{\cal M}^{u}{\cal M}^{u\ast}|}\nonumber
 \\ &&   ( 2 \ms^2 
         s /(u-\mc^2) /(u-\mg^2)  + 2 \ms^2 /(t-\mc^2) /(u-\mc^2) \mc^2  + 2 \ms^2 /(u-\mc^2) \nonumber \\ && 
- s /(u-\mc^2) 
         /(u-\mg^2) \mg^2 - s /(u-\mc^2) - s^2 /(u-\mc^2) /(u-\mg^2) \nonumber \\ && - 2 /(t-\mc^2) /(u-\mc^2) 
\mc^2 \mg^2 - 
         \mg^2 /(u-\mc^2)  )
      \nonumber \\ &+& i C_0(\mc^2,0,t;0,\ms,\mg) (u-\ms^2) g_s^2 (C_F-N_C/2) 2 
\overline{|{\cal M}^{t}{\cal M}^{u\ast}|}\nonumber
 \\ &&   (  - 1  + 2 
         \ms^2 /s - (u-\mg^2)/s  + \mc^2/s - 3 \mg^2/s )
      \nonumber \\ &+& i C_0(\mc^2,0,t;0,\ms,\mg) (u-\ms^2) g_s^2 (C_F-N_C/2) 2 
\overline{|{\cal M}^{u}{\cal M}^{u\ast}|}\nonumber
 \\ &&   ( 2 \ms^2 
         s /(u-\mc^2) /(u-\mg^2)  + 2 \ms^2 /(t-\mg^2) /(u-\mg^2) \mg^2  + 2 \ms^2 /(u-\mg^2) \nonumber \\ && 
- s /(u-\mc^2) 
         /(u-\mg^2) \mg^2 - s /(u-\mg^2) - s^2 /(u-\mc^2) /(u-\mg^2) \nonumber \\ && - 2 /(t-\mg^2) /(u-\mg^2) 
\mg^4 - \mg^2 
/(u-\mg^2)  )
      \nonumber \\
   &+& i C_0(0,0,s;\ms,\mg,\ms) (u-\ms^2) g_s^2 (C_F-N_C/2) 2 \overline{|{\cal M}^{t}{\cal M}^{u\ast}|} \nonumber \\
   &+& i C_0(0,0,s;\ms,\mg,\ms) (u-\ms^2) g_s^2 (C_F-N_C/2) 2 \overline{|{\cal M}^{u}{\cal M}^{u\ast}|}\nonumber \\ &&   (  - 2 
         \ms^2 s /(u-\mc^2) /(u-\mg^2)  + s /(u-\mc^2) /(u-\mg^2) \mg^2  + s^2 /(u-\mc^2) /(u-\mg^2) )
      \nonumber \\ &+& i D_0(\mc^2,\mg^2,0,0;\ms,0,\ms,\mg) (u-\ms^2) g_s^2 (C_F-N_C/2) 2 \overline{|{\cal M}^{t}{\cal M}^{u\ast}|}
\nonumber \\ &&   (  - 2
          \ms^2 (u-\mg^2)/s  + \ms^2 \mc^2/s - 5 \ms^2 \mg^2/s - 2 \ms^2  + 2 \ms^4 
         /s  + 2 (u-\mg^2)/s \mg^2 \nonumber \\ && - \mc^2/s \mg^2   + 3 /s \mg^4  + s  + (u-\mg^2) - \mc^2
           + 3 \mg^2 )
      \nonumber \\ &+& i D_0(\mc^2,\mg^2,0,0;\ms,0,\ms,\mg) (u-\ms^2) g_s^2 (C_F-N_C/2) 2 \overline{|{\cal M}^{u}{\cal M}^{u\ast}|}
\nonumber \\ &&   ( 
          - 2 \ms^2 s /(u-\mc^2) /(u-\mg^2) \mg^2 - \ms^2 s /(u-\mc^2) - 3 \ms^2 s /(u-\mg^2) \nonumber \\ && - 4 \ms^2
          s^2 /(u-\mc^2) /(u-\mg^2)  + 4 \ms^2 /(u-\mc^2) /(u-\mg^2) \mg^4  + \ms^2 /(u-\mc^2) \mg^2 \nonumber \\ && - 3 \ms^2
          /(u-\mg^2) \mg^2  + 4 \ms^4 s /(u-\mc^2) /(u-\mg^2) - 2 \ms^4 /(u-\mc^2) /(u-\mg^2) \mg^2  
\nonumber \\ && + 2 \ms^4
          /(u-\mg^2) - 2 s /(u-\mc^2) /(u-\mg^2) \mg^4 - s /(u-\mc^2) \mg^2  + 2 s /(u-\mg^2) \mg^2 
\nonumber \\ &&  + s^2 
         /(u-\mc^2) /(u-\mg^2) \mg^2  + s^2 /(u-\mg^2)  + s^3 /(u-\mc^2) /(u-\mg^2) 
\nonumber \\ && - 2 /(u-\mc^2) /(u-\mg^2) \mg^6
          - \mg^4 /(u-\mc^2)   +\mg^4 /(u-\mg^2)  ).
\eea
The corresponding $u$-channel contribution is 
obtained by exchanging $t$ and $u$.

For the third box diagram in Fig.\ref{gn_box} we find
\bea
 \overline{|{\cal M}^{\rm Box 3}|}^2 &=&
                     i B_0(t;\ms,0) (t-\ms^2) g_s^2 N_C 2 \overline{|{\cal M}^{t}{\cal M}^{u\ast}|} /s
      \nonumber \\ &+& i B_0(t;\ms,0) (t-\ms^2) g_s^2 N_C 2 \overline{|{\cal M}^{t}{\cal M}^{t\ast}|}\nonumber \\ &&
   (  -  \mg^2/((t-\mc^2)(t-\mg^2))
         - 1 /(t-\mc^2) )
      \nonumber \\ &+& i B_0(u;\mg,0) (t-\ms^2) g_s^2 N_C 2 \overline{|{\cal M}^{t}{\cal M}^{u\ast}|} /s
      \nonumber \\ &+& i B_0(u;\mg,0) (t-\ms^2) g_s^2 N_C 2 \overline{|{\cal M}^{t}{\cal M}^{t\ast}|}\nonumber \\ &&
   ( \mg^2/((t-\mc^2)(t-\mg^2))
           + \mc^2/((t-\mc^2)(u-\mc^2))   + 1/(t-\mc^2) \nonumber \\ && + \mg^2/((t-\mg^2)(u-\mg^2))  )
      \nonumber \\ &-& i B_0(\mg^2;\ms,0) (t-\ms^2) g_s^2 N_C 2 \overline{|{\cal M}^{t}{\cal M}^{u\ast}|} /s
      \nonumber \\ &+& i B_0(\mg^2;\ms,0) (t-\ms^2) g_s^2 N_C 2 \overline{|{\cal M}^{t}{\cal M}^{t\ast}|}\nonumber \\ &&
   (  -  \mc^2/((t-\mc^2) (u-\mc^2)) )
      \nonumber \\ &-& i B_0(\mc^2;\mg,0) (t-\ms^2) g_s^2 N_C 2 \overline{|{\cal M}^{t}{\cal M}^{u\ast}|} /s
      \nonumber \\ &+& i B_0(\mc^2;\mg,0) (t-\ms^2) g_s^2 N_C 2 \overline{|{\cal M}^{t}{\cal M}^{t\ast}|}  
 (  -  \mg^2/((t-\mg^2) (u-\mg^2)) )
      \nonumber \\ &+& i C_0(\mg^2,0,t;\ms,0,0) (t-\ms^2) g_s^2 N_C 2 \overline{|{\cal M}^{t}{\cal M}^{u\ast}|}\nonumber \\ &&
   (  - 1 - (t-\mg^2)/(2s)
           + \mc^2/(2s) - \mg^2/(2s) )
      \nonumber \\ &+& i C_0(\mg^2,0,t;\ms,0,0) (t-\ms^2) g_s^2 N_C 2 \overline{|{\cal M}^{t}{\cal M}^{t\ast}|}\nonumber \\ &&
   (  - 1/2  +  
         \ms^2 /(2(t-\mg^2)) - \mc^2/(2(t-\mg^2)) )
      \nonumber \\ &+& i C_0(\mg^2,0,u;0,\ms,\mg) (t-\ms^2) g_s^2 N_C 2 \overline{|{\cal M}^{t}{\cal M}^{u\ast}|}\nonumber \\ &&
   (  - 1/2  + \ms^2 
         /s -(t-\mg^2)/(2s) - \mg^2/s )
      \nonumber \\ &+& i C_0(\mg^2,0,u;0,\ms,\mg) (t-\ms^2) g_s^2 N_C 2 \overline{|{\cal M}^{t}{\cal M}^{t\ast}|}\nonumber \\ &&
   ( 1/2  + \ms^2
          s /(2(t-\mc^2) (t-\mg^2))  + \ms^2 \mc^2/((t-\mc^2) (u-\mc^2))  \nonumber \\ && + 1/2 \ms^2 /(t-\mc^2) 
 - s \mg^2/(2(t-\mc^2)
          (t-\mg^2))   + s /(2(t-\mg^2)) \nonumber \\ &&- \mc^2 \mg^2/((t-\mc^2) (u-\mc^2) )
  - \mg^2 /(2(t-\mc^2))  ) \nonumber 
\eea

\bea
      &+& i C_0(\mc^2,0,u;\mg,0,0) (t-\ms^2) g_s^2 N_C 2 \overline{|{\cal M}^{t}{\cal M}^{u\ast}|}\nonumber \\ &&
   ( 1/2  + (t-\mg^2)/(2s)
          - \mc^2/(2s)  + \mg^2/(2s) ) \nonumber \\
      &+& i C_0(\mc^2,0,u;\mg,0,0) (t-\ms^2) g_s^2 N_C 2 \overline{|{\cal M}^{t}{\cal M}^{t\ast}|}\nonumber \\ &&
   (  - 1/2 -
         \ms^2 s /(2(t-\mc^2) (t-\mg^2)) - \ms^2 /(2(t-\mg^2)) \nonumber \\ &&
 + s \mg^2/(2(t-\mc^2) (t-\mg^2)) 
 - s 
         /(2(t-\mg^2))  + \mc^2 /(2(t-\mg^2))  ) \nonumber \\
    &+& i C_0(\mc^2,0,t;0,\mg,\ms) (t-\ms^2) g_s^2 N_C 2 \overline{|{\cal M}^{t}{\cal M}^{u\ast}|}\nonumber \\ &&   (  - \ms^2 /s  +
         (t-\mg^2)/(2s)  + \mg^2/s )
      \nonumber \\ &+& i C_0(\mc^2,0,t;0,\mg,\ms) (t-\ms^2) g_s^2 N_C 2 \overline{|{\cal M}^{t}{\cal M}^{t\ast}|}\nonumber \\ &&
   (  - 1/2  + \ms^2 
         \mg^2/((t-\mc^2) (t-\mg^2))   + \ms^2 /(2(t-\mc^2)) -  \mg^4/((t-\mc^2) (t-\mg^2) )
\nonumber \\ && -\mg^2 /(2(t-\mc^2) )
          )
      \nonumber \\ &+& i D_0(\mc^2,0,\mg^2,0;0,\mg,\ms,0) (t-\ms^2) g_s^2 N_C 2 \overline{|{\cal M}^{t}{\cal M}^{u\ast}|}\nonumber \\ &&
   (  - s - (t-\mg^2)  + 
         \mc^2 - \mg^2 )
      \nonumber \\ &+& i D_0(\mc^2,0,\mg^2,0;0,\mg,\ms,0) (t-\ms^2) g_s^2 N_C 2 \overline{|{\cal M}^{t}{\cal M}^{t\ast}|}\nonumber \\ &&
   (  - s - (t-\mg^2)
           + \mc^2 - \mg^2 )
      \nonumber \\ &+& i D_0(\mc^2,0,\mg^2,0;0,\mg,\ms,0) (t-\ms^2)^2 g_s^2 N_C 2 \overline{|{\cal M}^{t}{\cal M}^{u\ast}|}\nonumber \\ &&
   (  - 1/2 - 
         (t-\mg^2)/(2s)  + \mc^2/(2s) - \mg^2/(2s) )
      \nonumber \\ &+& i D_0(\mc^2,0,\mg^2,0;0,\mg,\ms,0) (t-\ms^2)^2 g_s^2 N_C 2 \overline{|{\cal M}^{t}{\cal M}^{t\ast}|}\nonumber \\ &&
   ( 1  + 
          s /(2(t-\mc^2))  + s /(2(t-\mg^2)) - \mc^2/(2(t-\mg^2))   + \mg^2/(2(t-\mg^2))  )
      \nonumber \\ &+& i D_0(\mc^2,0,\mg^2,0;0,\mg,\ms,0) (t-\ms^2)^3 g_s^2 N_C 2 \overline{|{\cal M}^{t}{\cal M}^{t\ast}|}\nonumber \\ &&
   (  -  
         s /(2(t-\mc^2) (t-\mg^2)) - 1/(2(t-\mg^2)) ).
\eea
The corresponding $u$-channel contribution is 
obtained by exchanging $t$ and $u$.

The coefficients of the infrared singularities in the gluon
exchange box diagrams can be simplified considerably. The gluino-exchange box
diagram is infrared finite.
For the first box diagram in Fig.~\ref{gn_box} in the $t$-channel we find
\bea
 \overline{|{\cal M}^{\rm Box 1}_{\rm IR}|}^2 
 &=&-C_{\epsilon}\left[
      \frac{1}{\epsilon^2}
     -\frac{1}{\epsilon}\log\left(\frac{(t-\ms^2)^2 s}{(\mg^2-\ms^2)
      (\mc^2-\ms^2)Q^2}\right)
     +\frac{1}{\epsilon}\log\left(\frac{\ms^2-t}{\ms^2-\mg^2}\right)
      \frac{t-\ms^2}{t-\mg^2} \right. \nonu
 & &  \left.+\frac{1}{\epsilon}\log\left(\frac{\ms^2-t}{\ms^2-\mc^2}\right)
      \frac{t-\ms^2}{t-\mc^2} \right]
      4 \left(C_F-\frac{N_C}{2}\right) g_s^2
      [\overline{|{\cal M}^{t}{\cal M}^{t\ast}|}
      +\overline{|{\cal M}^{t}{\cal M}^{u\ast}|}] ,
\eea
and the $u$-channel contribution can be obtained by exchanging $t$ and $u$.
The result is completely symmetric under the exchange $\mg\leftrightarrow\mc$.

The infrared singular pieces of the third $t$-channel box diagram in 
Fig.~\ref{gn_box} are
\bea
 \overline{|{\cal M}^{\rm Box 3}_{\rm IR}|}^2 
 &=&-C_{\epsilon}\left[ 
      \frac{1}{2\epsilon^2}
     +\frac{1}{2\epsilon}\log\left(\frac{Q^2}{\mg^2}\right)
     -\frac{1}{\epsilon}\log\left(\frac{\ms^2-t}{\ms^2-\mc^2}\right)
      \frac{\ms^2-\mc^2}{t-\mc^2}
     -\frac{1}{\epsilon}\log\left(\frac{\mg^2-u}{\mg^2}\right)
     \right] \nonu
 & & \times 4 \frac{N_C}{2} g_s^2
     [\overline{|{\cal M}^{t}{\cal M}^{t\ast}|}
     +\overline{|{\cal M}^{t}{\cal M}^{u\ast}|}] .
\eea
For the $u$-channel result $t$ and $u$ have to be exchanged.

Finally we sum the infrared singularities encountered in the virtual
corrections and separate them into $C_F$ and $N_C$ color classes:
\bea
 \overline{|{\cal M}^{V}_{\rm IR}|}^2 
 &=&-C_{\epsilon}\left[\frac{1}{\epsilon^2}+\frac{3}{2\epsilon}
     +\frac{1}{\epsilon}\log\left(\frac{Q^2}{s}\right)\right]
     4 C_F g_s^2 \overline{|{\cal M}^B|}^2 \nonu
 & &+C_{\epsilon}\left[-\frac{1}{\epsilon}+\frac{1}{\epsilon}\log
     \left(\frac{(t-\mg^2)(u-\mg^2)}{s\mg^2}\right) \right] 
     4 \frac{N_C}{2} g_s^2 \overline{|{\cal M}^B|}^2 .
\eea

\newpage
\section{Scalar Integrals}
\label{scalarint}

The tensor integrals that occur in virtual loop diagrams can be reduced to one-,
\mbox{two-,} three-, and four-point integrals that are scalar functions of the
loop momentum $q$ \cite{pasvelt}. The general scalar one-point integral
is defined as
\bea
 A_0(m) &=& Q^{2\epsilon} \int\frac{\mbox{d}^nq}{(2\pi)^n}
 \frac{1}{q^2-m^2} .  
\eea
The general scalar two-point integral is
\bea
 B_0(p^2;m_1,m_2) &=& Q^{2\epsilon} 
 \int\frac{\mbox{d}^nq}{(2\pi)^n}
 \frac{1}{[q^2-m_1^2][(q+p)^2-m_2^2]},
\eea
and its derivative is 
\bea
 B_0'(p^2;m_1,m_2) &=& \frac{\partial}{\partial q^2} B_0(q^2;m_1,m_2)
 \Bigg|_{q^2=p^2} .
\eea
The general scalar three-point integral is
\bea
 C_0(p_1^2,p_2^2,(p_1+p_2)^2;m_1,m_2,m_3) &=& Q^{2\epsilon} \nonu
 &&\hspace*{-5cm} \times \int\frac{\mbox{d}^nq}{(2\pi)^n}
 \frac{1}{[q^2-m_1^2][(q+p_1)^2-m_2^2][(q+p_1+p_2)^2-m_3^2]},
\eea
and the general scalar four-point integral is
\bea
 D_0(p_1^2,p_2^2,p_3^2,(p_1+p_2+p_3)^2,(p_1+p_2)^2,(p_2+p_3)^2;m_1,m_2,
 m_3,m_4) &=& Q^{2\epsilon} \nonu
 &&\hspace*{-12cm} \times \int\frac{\mbox{d}^nq}{(2\pi)^n}
 \frac{1}{[q^2-m_1^2][(q+p_1)^2-m_2^2][(q+p_1+p_2)^2-m_3^2]
 [(q+p_1+p_2+p_3)^2-m_4^2]}.
\eea
The $p_i, i=1...4,$ are the four-momenta of the external particles, and
the $m_i, i=1...4,$ are the masses of the adjacent internal particles.

The scalar two- and three-point integrals relevant for the associated
production of gluinos and gauginos were calculated previously in a
different physical context \cite{hopker}. We recalculated them and
checked that the results in Ref.~\cite{hopker} are correct. The divergent
four-point integrals that contribute to the gluon exchange box diagrams in
Fig.~\ref{gn_box} were unknown, and they are presented here for the first
time. We calculate the absorptive parts with Cutkosky cutting rules
and the real parts with dispersion techniques.

The exchange of a massless gluon between the initial state quark and
antiquark in the first box diagram of Fig.~\ref{gn_box} leads to the
following divergent four-point integral:
\begin{eqnarray}
 D_0(0,0,\mg^2,\mc^2,s,t;0,0,0,m_{\tilde{q}}) &=&
 iC_{\epsilon} \frac{1}{s(t-\ms^2)}
 \le \frac{1}{\epsilon^2}-\frac{1}{\epsilon}
 \ln\frac{s(t-\ms^2)^2}{(\ms^2-\mg^2)(\ms^2-\mc^2)Q^2} \rp \nonumber \\
 &&\hspace*{-5cm}
 -2\li\lr 1+\frac{\ms^2-\mg^2}{t-\ms^2}\rr
 -2\li\lr 1+\frac{\ms^2-\mc^2}{t-\ms^2}\rr
 - \li\lr 1+\frac{(\ms^2-\mg^2)(\ms^2-\mc^2)}{s\ms^2}\rr \nonumber \\
 &&\hspace*{-5cm}
 -\frac{\pi^2}{4}
 +\frac{1}{2}\ln^2\lr\frac{s}{Q^2}\rr
 -\frac{1}{2}\ln^2\lr\frac{s}{\ms^2}\rr
 +2\ln\lr\frac{s}{Q^2}\rr\ln\lr\frac{t-\ms^2}{\ms^2}\rr \nonumber \\
 &&\hspace*{-5cm}
 \lp -\ln\lr\frac{\ms^2-\mg^2}{Q^2}\rr\ln\lr\frac{\ms^2-\mg^2}{\ms^2}\rr
 -\ln\lr\frac{\ms^2-\mc^2}{Q^2}\rr\ln\lr\frac{\ms^2-\mc^2}{\ms^2}\rr \re .
\end{eqnarray}
This integral also contributes to the pair production of gauginos of
unequal mass \cite{gaugino}. In the limit of two final state particles
of equal mass our result agrees with Ref.\ \cite{hopker}.

In the third box diagram of Fig.~\ref{gn_box}, the exchange of a massless
gluon between the the final state gluino and the initial state antiquark
gives rise to a second divergent four-point integral:
\begin{eqnarray}
 D_0(0,m_{\tilde{g}}^2,0,m_{\tilde{\chi}}^2,t,u;m_{\tilde{q}},m_{\tilde{g}},
 0,0) &=&
 iC_{\epsilon}\frac{1}{(t-\ms^2)(u-\mg^2)}
 \le\frac{1}{2\epsilon^2}
 -\frac{1}{\epsilon}\lr\ln\lr\frac{-t+\ms^2}{\ms^2-\mc^2}\rr\rp\rp\nonumber\\
 &&\hspace*{-5cm}
 \lp +\ln\lr\frac{-u+\mg^2}{\mg Q}\rr\rr
 -\frac{\pi^2}{8}
 +\ln^2\lr\frac{-u+\mg^2}{\mg Q}\rr
 +2\ln\lr\frac{-u+\mg^2}{\mg Q}\rr\ln\lr\frac{-t+\ms^2}{\ms^2-\mc^2}\rr
  \nonumber \\
 &&\hspace*{-5cm}
 \lp +\li\lr 1+\frac{u-\mg^2}{\ms^2-\mc^2}\rr
 +\li\lr 1+\frac{\ms^2(u-\mg^2)}{\mg^2(\ms^2-\mc^2)}\rr
 -2\li\lr 1+\frac{\ms^2-\mc^2}{t-\ms^2}\rr \re .
\end{eqnarray}
Again, our result agrees with Ref.\ \cite{hopker} for the case of two
final state particles of equal mass.

The supersymmetric counterpart of the first box diagram is the second box
diagram in Fig.~\ref{gn_box}. Since supersymmetry is broken and the gluino
propagator and the two squark propagators are massive, this diagram does not
have infrared divergences. The corresponding four-point integral can be
expressed in terms of 16 dilogarithms \cite{scharf}.
\newpage
\section{Soft Gluon Emission Contribution}
\label{softgapp}
In this Appendix, we collect the expressions for the finite pieces
of the soft gluon emission contribution.  The finite pieces remain after 
mass-factorization and cancellation of soft poles between soft gluon
emission and virtual contributions, as described in the text.
We begin with the $C_F$ color class,
\bea
   \frac{d^2\hat{\sigma}^S}{dt_2 \, du_2} &=& 
   \frac{d^2 {\hat{\sigma}}^B}{dt_2 \, du_2} 
   \left( \frac{C_F \, \alphas}{\pi} \right) \left\{
     {\rm Li}_2 \left( \frac{ u_2 \, t_2 - s \, m_2^2}{(s+t_2)(s+u_2)} \right) 
  \right. \\[0.3cm] & & \left.
   + \frac{1}{2} \log^2 \left( \frac{ \mu^2}{m_1^2 \delta^2} \right)
   + \log \left(\frac{(s+t_2)(s+u_2)}{s \, m_1^2} \right)
     \log \left( \frac{ \mu^2}{m_1^2 \delta^2} \right) \right. 
\nonumber \\[0.3cm]
& & \left.
   + \frac{1}{2}\log^2 \left( \frac{(s+t_2)(s+u_2)}{s \, m_1^2} \right)
   \right\} , \nonumber
\eea
where $\delta = \Delta / m_1^2$ is the cut-off between hard and soft
emission mentioned in the text.  Its appearance in these terms
is the explicit cut-off dependence that matches the implicit
logarithmic behavior of the hard real emission contributions.

The soft terms associated with the $N_C$ color class are 
\bea
   \frac{d^2\hat{\sigma}^S}{dt_2 \, du_2} &=& 
   -\frac{d^2 {\hat{\sigma}}^B}{dt_2 \, du_2} 
   \left( \frac{N_C \, \alphas}{2 \pi} \right) \left\{
     -2 + 
     {\rm Li}_2 \left( \frac{ u_2 \, t_2 - s \, m_2^2}{(s+t_2)(s+u_2)} \right) 
  \right. \\[0.3cm] & & \left.
   + \frac{1}{2} \log^2 \left( \frac{ \mu^2}{m_1^2 \delta^2} \right)
  \right. \nonumber \\[0.3cm] & & \left.
      + \left[ \log \left(\frac{(s+t_2)(s+u_2)}{s \, m_1^2} \right)
     - 1 \right]
     \log \left( \frac{(s+t_2)(s+u_2)}{s \, m_1^2} \right) \right\} .
\nonumber
\eea
Again logarithmic dependence on the hard/soft cut-off $\delta$ is apparent.

\newpage
\section{Hard Gluon Emission Contribution}
\label{hardgapp}

In this Appendix, we collect explicit expressions for the thirty-six 
matrix elements that contribute to the real emission of a gluon 
in the 2-to-3 partonic subprocess 
$q \, \bar{q} \ra g \, \gluino \, \gaugino$.

The hard gluon emission cross section (in four dimensions) is 
\bea
  \frac{d^3 \hat{\sigma}^{h}}{ds_4 \, dt_2 \, du_2} &=&
  \frac{d^3 \hat{\sigma}_1^{g}}{ds_4 \, dt_2 \, du_2} +
  \frac{\alphas \, \alphash}{16 \, \pi^2} \,
  \frac{ s_4 \, \delta \, (s + t_2 + u_1 - s_4)}{36 \, s^2 \,
  (s_4 + m_1^2)}
  \, \sum_{i=1..8} \, \sum_{j=i..8} \, 
  \hat{M}^{g}_{i j} ,
\eea
where $i$ and $j$ label the diagrams in Fig.~\ref{realglu}.
The remainder of the factorization process is 
\bea
  \frac{d^3 \hat{\sigma}_1^{g}}{ds_4 \, dt_2 \, du_2} &=&
  \frac{ C_F \, \alphas \, \alphash \, \delta \, (s + t_2 + u_1 - s_4)}
  {36 \, \pi \, s^2}
  \log \left( \frac{\mu^2 (s_4 + m_1^2)}{s_4^2} \right) \\[0.3cm]
  & & \times \, \left\{
  \left( \frac{s_4^2 - 2 \, s_4 (s+u_2) + 2 (s+u_2)^2}{s_4 (s+u_2)} \right) 
  \, \left( \frac{X_t \, t_2}{(t - \msqu{t}^2)^2}
  \right. \right. \nonumber \\[0.3cm]
  & & \left. \left.
  + \frac{2 \, X_{t u} \, s \, m_1 \, m_2}
  {(t- \msqu{t}^2) [(\delu-s-t_2)(s+u_2) + s \, s_4]} 
  + \frac{X_u \, u_2 \, [s_4 \, u_2 - u_1 \, (s+u_2)]}
  {[(\delu-s-t_2)(s+u_2) + s \, s_4]^2} \right) \right. \nonumber \\[0.3cm]
   & & + \left.
  \left( \frac{s_4^2 - 2 \, s_4 (s+t_2) + 2 (s+t_2)^2}{s_4 (s+t_2)} \right) 
  \,\left( \frac{X_t \, t_2 \, [s_4 \, t_2 - t_1 \, (s+t_2)]}
  {[(\delt-s-u_2)(s+t_2) + s \, s_4]^2}
  \right. \right. \nonumber \\[0.3cm]
  & & + \left. \left.
  \frac{2 \, X_{t u} \, s \, m_1 \, m_2}
  {(u- \msqu{u}^2) [(\delt-s-u_2)(s+t_2) + s \, s_4]}
  + \frac{X_u \, u_2}{ (u - \msqu{u}^2)^2 }
  \right) \right\} . \nonumber
\eea
The elements $\hat{M}^{g}_{i j}$ are 
\bea
  \hat{M}^{g}_{1 1} &=& \frac{-16 \, C_F \, \pi \, X_t \, t_2 \,
  (s_4 + m_1^2)}
  {(t - \msqu{t}^2)^2 \, (s + u_2)} ,  \\[0.4cm]
  \hat{M}^{g}_{1 2}&=& \left( \frac{8 \, (C_F - N_C / 2) \, X_t}
  {(t - \msqu{t}^2) \, (\delt - s - u_2)}
  \right) \nonumber \\
  & & \times 
  \left\{ s\, [-2 \, t_2 \, (s+u_2) + s_4 \, (s+t_2+u_2)] \ih{1}{t' \, u'}
  + [s - t_2 ] \ih{u'}{t'} \nonumber \right. \\
  & & + \, [\delt \, s \, (\delt + t_1)
  + t_2 \, (s \, (m_2^2 - m_1^2) + \delt \, 
  (t_2 + u_2 - \delt) - u_2 t_2)] \ih{1}{t' \, \ud7} \nonumber \\
  & & + \, \left. [\delt \, (s + u_2 - t_2) + t_2 \, 
  (\delt - s - u_2)]\ih{1}{\ud7}
  + [s - t_2] \ih{\ud7}{t'} \right\} , \nonumber
\eea
\bea
  \hat{M}^{g}_{1 3}&=& \frac{8 \, N_C \, X_t \, t_2 
  \, (s + u_2 - m_1^2)}
  {s_4 \, (t - \msqu{t}^2)^2}
  \, \hat{I}(1) , \nonumber \\[0.4cm]
  \hat{M}^{g}_{1 4}&=& \left( \frac{4 \, (C_F - N_C / 2) \, X_t \, t_2}
  {(t - \msqu{t}^2)^2} \right)
  \left\{ [4 \, t_1 \, (s+u_2)]\ih{1}{u' \, \ud7} \nonumber \right. \\
  & & + \, \left. [2 \, (2 \, t_2 + 2 m_2^2 - s - u_2)] \ih{1}{\ud7} 
  - 2 \ih{u'}{\ud7} \right\} , \nonumber \\[0.4cm]
  \hat{M}^{g}_{1 5}&=& \frac{-16 \, (C_F - N_C / 2) 
   \, X_{t u} \, s^2 \, m_1 \, m_2}
  {(t - \msqu{t}^2) \, (u - \msqu{u}^2)}
  \ih{1}{t' \, u'} , \nonumber \\[0.4cm]
  \hat{M}^{g}_{1 6}&=& \left( \frac{-16 \, C_F \, X_{t u} \, m_1 \, m_2}
  { (t - \msqu{t}^2)} \right) \left\{ [\delu - s - t_2]
  \, \ih{1}{u' \, \u6d} \right. \nonumber \\
  & & - \, \left. \frac{2 \, \pi (s_4 + m_1^2) 
  (\delu - s - t_2)}{s_4 \, [(\delu-s-t_2)(s+u_2) + s \, s_4]}
  + \frac{2 \, \pi (s_4 + m_1^2)}{s_4 \, (s+u_2)} \right\} , \nonumber
  \\[0.4cm]
  \hat{M}^{g}_{1 7}&=& \left( \frac{4 \, N_C \, X_{t u} \, m_1 \, m_2}
  {s_4 \, (t - \msqu{t}^2) \, (u - \msqu{u}^2)} \right) \left\{
  [s+u_2]\ih{t'}{u'} + [s-t_2] \, \hat{I}(1) 
  \right\}, \nonumber \\[0.4cm]
  \hat{M}^{g}_{1 8}&=& \left( \frac{-4 \, C_F \, X_{t u} \, m_1 \, m_2}
  {(t - \msqu{t}^2)(u - \msqu{u}^2)} \right) \left\{
  2\,[s\,(\delu + m_2^2 - m_1^2)- u_2 \, (\delu -t_2)]
  \ih{1}{u' \, \u6d} \right. \nonumber \\
  & & \left. + \, [2 \, t_2]\ih{1}{\u6d} \right\} , \nonumber \\[0.4cm]
  \hat{M}^{g}_{2 2}&=& \left( 8 \, C_F \, X_t \right) \left\{
  [\delt (\delt+t_1)]\ih{1}{t' \, \ud7^2} 
  - [2 \delt + t_1]\ih{1}{t' \ud7} + \delt \ih{1}{\ud7^2}
  \right. \nonumber \\
  & & \left. 
  - \ih{1}{\ud7}
  - \, \left( \frac{2 \, \pi (s_4 + m_1^2)}{s_4} \right)
  \left( \frac{ (s+t_2)\, \delt \, (\delt + t_1)}
  {[(\delu-s-t_2)(s+u_2) + s \, s_4]^2} \right. \right. \nonumber \\
  & & \left. \left. \, - \, \frac{2\, \delt + t_1}
  {[(\delu-s-t_2)(s+u_2) + s \, s_4]} + \frac{1}{(s+t_2)} \right)
  \right\} , \nonumber
\eea
\bea
  \hat{M}^{g}_{2 3}&=& \left( \frac{-4 \, N_C \, X_t }
  {s_4 \, (t - \msqu{t}^2) } \right) \times \nonumber \\
  & & \, \left\{
  [\delt ((s+t_2)(\delt + m_2^2 - m_1^2) + s_4 t_2) + 
  t_2 (s s_4 - s^2 - s t_2 - s u_2 - t_2 u_2)]\ih{1}{t' \, \ud7}
  \right. \nonumber \\
  & & \left. \, + \, [s+t_2]\ih{\ud7}{t'}
  + [\delt (m_1^2 + m_2^2 + s_4) - t_2 (s + u_2 - \delt)]\ih{1}{\ud7}
  \right . \nonumber \\
  & & \left. \, - \, [t + m_1^2 + s_4] \, 
  \hat{I}(1) \right\} , \nonumber  \\[0.4cm]
  \hat{M}^{g}_{2 4}&=& \left( \frac{-4 \, C_F \, X_t}{(t- \msqu{t}^2)}
  \right)
  \left\{ [4 \, t_2 \, \delt \, (\delt+m_2^2-m_1^2)]
  \ih{1}{t' \, \ud7^2} + [2 \, \delt (t_2 + 2 \, m_2^2)]\ih{1}{\ud7^2}
  \right. \nonumber \\
  & & - \, \left. [4\, t_2 \, (2 \delt + m_2^2 - m_1^2)]
  \ih{1}{t' \, \ud7} - 2 \, [t_2 + 2 \, m_2^2]\ih{1}{\ud7} \right\} ,
  \nonumber \\[0.4cm]
  \hat{M}^{g}_{2 5}&=& \left( \frac{-16 \, C_F \, X_{t u} \, m_1 \, m_2}
  {(u - \msqu{u}^2)} \right) \left\{
  [\delt-s-u_2]\ih{1}{t' \, \ud7} \right. \nonumber \\
  & & - \, \left. \frac{2 \, \pi (\delt-s-u_2)(s_4 + m_1^2)}
  {s_4 \, [(\delt-s-u_2)(s+t_2) + s \, s_4]}
  + \frac{2 \, \pi (s_4 + m_1^2)}{s_4 \, (s+t_2)} 
  \right\} , \nonumber \\[0.4cm]
  \hat{M}^{g}_{2 6}&=& \left( \frac{-16 \, (C_F - N_C / 2) \, X_{t u} \, s 
            \, m_1 \, m_2}
  {(\delt-s-u_2)(\delu-s-t_2)} \right) \left\{
  [s] \ih{1}{t' \, u'} + [\delt-u_2]\ih{1}{t' \, \ud7} 
  \right. \nonumber \\
  & & + \, \left. [\delu - t_2]\ih{1}{u' \, \u6d}
  + [\delu + \delt -s -t_2 - u_2]\ih{1}{\u6d \ud7}
  \right\} , \nonumber \\[0.4cm]
  \hat{M}^{g}_{2 7}&=& \left( \frac{4 \, N_C \, X_{t u} \, \, m_1 \, m_2}
  {s_4 \, (u - \msqu{u}^2)} \right) \left\{
  [\delt s + s^2 - s s_4 + \delt t_2 + s t_2 - s u_2 - t_2 u_2]
  \ih{1}{t' \, \ud7} \right. \nonumber \\
  & & + \, \left. [s+u_2 - 2 \delt]\ih{1}{\ud7} +2 \hat{I}(1)
  \right\} , \nonumber \\[0.4cm]
  \hat{M}^{g}_{2 8}&=& \left( \frac{4 \, (C_F - N_C / 2) \, X_{t u} 
       \, \, m_1 \, m_2}
  {(u - \msqu{u}^2)(\delu-s-t_2} \right) \left\{
  2 \, [s u_1 - (s+t_2)(\delt-u_2)]\ih{1}{t' \, \ud7} 
  \right. \nonumber \\ & & \, \left.
  + 2 \, [s+t_2] \ih{1}{\u6d}
  + [-2 \, \delt \, (s + t_2) + 2 \, \delu \, u_2 + 2 \, s \, u_1] 
  \ih{1}{\u6d \, \ud7}
  \right\} , \nonumber \\[0.4cm]
  \hat{M}^{g}_{3 3}&=& \left( \frac{8 \, N_C \, X_{t} \, t_2}
  {s_4^2 \, (t - \msqu{t}^2)^2} \right) \left\{
  2 \, m_1^2 \, (s+u_2)\, \hat{I}(1) + s_4 \, \hat{I}(\, u' \,)
  \right\} , \nonumber \\[0.4cm]
  \hat{M}^{g}_{3 4}&=& \left( \frac{-4 \, N_C \, X_{t} \, t_2}
  {s_4 \, (t - \msqu{t}^2)^2} \right) \left\{
  [\delt (s_4 - 2 \, t) - 2 \, m_1^2 (u_2 + 2 s)]\ih{1}{\ud7}
  \right. \nonumber \\ & & \, \left.
  + [m_2^2 + m_1^2 - s + t_2 - u_2]\, \hat{I}(1)
  \right\} , \nonumber
\eea
\bea
  \hat{M}^{g}_{3 5}&=& \left( \frac{4 \, N_C \, X_{t u} \, m_1 \, m_2}
  {s_4 \, (t - \msqu{t}^2) (u - \msqu{u}^2)} \right) \left\{
  [s+t_2]\ih{u'}{t'} + [s - u_2]\, \hat{I}(1)
  \right\} , \nonumber \\[0.4cm]
  \hat{M}^{g}_{3 6}&=& \left( \frac{4 \, N_C \, X_{t u} \, m_1 \, m_2}
  {s_4 \, (t - \msqu{t}^2)} \right) \left\{
  [\delu(s+u_2) + s^2 - s \, s_4 - s \, t_2 + s \, u_2 - t_2 \, u_2]
  \ih{1}{u' \, \u6d} \right. \nonumber \\
  & & \left. \, + [s+t_2 - 2 \delu]\ih{1}{\u6d} + 2 \, \hat{I}(1)
  \right\} , \nonumber  \\[0.4cm]
  \hat{M}^{g}_{3 7}&=& \frac{16 \, N_C \, X_{t u} \, s \, m_1 \, m_2
  \, (s_4 + 2 \, m_1^2) }
  {s_4^2 \, (t - \msqu{t}^2) \, (u - \msqu{u}^2)} 
  \, \hat{I}(1) , \nonumber \\[0.4cm]
  \hat{M}^{g}_{3 8}&=& \left( \frac{-4 \, N_C \, X_{t u} \, m_1 \, m_2}
  {s_4 \, (t - \msqu{t}^2) (u - \msqu{u}^2)} \right) \left\{
  [u_2 \, (t_2 - \delu) + s (\delu - s_4 - m_2^2 - 3 m_1^2)]
  \ih{1}{\u6d} \right. \nonumber \\
  & & + \, \left. [s + t_2] \ih{u'}{\u6d} + [u_2 - s]\, \hat{I}(1)
  \right\} , \nonumber \\[0.4cm]
  \hat{M}^{g}_{4 4}&=& \left( \frac{-8 \, C_F \, X_{t} \, t_2}
  {(t - \msqu{t}^2)^2} \right) \left\{
  [\delt (\delt - t - m_1^2)] \ih{1}{\ud7^2}
  +[t + m_1^2 - 2 \delt]\ih{1}{\ud7} + \hat{I}(1)
  \right\} , \nonumber \\[0.4cm]
  \hat{M}^{g}_{4 5}&=& \left( \frac{-8 \, C_F \, X_{t u} \, m_1 \, m_2}
  {(t - \msqu{t}^2)\, (u - \msqu{u}^2)} \right) \left\{
  [t_2 (u_2 - \delt) + s (\delt + m_2^2 - m_1^2)]
  \ih{1}{t' \, \ud7} + u_2 \ih{1}{\ud7}
  \right\} , \nonumber \\[0.4cm]
  \hat{M}^{g}_{4 6}&=& \left( \frac{4 \, (C_F - N_C / 2) 
    \, X_{t u} \, m_1 \, m_2}
  {(t - \msqu{t}^2)\, (\delt - s -u_2)} \right) \left\{
  2\,[(t_2-\delu)(s+u_2) + s \, t_1]\ih{1}{u' \, \u6d}
  \right. \nonumber \\ & & \, \left. 
  + 2\,[s+u_2] \ih{1}{\ud7} 
  + [-2 \, \delu \, (s+u_2) + 2 \, s \, t_1 
  + 2 \, \delt \, t_2] \ih{1}{\u6d \ud7}
  \right\} , \nonumber \\[0.4cm]
  \hat{M}^{g}_{4 7}&=& \left( \frac{-4 \, N_C \, X_{t u} \, m_1 \, m_2}
  {s_4 \, (t - \msqu{t}^2)\, (u - \msqu{u}^2)} \right) \left\{
  [t_2 \, (u_2 - \delt) + \delt \, s - s \,
  (s_4 + m_2^2 + 3 m_1^2)] \ih{1}{\ud7} \right. \nonumber \\
  & & + \, \left. [s + u_2] \ih{t'}{\ud7} + [t_2 - s]
  \, \hat{I}(1)
  \right\} , \nonumber \\[0.4cm]
  \hat{M}^{g}_{4 8}&=& \left( \frac{-8 \, (C_F - N_C / 2) 
   \, X_{t u} \, s \, m_1 \, m_2}
  {(t - \msqu{t}^2)\, (u - \msqu{u}^2)} \right) \left\{
  [2 \, s + 2 \, m_1^2 - \delt - \delu + t + u] \ih{1}{\u6d \, \ud7}
  \right. \nonumber \\
  & & + \, \left. \ih{1}{\u6d} + \ih{1}{\ud7}
   \right\} , \nonumber \\[0.4cm]
  \hat{M}^{g}_{5 5}&=& \frac{-16 \, C_F \, \pi \,
  X_u \, u_2 \, (s_4 + m_1^2)}
  {(u - \msqu{u}^2)^2 \, (s + t_2)} , \nonumber 
\eea
\bea
  \hat{M}^{g}_{5 6}&=& \left( \frac{8 \, (C_F - N_C / 2) \, X_u }
  {(u - \msqu{u}^2)\, (\delu - s - t_2)} \right) \nonumber \\
  & & \times \, \left\{
  [\delu^2(s-u_2) + \delu(s u_1 + u_2(t_2 + u_2)) + s \, u_2 
  (m_2^2 - m_1^2) - t_2 u_2^2]\ih{1}{u' \, \u6d} \right. 
  \nonumber \\
  & & + \, \left. [s-u_2]\ih{\u6d}{u'} + 
  [\delu(s+t_2) - u_2 (s+t_2)]\ih{1}{\u6d}
  \right. \nonumber \\
  & & + \, \left. s \, [(s+t_2)(s+t_1) + u_2 \, u_1]\ih{1}{u' \, t'}
  +[s-u_2]\ih{t'}{u'}
   \right\} , \nonumber \\[0.4cm]
  \hat{M}^{g}_{5 7}&=& \left( \frac{8 \, N_C \, X_u \, u_2 \,
  [s + t_2 - m_1^2] }
  {s_4 \, (u - \msqu{u}^2)^2} \right) \hat{I}(1) , \nonumber \\[0.4cm]
  \hat{M}^{g}_{5 8}&=& \left( \frac{4 \, (C_F - N_C / 2) \, X_u \, u_2}
  {(u - \msqu{u}^2)^2} \right) \left\{
  4 \, u_1 \, [s+t_2]\ih{1}{t' \, \u6d}
  + 2 \, [3 \, u_1 + 2 m_1^2 - s_4]\ih{1}{\u6d}
  \right. \nonumber \\ & & \, \left.
  - 2 \ih{t'}{\u6d}
   \right\} , \nonumber \\[0.4cm]
  \hat{M}^{g}_{6 6}&=& \left( 8 \, C_F \, X_u
  \right) \left\{
  \delu [\delu + u_1]\ih{1}{u' \, \u6d^2}
  -[2 \, \delu + u_1]\ih{1}{u' \, \u6d}
  \right. \nonumber \\
  & & + \, \left.
  \delu \ih{1}{\u6d^2} - \ih{1}{\u6d} \right. \nonumber \\
  & & - \, \left. \left( \frac{2 \, \pi (s_4 + m_1^2)}{s_4} 
  \right) \left( \frac{\delu (\delu + u_1)(s+u_2)}
  {[(\delu-s-t_2)(s+u_2)+ s \, s_4]^2} \right. \right. \nonumber \\
  & & - \, \left. \left. \frac{2 \, \delu + u_1}
  {[(\delu-s-t_2)(s+u_2)+ s \, s_4]} + \frac{1}{s+u_2} \right)
   \right\} , \nonumber \\[0.4cm]
  \hat{M}^{g}_{6 7}&=& \left( \frac{-4 \, N_C \, X_u}
  {s_4 \, (u - \msqu{u}^2)} \right) \left\{
  [s+u_2]\ih{\u6d}{u'} - [s_4 + m_1^2 + u]\, \hat{I}(1) 
  \right. \nonumber \\
  & & + \, \left. [\delu^2 (s+u_2) + \delu (s \, (m_2^2-m_1^2)
  \right. \nonumber \\
  & & + \, \left. 
  u_2 \, (m_2^2 - m_1^2)) 
  - t_2 \, u_2) ]\ih{1}{u' \, \u6d}
   \right. \nonumber \\
  & & + \, \left.
  [\delu (s + t + u) + u_2 (\delu -s -t_2)]\ih{1}{\u6d}
   \right\} , \nonumber \\[0.4cm]
  \hat{M}^{g}_{6 8}&=& \left( \frac{-4 \, C_F \, X_u}
  {(u - \msqu{u}^2)} \right) \left\{
  4 \, \delu \, u_2 \, [\delu + m_2^2 - m_1^2] \ih{1}{u' \, \u6d^2}
  + 2 \, \delu [u_2 + 2 \, m_2^2] \ih{1}{\u6d^2}
  \right. \nonumber \\
  & & + \, \left. 4 \, u_2 [-2 \delu - m_2^2 + m_1^2]\ih{1}{u' \, \u6d}
  -2 \, [u_2 + 2 m_2^2] \ih{1}{\u6d}
   \right\} , \nonumber
\eea
\bea
  \hat{M}^{g}_{7 7}&=& \left( \frac{4 \, N_C \, X_u \, u_2}
  {s_4^2 \, (u - \msqu{u}^2)^2} \right) \left\{
  4 m_1^2 [s + t_2] \, \hat{I}(1)
  + 2 \, s_4 \, \hat{I}(t')
   \right\} , \nonumber \\[0.4cm]
  \hat{M}^{g}_{7 8}&=& \left( \frac{-2 \, N_C \, X_u \, u_2}
  {s_4 \, (u - \msqu{u}^2)^2} \right) \left\{
  2 \, [m_2^2 + m_1^2 - s - t_2 + u_2] \, \hat{I}(1)
  \right. \nonumber \\
  & & + \, \left.
  [s_4(s+t_2) +\delu (\delu - 3 m_2^2 - 5 m_1^2 - 3 u_2)
  \right. \nonumber \\
  & & + \, \left.
  (\delu - s - t_2)(-\delu + m_2^2 + 3 m_1^2 + s + t_2 + u_2)]
  \ih{1}{\u6d}
   \right\} , \nonumber  \\[0.4cm]
  \hat{M}^{g}_{8 8}&=& \left( \frac{-4 \, C_F \, X_u \, u_2}
  {(u - \msqu{u}^2)^2} \right) \left\{
  -2 \, \delu \, [-\delu + m_2^2 + m_1^2 + u_2] \ih{1}{\u6d^2}
  \right. \nonumber \\
  & & + \, \left.
  2 \, [u_2 + m_2^2 + m_1^2 - 2 \, \delu] \ih{1}{\u6d}
  + 2 \, \hat{I}(1)
   \right\} . \nonumber
\eea


\newpage
\section{(Anti-) Quark Emission Contributions}
\label{qapp}

\def\nl{\right. \nonumber \\ && \: \left.}
\def\square1{[(s+u_2)(\delu-s-t_2) + s \, s_4]}

\indent \indent
In this Appendix we present the matrix elements for the contributions 
from emission of an additional light quark or anti-quark in the final state.  
These matrix elements may be obtained from those for gluon emission upon 
crossing one of the initial state partons and the final state gluon. However,
we compute them {\it ab initio} in order to provide
a check on the validity of the results.  In all cases, the result of 
the crossing agrees with the explicit computation.  We limit our 
presentation to the quark emission contributions since the 
anti-quark emission contributions may be obtained from these
expressions by replacements of $t \leftrightarrow u$,  
$t' \leftrightarrow u'$, and
$u_6 \leftrightarrow u_7$ everywhere.

The quark emission contribution has two parts, consisting of the remainder 
of terms that are collinear singular, and thus have the angular integrations  
done analytically, as well as a large class of terms which are collinear 
finite and for which the angular integrations are done numerically.
The matrix elements with collinear singularities may be written
in a form very similar to the gluon emission cross section, with
\bea 
  \frac{d^3 \hat{\sigma}^{h}}{ds_4 \, dt_2 \, du_2} &=& 
  \frac{d^3 \hat{\sigma}_1^{q}}{ds_4 \, dt_2 \, du_2} + 
  \frac{\alphas \, \alphash}{16 \, \pi^2} \, 
  \frac{ s_4 \, \delta \, (s + t_2 + u_1 - s_4)}{96 \, s^2 \, 
  (s_4 + m_1^2)} 
  \, \sum_{i=1..2} \, \sum_{j=i..8} \,  
  \hat{M}^{q}_{i j} . 
\eea 
The remainder of the factorization is  
\bea
\frac{d^3 \hat{\sigma}_1^{q}}{ds_4 \, dt_2 \, du_2} &=&
\frac{\alphas \, \alphash \, C_F}
     {48 \, \pi s^2} 
\left(1 + \log \left[ \frac{\mu^2 (s_4+m_1^2)}{s_4^2 } \right] \right)
\frac{1}{2} \left(\frac{2 s_4^2 - 2 s_4 (s+u_2) + (s+u_2)^2}{(s+u_2)^2} \right)
\nonumber \\ &&
\left\{
\frac{X_t \, t_2}{(t - \msqu{t}^2)^2}
+ \frac{2 \, X_{tu} \, s \, m_1 \, m_2}{(t - \msqu{t}^2) \square1} \nl
+ \frac{X_u \, u_2 (u_2 s_4 - u_1 (s+u_2))}{ \square1^2}
\right\} .
\eea
After partial-fractionation, the collinearly-divergent pieces of the 
hard matrix elements are 
\bea
\hat{M}^{q}_{11} &=&
 \left( -8 C_F X_u u_2 \right)  
 \left\{  -\delu \ih{1}{u' \u6d^2} 
        + \ih{1}{u' \u6d}
        + \frac{\delu 2 \pi (s_4+m_1^2)(s+u_2)}{s_4 \square1^2} 
\right. \\ && \: \left.
        - \frac{2 \pi (s_4+m_1^2)}{s_4 \square1} \right\} , \nonumber \\[0.4cm]
\hat{M}^{q}_{12} &=& 
\left( \frac{16 C_F X_{tu} s m_1 m_2}{(t - \msqu{t}^2)} \right)  
\left\{  \ih{1}{u' \u6d} 
       - \frac{2 \pi (s_4+m_1^2)}{s_4 \square1} \right\} , \nonumber
\eea
\bea
\hat{M}^{q}_{13} &=& 
\left( \frac{4 (C_F - N_C/2) X_{tu} m_1 m_2 \s4d}
            {(\s4d^2 + \msqu{t}^2 \Gamma^2) (t-\msqu{t}^2)} \right)
\left\{ [-2 \delu m_2^2 + 2  \delu m_1^2 + s m_2^2 - s m_1^2 \nl
       - s s - 2  \delu s_4 + s s_4 - 2  \delu t_2
       + 2 m_2^2 t_2 - 2  m_1^2 t_2 + s t_2 + 2  s_4 t_2
       + 2  t_2 t_2 - s u_2] \ih{1}{u' \u6d} \nl
       - [2 \delu] \ih{1}{\u6d} 
       + 2  \hat{I}(1) \right\} , \nonumber \\[0.4cm]
\hat{M}^{q}_{14} &=& 
\left( \frac{4 C_F X_u}{u_2^2} \right)  
\left\{  
      [ 4 (u_2-\delu ) (-\delu - m_2^2 + m_1^2 )  
             (-\delu - m_2^2 + m_1^2 + u_2) ] \ih{1}{u' \sd3} \nl
      + 2 [-6 \delu + 4  m_1^2 - 4  m_2^2 + 4  u_2 ] \ih{\sd3}{u'}
      + 4 \ih{\sd3^2}{u'} \nl
      + [2 (-\delu + m_1^2 - 3  m_2^2 )(-\delu + u_2 ) \nl
         + (-\delu + m_1^2 - m_2^2 + u_2)(-2 \delu + m_1^2 - m_2^2
            + s - s_4 + t_2 + u_2) \nl
         + (-\delu + m_1^2 - m_2^2) (-2  \delu + m_1^2 - m_2^2
           + s - s_4 + t_2 + 3 u_2) ] \ih{1}{\sd3} \nl
      + 2 [-6 \delu + 4 m_1^2 - 6 m_2^2 + s - s_4 + t_2 + 4 u_2] \hat{I}(1)
      + 6 \hat{I}(\sd3) \nl
      + 2 [-\delu + m_1^2 - 3  m_2^2] \ih{u'}{\sd3}
      + 2 \hat{I}(u') \right\} \nonumber \\
&+& \left(\frac{-4  C_F X_u}{u_2} \right)  
    \left\{
     -\delu [ (\delu-m_1^2+m_2^2)(2 \delu - m_1^2 + m_2^2 - s
                                       + s_4 - t_2 + u_2) \nl
            + (-\delu+m_1^2-m_2^2-u_2)(-2  \delu + m_1^2
            - m_2^2 + s - s_4 + t_2 + u_2)] \ih{1}{u' \u6d^2} \nl
     + [ (\delu-m_1^2+m_2^2)
              (2 \delu - m_1^2 + m_2^2 - s + s_4 - t_2 + u_2) \nl
             + 2  \delu 
               (4 \delu - 3  m_1^2 + 3  m_2^2 - s + s_4 - t_2 + u_2) \nl
             + (-\delu + m_1^2 - m_2^2 - u_2)  
               (-2  \delu + m_1^2 - m_2^2 + s - s_4 + t_2 + u_2)
            ] \ih{1}{u' \u6d} \nl
     + 4 \ih{\u6d}{u'} 
     + 2 \delu [-3 \delu + 2  m_1^2 + s - s_4 + t_2] \ih{1}{\u6d^2} \nl
     + 2 [6  \delu - 2  m_1^2 - s + s_4 - t_2] \ih{1}{\u6d}
     - 6 \hat{I}(1) 
     - 2  \delu \ih{u'}{\u6d^2}
     + 2  \ih{u'}{\u6d} \right\} \nonumber
\eea
\bea
&+& \left(\frac{-4  C_F X_u}{u_2^2} \right)  
\left\{ 
     -\delu [ (\delu-m_1^2+m_2^2)(2 \delu - m_1^2 + m_2^2 - s
                 + s_4 - t_2 + u_2) \nl
                + (-\delu+m_1^2-m_2^2-u_2)(-2  \delu + m_1^2
                - m_2^2 + s - s_4 + t_2 + u_2)] \ih{1}{u' \u6d} \nl
     + 2 [-6  \delu + 3  m_1^2 - 3  m_2^2 + s - s_4 + t_2 - u_2] 
       \ih{\u6d}{u'}
     + 4 \ih{\u6d^2}{u'} \nl
     + 2 \delu [-3  \delu + 2  m_1^2 + s - s_4 + t_2] \ih{1}{\u6d}
     + 2 [6 \delu - 2 m_1^2 - s + s_4 - t_2] \hat{I}(1) \nl
     - 6 \hat{I}(\u6d) 
     - 2 \delu \ih{u'}{\u6d}
     + 2 \hat{I}(u') \right\} , \nonumber \\[0.4cm]
\hat{M}^{q}_{15} &=& 
\left( \frac{16 (C_F - N_C/2) X_{tu} m_1 m_2 \s4d}
            {s (\s4d^2 + \msqu{t}^2 \Gamma^2)} \right)  
\left\{   [(-\delu+t_2)(-\delu +s+t_2)] \ih{1}{u' \u6d} \nl
        + [\delu - s - t_2] \ih{1}{\u6d}
        - \hat{I}(1) 
        + \ih{\u6d}{u'} \right\} , \nonumber \\[0.4cm]
\hat{M}^{q}_{16} &=& 
\left( \frac{8 (C_F - N_C/2) X_u}{s u_2} \right)  
\left\{ [(-\delu + m_1^2 - m_2^2) s s_4 + (\delu - m_1^2 + m_2^2) s 
         (-\delu + u_2) \nl
         + 2 (-\delu + m_1^2 - m_2^2)(-\delu + u_2)
         (-\delu + s + t_2 + u_2)
         + u_2 (-\delu + u_2)(-\delu + s + t_2 + u_2) \nl
         + (\delu - m_1^2 + m_2^2)(s + u_2)(-\delu + s + t_2 + u_2)
          ] \ih{1}{u' \sd3} \nl
       + 2 [-3  \delu + m_1^2 - m_2^2 + t_2 + 2 u_2] \ih{\sd3}{u'}
       + 2 \ih{\sd3^2}{u'} \nl
       + [2 \delu \delu - 2 \delu m_1^2 + 2 \delu m_2^2
          + \delu s - m_1^2 s + m_2^2 s - 2  \delu t_2
          + m_1^2 t_2 - m_2^2 t_2 - 3 \delu u_2 \nl
          + m_1^2 u_2 - m_2^2 u_2 + s u_2 + 2  t_2 u_2 
          + 2 u_2 u_2] \ih{1}{\sd3} \nl
       + [-4 \delu + 2 m_1^2 - 2 m_2^2 - s + 2 t_2 + 3 u_2] \hat{I}(1) 
       + 2 \hat{I}(\sd3) 
       + [t_2 + u_2] \ih{u'}{\sd3} \right\} \nonumber \\
&+& \left(\frac{-8 (C_F - N_C/2) X_u}{s u_2} \right)  
\left\{ 
      [s s_4 (-\delu + m_1^2 - m_2^2 - u_2) 
          + \delu u_2 (\delu - s - t_2) \nl
          - \delu s (\delu - m_1^2 + m_2^2 + u_2)
          + 2 \delu (-\delu + s + t_2)(\delu - m_1^2 + m_2^2 + u_2) \nl
          + (s+u_2)(-\delu + s + t_2)(\delu + u_1)] \ih{1}{u' \u6d}
      + 2 [-3 \delu + m_1^2 - m_2^2 + t_2 - u_2] \ih{\u6d}{u'} \nl
      + 2 \ih{\u6d^2}{u'}
      + [-4 \delu \delu - 2 \delu m_2^2 + \delu s
          + 2 \delu m_1^2 - m_1^2 s + m_2^2 s + s^2 \nl
          - s s_4 + 2 \delu t_2 
          - m_1^2 t_2 + m_2^2 t_2 + s t_2 - 3 \delu u_2 + 2 s u_2
          + 2 t_2 u_2] \ih{1}{\u6d} \nl
      + [8 \delu - 2  m_1^2 + 2 m_2^2 - s - 2 t_2 + 3 u_2] \hat{I}(1)
      - 4 \hat{I}(\u6d)  \nl
      + [-2 \delu + s + t_2] \ih{u'}{\u6d}
      + 2 \hat{I}(u') \right\} , \nonumber
\eea
\bea
\hat{M}^{q}_{17} &=& 
\left( \frac{-4 N_C X_u}{u_2 (s+u_2)} \right)  
\left\{ 
      [(\delu - m_1^2 + m_2^2) s s_4 + s_4 u_2 (\delu - u_2)
         + 2 (\delu - m_1^2 + m_2^2) s_4 (-\delu + u_2) \nl
         + (-\delu + m_1^2 - m_2^2)(-\delu + u_2)(s+u_2) \nl
         + (-\delu + m_1^2 - m_2^2)(s+u_2)
         (-\delu + s + t_2 + u_2)] \ih{1}{u' \sd3} \nl
      + 2 [s + u_2 - s_4] \ih{\sd3}{u'}
      + [2 \delu m_2^2 - 3 \delu s + 3 m_1^2 s - 3 m_2^2 s 
         + 2 \delu s_4 - 2 m_1^2 s_4 \nl
         + 2 m_2^2 s_4 + m_1^2 t_2 
         - m_2^2 t_2 - \delu u_2 + 2 m_1^2 u_2 - 4  m_2^2 u_2 - s_4 u_2
         - t_2 u_2 - u_2 u_2] \ih{1}{\sd3} \nl
      + [-2 m_2^2 + 3 s - 2 s_4 + u_2] \hat{I}(1)
      + [-2 m_2^2 - t_2 - u_2] \ih{u'}{\sd3} \right\} \nonumber \\
&+& \left( \frac{4 N_C X_u}{u_2 (s+u_2)} \right)
\left\{ 
     [2 s_4 \delu (-\delu + m_1^2 - m_2^2 - u_2)
         + \delu s_4 u_2 + s s_4 (\delu - m_1^2 + m_2^2 + u_2) \nl
         + \delu (\delu + u_1)(s + u_2)
         + (\delu - s - t_2)(\delu + u_1)(s+u_2)] \ih{1}{u' \u6d} \nl
     + 2 [s + u_2 - s_4] \ih{\u6d}{u'}
     + [2 \delu m_2^2 + \delu s + m_1^2 s - m_2^2 s - s s - 2 \delu s_4 
        + s s_4 \nl
        + m_1^2 t_2 - m_2^2 t_2 - s t_2 + 3  \delu u_2
        - 2 s u_2 - 2 t_2 u_2] \ih{1}{\u6d} \nl
     + [-2 m_2^2 - s + 2 s_4 - 3 u_2] \hat{I}(1)
     - [s+t_2] \ih{u'}{\u6d} \right) , \nonumber \\[0.4cm]
 \hat{M}^{q}_{18} &=& 
\left( \frac{-4 N_C X_{tu} m_1 m_2}{(t - \msqu{t}^2) (s+u_2)} \right) 
     \times \nonumber \\ &&
\left\{ 
      [-\delu s + s s + 2 \delu s_4 - s s_4 + s t_2
         - 2 s_4 t_2 - \delu u_2 + s u_2
         + t_2 u_2] \ih{1}{u' \u6d} \nl
      + [2 \delu + s - t_2] \ih{1}{\u6d} 
      - 2 \hat{I}(1) \right\} , \nonumber \\[0.4cm]
\hat{M}^{q}_{22} &=& 
\left( \frac{8 C_F X_t t_2}{(t - \msqu{t}^2)(t - \msqu{t}^2)} \right)  
\left\{ \hat{I}(1) 
      - \frac{2 \pi (s_4+m_1^2)}{s_4} \right\} , \nonumber \\[0.4cm]
\hat{M}^{q}_{23} &=& 
\left( \frac{4 (C_F - N_C/2) X_t t_2 \s4d}
       {(t-\msqu{t}^2)(t-\msqu{t}^2)(\s4d^2 + \msqu{t}^2 \Gamma^2)} \right)  
\left\{ -2 [m_1^2 + m_2^2 + s + t_2 + u_2] \hat{I}(1) \right\} ,
 \nonumber
\eea
\bea
\hat{M}^{q}_{24} &=& 
\left( \frac{4 C_F X_{tu} m_1 m_2}{u_2 (t - \msqu{t}^2)} \right)  
\left\{  4 \ih{\sd3}{u'} 
       + 6 \hat{I}(1) 
       + 2 \ih{u'}{\sd3} \nl
       + 2 [-3 \delu + m_1^2 - m_2^2 + 2 s - s_4 + 3 t_2 + 3 u_2] \ih{1}{\sd3}
       \nl
       + [4 \delu^2 - 2 \delu m_1^2 + 2 \delu m_2^2
          - 4 \delu s + m_1^2 s - m_2^2 s + s^2
          + 2 \delu s_4 - s s_4 - 6 \delu t_2
          + 2 m_1^2 t_2 \nl
          - 2 m_2^2 t_2 + 3 s t_2
          - 2 s_4 t_2 + 2 t_2^2 - 8 \delu u_2
          + 2 m_1^2 u_2 + 4 u_2^2 - 2 m_2^2 u_2
          + 5 s u_2 \nl
          - 2 s_4 u_2 + 6 t_2 u_2] \ih{1}{u' \sd3} 
\right\} \nonumber \\
&+& 
\left( \frac{-4 C_F X_{tu} m_1 m_2}{u_2 (t - \msqu{t}^2)} \right)  
\left\{  2 \delu \ih{1}{\u6d} 
       - 2 \hat{I}(1) 
       + 4 \ih{\u6d}{u'} \nl
       + [4 \delu^2 - 2 \delu m_1^2 + 2 \delu m_2^2 - 4 \delu s 
          + m_1^2 s - m_2^2 s + s^2 + 2 \delu s_4 - s s_4 - 6 \delu t_2 \nl
          + 2 m_1^2 t_2 - 2 m_2^2 t_2 + 3 s t_2
          - 2 s_4 t_2 + 2 t_2 t_2 + s u_2] \ih{1}{u' \u6d} 
\right\} , \nonumber \\[0.4cm]
  \hat{M}^{q}_{25} &=& 
\left( \frac{8 (C_F - N_C/2) X_t \s4d}
       {s (\s4d^2 + \msqu{t}^2 \Gamma^2) (t-\msqu{t}^2)} \right)  
\left\{  
   [-(s+t_2)(s_4+t_2)] \hat{I}(1) \nl
   + [-s s_4 + s t_2 - 2 s_4 t_2 - s_4 u_2 + t_2 u_2]
        \ih{t'}{u'} \right\} , \nonumber \\[0.4cm]
 \hat{M}^{q}_{26} &=& 
\left( \frac{16 (C_F - N_C/2) X_{tu} m_1 m_2}
            {s (t - \msqu{t}^2)} \right)  
\left\{ 
   [(-\delu+t_2+u_2)(-\delu+s+t_2+u_2)] \ih{1}{u' \sd3} \nl
   + \hat{I}(1) 
   + [-\delu + t_2 + u_2] \ih{1}{\sd3} 
   + \ih{\sd3}{u'} \right\} , \nonumber \\[0.4cm]
 \hat{M}^{q}_{27} &=& 
\left( \frac{-4 N_C X_{tu} m_1 m_2}{(s+u_2) (t - \msqu{t}^2)} \right)  
\left\{ 
  [2 s - 2 s_4 + t_2 + u_2 ] \ih{1}{\sd3} \nl
  + [s s_4 - 2 s_4 (-\delu + s + t_2 + u_2)
   + (s+u_2) (-\delu + s + t_2 + u_2)] \ih{1}{u' \sd3} \right\} , 
  \nonumber \\[0.4cm]
 \hat{M}^{q}_{28} &=& 
\left\{ \frac{8 N_C X_t t_2}{(t - \msqu{t}^2) (t - \msqu{t}^2)} \right)  
   \left( -s_4^2 \ih{1}{u' u_7}
     + (m_1^2 - s_4) \ih{1}{u_7} - \hat{I}(1) \right\} . \nonumber
\eea
A squark width $\Gamma$ is included to regularize a possible
on-shell squark pole.  As discussed in the text, $\Gamma$
serves only to regulate the divergence of these interference terms,
and it is taken to be very small compared to all physical masses and momenta.

The finite pieces of the quark emission terms are evaluated directly from 
Eq.~(\ref{ph3}) with the angular integrals 
integrated numerically.  They may be expressed as
\bea
   \overline{|{\cal M}^q|}^2 &=&
   \, \left( \frac{\pi^2 \, \alphas \alphash}{3} \right)
   \sum_{i=1..8} \, \sum_{j=i..8} \, M^{q}_{i j} ,
\eea
with 
\bea
M^{q}_{1 1} = M^{q}_{1 2} = M^{q}_{1 3} =
M^{q}_{1 5} = M^{q}_{2 2} = M^{q}_{2 3} =
M^{q}_{2 5} = M^{q}_{2 6} = M^{q}_{2 8} &=& 0 ,
\eea
\bea
 M^{q}_{1 4} &=&
 \left( \frac{-4 \, C_F \, X_{t} \, u_7 \, \sd3}
    {t_2^2 \, (\sd3^2 + \msqu{t}^2 \Gamma^2) \, \ud7} \right)
 \left\{ 2 \, s_{32}^2 - s \, s_{32} - 2 \, s_4 \, s_{32} 
    + 2 \, t_2 \, s_{32} 
      - 2 \, s_{32} \, u' - u6 \, s_{32} + t' \, s_5 - t' \, m_1^2 
  \right. \nonumber \\ & & \left.
     - 3 \, t' \, m_2^2 + t' \, u_2 - s_4 \, t_2 - t_2 \, u' \right\}
  \nonumber \\
&+&
 \left( \frac{-4 \, C_F \, X_{t} \, u_7 \, \sd3} 
    {t_2 \, (\sd3^2 + \msqu{t}^2 \Gamma^2) \, \ud7^2} \right) 
 \left\{ 2 \, s_{32}^2 - s \, s_{32} - 2 \, s_4 \, s_{32}  
    + 2 \, t_2 \, s_{32}  
      - 2 \, s_{32} \, u' - u6 \, s_{32} + t' \, s_5 - t' \, m_1^2  
  \right. \nonumber \\ & & \left. 
     - 3 \, t' \, m_2^2 + t' \, u_2 - s_4 \, t_2 - t_2 \, u' \right\} ,
  \nonumber \\[0.4cm]
 M^{q}_{1 6} &=&
 \left( \frac{8 \, (C_F - N_C / 2) \, X_{t} \, \sd3}
        {s \, (\sd3^2 + \msqu{t}^2 \Gamma^2) \, \ud7 \, t_2} \right)
 \left\{ -s \, s_{32} \, s_4 + s_{32} \, u' \, u_6 + s \, s_{32} \, u_7 
         + s_{32} \, t' \, u_7
 \right. \nonumber \\ & & \left.
         + 2 \, s_{32} \, u' \, u_7 + t_2 \, u' \, u_7 
         - t' \, u_2 \, u_7  \right\} ,
 \nonumber \\[0.4cm]
 M^{q}_{1 7} &=&
 \left( \frac{4 \, N_C \, X_{t} \, \sd3}
        {(s+t_2) \, (\sd3^2 + \msqu{t}^2 \Gamma^2) \, \ud7 \, t_2} \right)
 \left\{ -s \, s_{32} \, s_4 + s_{32} \, u' \, u_6 - 2 \, s_4 \, s_{32} u_7 
        - t' \, u_7 \, m_2^2
 \right. \nonumber \\ & & \left.
         - m_1^2 \, t' \, u_7 - s_{32} \, t' \, u_7 + s_5 \, t' \, u_7
         - s_4 \, t_2 \, u_7 - s_{32} \, u_6 \, u_7  \right\}
 \nonumber \\
&+&
 \left( \frac{-4 \, N_C \, X_{t} \, \sd3} 
        {u_6 \, (\sd3^2 + \msqu{t}^2 \Gamma^2) \, \ud7 \, (s+t_2)} \right) 
 \left\{ -s \, s_{32} \, s_4 + s_{32} \, u' \, u_6 - 2 \, s_4 \, s_{32} u_7  
        - t' \, u_7 \, m_2^2 
 \right. \nonumber \\ & & \left. 
         - m_1^2 \, t' \, u_7 - s_{32} \, t' \, u_7 + s_5 \, t' \, u_7 
         - s_4 \, t_2 \, u_7 - s_{32} \, u_6 \, u_7  \right\} 
 \nonumber \\[0.4cm]
 M^{q}_{1 8} &=&
 \left( \frac{4 \, N_C \, X_{tu} \, m_1 \, m_2 }
        { u_6 \, (u - \msqu{u}^2) \, \ud7 \, (s+t_2)} \right)
 \left\{ -s \, s_4 - 2 s \, t' - 2 \, s_4 \, u' 
        - 2 \, t' \, u' - u' \, u_6
         + t' \, u_7 \right\} ,
 \nonumber \\[0.4cm]
 M^{q}_{2 4} &=&
 \left( \frac{-4 \, C_F \, X_{tu} \, m_1 \, m_2 \, \sd3}
        { \ud7 \, (\sd3^2 + \msqu{t}^2 \Gamma^2) 
          \, (u - \msqu{u}^2) \, t_2} \right)
 \left\{ -s \, s_{32} + s \, s_4 + 2 \, s \, u' - 2 \, s_{32} \, u' 
        + 2 \, s_4 \, u' - t_2 \, u'
 \right. \nonumber \\ & & \left.
        + 2 \, {u'}^2 + t' \, u_2 + u' \, u_6 - t' \, u_7 \right\} ,
 \nonumber \\[0.4cm]
 M^{q}_{2 7} &=&
 \left( \frac{-4 \, N_C \, X_{tu} \, m_1 \, m_2 \sd3}
        { u_6 \, (\sd3^2 + \msqu{t}^2 \Gamma^2) 
          \, (u - \msqu{u}^2) \, (s+t_2)} \right)
 \left\{ -s \, s_4 - 2 \, s_4 \, u' - u' \, u_6 + t' \, u_7 \right\} ,
 \nonumber \\[0.4cm]
 M^{q}_{3 3} &=&
 \left( \frac{4 \, C_F \, X_{u} \, s_4 \, u_2}
        { (u - \msqu{u}^2)^2 \, (\s4d^2 + \msqu{t}^2 \Gamma^2)} \right)
 \left\{ 2 \, m_2^2 + 2 \, m_1^2 - s_{32} + s_4 - s_5 - u' + u_2 - u_7 \right\}
,  \nonumber \\[0.4cm]
 M^{q}_{3 4} &=&
 \left( \frac{-8 \, (C_F - N_C / 2) \, X_{tu} \, \, \s4d \, 
              \sd3 \, m_1 \, m_2 \, u'}
        { (u - \msqu{u}^2)^2 \, (\s4d^2 + \msqu{t}^2 \Gamma^2) \, \ud7 
         \, (\sd3^2 + \msqu{t}^2 \Gamma^2) } \right)
 \left\{ s_5 - u' - 2 \, m_2^2 - 2 \, m_1^2 \right\} ,
 \nonumber
\eea
\bea
 M^{q}_{3 5} &=&
 \left( \frac{-4 \, C_F \, X_{u} \, s_4}
        { (u - \msqu{u}^2) \, (\s4d^2 + \msqu{t}^2 \Gamma^2) \, s } \right)
 \left\{ 3 \, m_2^2 \, s + m_1^2 \, s - s \, s_{32} - s \, s_5 + t_2 \, u' 
        + t' \, u_2
 \right. \nonumber \\ & & \left.
        - 2 \, t_2 \, u_2 + 2 \, u' \, u_2 - 2 \, u_2^2 + u_2 \, u_6 
        + t_2 \, u_7
        + 2 \, u_2 \, u_7 \right\} ,
 \nonumber \\[0.4cm]
 M^{q}_{3 6} &=&
 \left( \frac{4 \, C_F \, X_{tu} \, \s4d \, \sd3 \, m_1 \, m_2}
        { (u - \msqu{u}^2) \, (\s4d^2 + \msqu{t}^2 \Gamma^2) 
         \, (\sd3^2 + \msqu{t}^2 \Gamma^2) 
         \, s } \right)
 \left\{ s \, s_{32} - s \, s_4 + 2 \, t' \, u' - t_2 \, u' 
        + 2 \,{u'}^2 - t' \, u_2
 \right. \nonumber \\ & & \left.
        - 2 \, u' \, u_2 + u' \, u_6 + t' \, u_7 + 2 \, u' \, u_7 \right\} ,
 \nonumber \\[0.4cm]
 M^{q}_{3 7} &=&
 \left( \frac{2 \, N_C \, X_{tu} \, \s4d \, \sd3 \, m_1 \, m_2}
        { (u - \msqu{u}^2) \, (\s4d^2 + \msqu{t}^2 \Gamma^2)
         \, (\sd3^2 + \msqu{t}^2 \Gamma^2) \, u_6 } \right)
 \left\{ s \, s_{32} - s \, s_4 - 2 \, m_2^2 \, u' - 6 \, m_1^2 \, u' 
 \right. \nonumber \\ & & \left.
        + 2 \, s \, u' - 2 \, s_4 \, u'
        + 2 \, s_5 \, u' + t_2 \, u' 
        - t' \, u_2 - u' \, u_6 + t' \, u_7 + 2 \, u' \, u_7 \right\} ,
 \nonumber \\[0.4cm]
 M^{q}_{3 8} &=&
 \left( \frac{-2 \, N_C \, X_{u} \, \s4d \, u_2}
        { (u - \msqu{u}^2)^2 \, (\s4d^2 + \msqu{t}^2 \Gamma^2) \, u_6 } \right)
 \left\{ -2 \, m_2^2 \, s_4 - 6 \, m_1^2 \, s_4 + s \, s_4 
        - 2 \, s_4^2 + 2 \, s_4 \, s_5
 \right. \nonumber \\ & & \left.
        - m_2^2 \, t' - 3 \, m_1^2 \, t' - 2 \, s_4 \, t' + s_5 \, t' 
        + s_4 \, t_2
        - s_{32} \, u_6 - u' \, u_6 + 2 \, s_4 \, u_7 + t' \, u_7 \right\} ,
 \nonumber \\[0.4cm]
 M^{q}_{4 4} &=&
 \left( \frac{4 \, C_F \, X_{t} \, s_{32} \, u_7}
        { \ud7^2 \, (\sd3^2 + \msqu{t}^2 \Gamma^2) } \right)
 \left\{ 2 \, m_2^2 + 2 \, m_1^2 + s_{32} - s_4 - s_5 - u' - u_2 + u_7 \right\}
, \nonumber \\[0.4cm]
 M^{q}_{4 5} &=&
 \left( \frac{4 \, (C_F - N_C / 2) \, X_{tu} \, \s4d \, \sd3 \, m_1 \, m_2}
        { \ud7 \, (\sd3^2 + \msqu{t}^2 \Gamma^2) \, s 
         \, (\s4d^2 + \msqu{t}^2 \Gamma^2) } \right)
 \left\{ -s \, s_{32} + s \, s_4 + 2 \, t' \, u' + t_2 \, u' 
        + 2 \, {u'}^2 + t' \, u_2
 \right. \nonumber \\ & & \left.
        + 2 \, u' \, u_2 - u' \, u_6 - t' \, u_7 - 2 \, u' \, u_7 \right\} ,
 \nonumber \\[0.4cm]
 M^{q}_{4 6} &=&
 \left( \frac{-4 \, (C_F - N_C / 2) \, X_{t} \, s_{32} }
        { \ud7 \,  (\sd3^2 + \msqu{t}^2 \Gamma^2) \, s } \right)
 \left\{  s \, m_2^2 + 3 \, s \, m_1^2 - s \, s_4 
        - s \, s_5 + u' \, u_6 + u_2 \, u_6
 \right. \nonumber \\ & & \left.
        + t' \, u_7 + t_2 \, u_7 + 2 \, u' \, u_7 
        + 2 \, u_2 \, u_7 - 2 \, u_6 \, u_7 - 2 \, u_7^2 \right\} ,
 \nonumber \\[0.4cm]
 M^{q}_{4 7} &=&
 \left( \frac{2 \, N_C \, X_{t} \, s_{32} }
        { \ud7 \, (\sd3^2 + \msqu{t}^2 \Gamma^2) \, u_6 } \right)
 \left\{  s \, m_2^2 + 3 \, s \, m_1^2 - s \, s_4 - s \, s_5 
        + 2 \, u_7 \, m_2^2
        + u' \, u_6 + u_2 \, u_6 
 \right. \nonumber \\ & & \left.
        + 6 \, u_7 \, m_1^2 + 2 \, s \, u_7 - 2 \, s_4 \, u_7
        - 2 \, s_5 \, u_7 - t' \, u_7 - t_2 \, u_7 + 2 \, u_7^2 \right\} ,
 \nonumber \\[0.4cm]
 M^{q}_{4 8} &=&
 \left( \frac{-2 \, N_C \, X_{tu} \, \sd3 \, m_1 \, m_2 }
        { \ud7 \, (\sd3^2 + \msqu{t}^2 \Gamma^2) \, u_6 
        \, (u - \msqu{u}^2) } \right)
 \left\{  s \, s_{32} - s \, s_4 + 2 \, u' \, m_2^2 + 6 \, u' \, m_1^2 
        - 2 \, s_4 \, u'
 \right. \nonumber \\ & & \left. 
        - 2 \, s_5 \, u' - 2 \, t' \, u'
        - t_2 \, u' - t' \, u_2 
        + u' \, u_6 + t' \, u_7 + 2 \, u' \, u_7 \right\} ,
 \nonumber
\eea
\bea
 M^{q}_{5 5} &=&
 \left( \frac{-8 \, C_F \, X_{u} \, s_4 \, t_2 }
        { s \, (\s4d^2 + \msqu{t}^2 \Gamma^2) } \right) ,
 \nonumber \\[0.4cm]
 M^{q}_{5 6} &=&
 \left( \frac{16 \, C_F \, X_{tu} \, \s4d \, \sd3 \, m_1 \, m_2 \, t'}
        { s \, (\sd3^2 + \msqu{t}^2 \Gamma^2) \, 
         (\s4d^2 + \msqu{t}^2 \Gamma^2) } \right) ,
 \nonumber \\[0.4cm]
 M^{q}_{5 7} &=&
 \left( \frac{4 \, N_C \, X_{tu} \, \s4d \, \sd3 \, m_1 \, m_2 }
        { s \, (\sd3^2 + \msqu{t}^2 \Gamma^2) 
         \, u_6 \, (\s4d^2 + \msqu{t}^2 \Gamma^2) } \right)
 \left\{  -s \, s_4 + 2 \, s \, t' + 2 \, s \, u' + u' \, u_6
        + t' \, u_7 + 2 \, u' \, u_7 \right\} ,
 \nonumber \\[0.4cm]
 M^{q}_{5 8} &=&
 \left( \frac{4 \, N_C \, X_{u} \, \s4d }
        { s \, (u - \msqu{u}^2) \, u_6 
         \,  (\s4d^2 + \msqu{t}^2 \Gamma^2)} \right)
 \left\{ -s \, s_4 \, m_2^2 - s \, s_4 \, m_1^2 + s \, s_4 \, s_5 
        - s \, s_4 \, u_2
        - s_4 \, u_2 \, u_6 
 \right. \nonumber \\ & & \left.
        + u' \, u_2 \, u_6 - s_4 \, t_2 \, u_7
        - 2 \, s_4 \, u_2 \, u_7 - t' \, u_2 \, u_7 \right\} ,
 \nonumber \\[0.4cm]
 M^{q}_{6 6} &=&
 \left( \frac{-8 \, C_F \, X_{t} \, s_{32} \, s \, u_6 }
        { s^2 \, (\sd3^2 + \msqu{t}^2 \Gamma^2) } \right) ,
 \nonumber \\[0.4cm]
 M^{q}_{6 7} &=&
 \left( \frac{-2 \, N_C \, X_{t} }
        { s \, u_6 \, (\sd3^2 + \msqu{t}^2 \Gamma^2) } \right)
 \left\{ -4 \, s \, m_1^2 + 4 \, s \, u_6 + 4 \, s \, u_7
        + 4 \, u_6 \, u_7 + 4 \, u_7^2 \right\} ,
 \nonumber \\[0.4cm]
 M^{q}_{6 8} &=&
 \left( \frac{2 \, N_C \, X_{tu} \, \sd3 \, m_1 \, m_2}
        { s \, u_6 \, (\sd3^2 + \msqu{t}^2 \Gamma^2) 
         \, (u - \msqu{u}^2) } \right)
 \left\{ -s \, s_4 + u' \, u_6 + t' \, u_7 + 2 \, u' \, u_7 \right\} ,
 \nonumber \\[0.4cm]
 M^{q}_{7 7} &=&
 \left( \frac{-4 \, N_C \, X_{t} \, s_{32} }
        { u_6^2 \, (\sd3^2 + \msqu{t}^2 \Gamma^2) } \right)
 \left\{ -4 \, m_1^2 \, s + 2 \, s \, u_6 
        - 4 \, m_1^2 \, u_7 \right\} ,
 \nonumber \\[0.4cm]
 M^{q}_{7 8} &=&
 \left( \frac{-16 \, N_C \, X_{tu} \, \sd3 \, m_1 \, m_2 \, u' \,
        (2 \, m_1^2 + u_6) }
        { u_6^2 \, (\sd3^2 + \msqu{t}^2 \Gamma^2) \, (u - \msqu{u}^2) } \right)
, \nonumber \\[0.4cm]
 M^{q}_{8 8} &=&
 \left( \frac{4 \, N_C \, X_{u} \, u_2 }
        { u_6^2 \, (u - \msqu{u}^2)^2 } \right)
 \left\{ 4 \, m_1^2 \, s_4 + 4 \, m_1^2 \, t' 
        - 2 \, t' \, u_6 \right\} .
 \nonumber
\eea


\end{appendix}




\begin{table}
\label{sugratab}
\begin{tabular}{ccccc}
$m_{1/2}$~(GeV) & $m_{\tilde{g}}$~(GeV) & $m_{\tilde{\chi}}$~(GeV) &
Tevatron (pb) & LHC (pb) \\ \hline \hline
  100  & 284.9  & 31.1   & 0.165     & 2.82  \\
  100  & 284.9  & 62.6   & 0.233     & 4.50  \\
  100  & 284.9  & -211.3 & 0.00588   & 0.176 \\
  100  & 284.9  & 240.9  & 0.00424   & 0.364 \\
  100  & 284.9  & 56.8   & 0.329     & 9.61  \\
  100  & 284.9  & 240.2  & 0.0100    & 0.928 \\
  200  & 533.7  & 77.6   & 0.00283   & 0.212   \\
  200  & 533.7  & 146.6  & 0.00231   & 0.354   \\
  200  & 533.7  & -358.2 & 1.22$\cdot10^{-5}$  & 0.00452 \\
  200  & 533.7  & 380.6  & 9.73$\cdot10^{-6}$  & 0.0151  \\
  200  & 533.7  & 145.2  & 0.00238   & 0.720   \\
  200  & 533.7  & 379.5  & 1.680$\cdot10^{-5}$ & 0.0379  \\
  300  & 775.5  & 120.3  & 5.51$\cdot10^{-5}$  & 0.0400   \\
  300  & 775.5  & 232.3  & 2.07$\cdot10^{-5}$  & 0.0605   \\
  300  & 775.5  & -510.9 & 1.46$\cdot10^{-8}$  & 0.000385 \\
  300  & 775.5  & 528.2  & 7.71$\cdot10^{-9}$  & 0.00144  \\
  300  & 775.5  & 231.7  & 1.58$\cdot10^{-5}$  & 0.128    \\
  300  & 775.5  & 527.2  & 1.06$\cdot10^{-8}$  & 0.00368  \\
  400  & 1012.4 & 162.2  & 6.58$\cdot10^{-7}$  & 0.0110   \\
  400  & 1012.4 & 317.1  & 6.85$\cdot10^{-8}$  & 0.0151   \\
  400  & 1012.4 &-665.3  & 5.23$\cdot10^{-13}$  & 5.53$\cdot10^{-5}$ \\
  400  & 1012.4 & 679.2  & 9.85$\cdot10^{-14}$  & 0.000207 \\
  400  & 1012.4 & 316.77 & 3.52$\cdot10^{-8}$  & 0.0327   \\
  400  & 1012.4 & 678.4  & 1.16$\cdot10^{-13}$  & 0.000550 \\
\end{tabular}
\caption{Total cross sections in pb at the Tevatron and at the LHC 
 for the parameters of the SUGRA model.  Shown in column 1 are four 
 chosen values of the common fermion mass.  In column 2 are the 
 derived values of the gluino masses, and in column 3 are the masses 
 of the four neutralinos and the two charginos. See also 
 Figs.~\ref{xsectev} and~\ref{xseclhc}.}    
\end{table}



\begin{references}

\bibitem{susy}
J. Wess and B. Zumino, Nucl. Phys. {\bf B 70}, 39 (1974).

\bibitem{hierarchy}
E. Witten, Nucl. Phys. {\bf B 188}, 513 (1981);
J. Polchinski and L. Susskind, Phys. Rev. {\bf D 26}, 3661 (1982);
R.K. Kaul and P. Majumdar, Nucl. Phys. {\bf B 199}, 36 (1982).

\bibitem{radbreak}
L.E. Iba\~nez and G.G. Ross, Phys. Lett. {\bf B 110}, 215 (1982);
J. Ellis, D.V. Nanopoulos, and K. Tamvakis, Phys. Lett. {\bf B 121}, 123
 (1983).

\bibitem{unif}
L.E. Iba\~nez and G.G. Ross, Phys. Lett. {\bf B 105}, 439 (1981);
S. Dimopoulos, S. Raby, and F. Wilczek, Phys. Rev. {\bf D 24}, 1681 (1981);
P. Langacker and M. Luo, Phys. Rev. {\bf D 44}, 817 (1991);
U. Amaldi, W. de Boer, and H. F\"urstenau, Phys. Lett. {\bf B 260}, 447 (1991).

\bibitem{searches}
Recent reviews include M. Carena, R.L. Culbertson, S. Eno, H.J. Frisch, 
and S. Mrenna, Rev.~Mod.~Phys. {\bf 71}, 937 (1999) hep-ex/9712022 and 
hep-ex/9802006.

\bibitem{mssm}
H.P. Nilles, Phys. Rep. {\bf 110}, 1 (1984);
H. Haber and G.L. Kane, Phys. Rep. {\bf 117}, 75 (1985). 
For a recent review, consult S. Dawson, TASI-97 lectures, hep-ph/9712464.

\bibitem{squarkgluino}
W. Beenakker, R. H\"opker, M. Spira, and P.M. Zerwas, Nucl. Phys. {\bf B 492},
 51 (1997).

\bibitem{stop}
W. Beenakker, M. Kr\"amer, T. Plehn, M. Spira, and P.M. Zerwas, Nucl. Phys.
 {\bf B 515}, 3 (1998).

\bibitem{slepton}
H. Baer, B.W. Harris, and M. Hall Reno, Phys. Rev. {\bf D 57}, 5871 (1998).

\bibitem{gaugino}
W. Beenakker, M. Klasen, M. Kr\"amer, T. Plehn, M. Spira, and P.M. Zerwas, 
 Phys. Rev. Lett. {\bf 83}, 3780 (1999).  

\bibitem{sletter}
E.L. Berger, M. Klasen, and T. Tait, Phys. Lett. B {\bf 459}, 165 (1999).   

\bibitem{sugra}
R. Arnowitt and P. Nath, to appear in ``Perspectives on Supergravity", 
World Scientific, Editor G. Kane, hep-ph/9708254, and references therein; 
M. Drees and S.~P. Martin, Report of the DPF Working Group 
on Electroweak Symmetry Breaking and Beyond the Standard Model,  
hep-ph/9504324.  

\bibitem{susybreak}
C. Kolda, Nucl.~Phys.~Proc.~Suppl. {\bf 62}, 266 (1998) hep-ph/9707450, and 
original references therein.  

\bibitem{lo2}
H. Baer, D.D. Karatas, and X. Tata, Phys. Rev. {\bf D 42}, 2259 (1990).

\bibitem{lo1}
S. Dawson, E. Eichten, and C. Quigg, Phys. Rev. {\bf D 31}, 1581 (1985).

\bibitem{gaugino1}
J. Ellis, C. Kounnas, and D.~V. Nanopoulos, Nucl. Phys. {\bf B 247}, 373 (1984);
A.B. Lahanas and D.~V. Nanopoulos, Phys. Rep. {\bf 145}, 1 (1987).

\bibitem{gaugino2}
D.~E. Kaplan, G.~D. Kribs, and M. Schmaltz, hep-ph/9911293;
Z. Chacko, M.~A. Luty, A.~E. Nelson, and E. Ponton, JHEP {\bf 0001}, 003 (2000);
M. Schmaltz and W. Skiba, hep-ph/0001172;
M. Schmaltz and W. Skiba, hep-ph/0004210.

\bibitem{gaugino3}
D.~E.~Kaplan and T.~M.~P.~Tait, hep-ph/0004200.

\bibitem{amsb}
L.~Randall and R.~Sundrum, Nucl.~Phys. {\bf B 557}, 79 (1999) hep-ph/9810155; 
G.~F.~Giudice, M.~A.~Luty, H.~Murayama, and R.~Rattazzi, JHEP {\bf 9812}, 027 
(1998); T.~Gherghetta, G.~F.~Giudice, and J.~D.~Wells, 
Nucl.~Phys. {\bf B 559}, 27 (1999) hep-ph/9904378.  

\bibitem{Abel:2000}
 S.~Abel {\it et al.}  [SUGRA Working Group Collaboration],
Report of the SUGRA Working group, 
Fermilab workshop in preparation for Run II at the Fermilab Tevatron, 
November, 1998, hep-ph/0003154.

\bibitem{lepdata}
ALEPH Collaboration, R. Barate {\it et al}, Eur. Phys. J. {\bf C 11}, 193 (1999) 
and G.~Ganis {\it et al}, ALEPH 2000--012 (CONF 2000--009); 
DELPHI Collaboration, P.~Abreu {\it et al}, Phys. Lett. {\bf B466}, 61 (1999); 
L3 Collaboration, M. Acciarri {\it et al}, Phys. Lett. {\bf B459}, 354 (1999);
OPAL Collaboration, G. Abbiendi {\it et al}, CERN report CERN-EP-99-123 (1999) 
hep-ex/9909051 and references therein.  
 
\bibitem{raby}
S.~Raby, Phys. Rev. {\bf D56}, 2852 (1997); Phys.~Lett.~B {\bf 422}, 158 (1998).

\bibitem{raby2} A. Mafi and S. Raby, hep-ph/9912436.

\bibitem{foot1}
Higgsino couplings to down-type quarks may be enhanced for
large $\tan \beta$ and can be appreciable for bottom ($b$) quarks.
Thus, associated production modes involving $b$ quarks and Higgsinos,
ignored in our calculations, could play an important role in
searches for supersymmetry at hadron colliders.

\bibitem{denner}
A. Denner, H. Eck, O. Hahn, and J. K\"ublbeck, Nucl. Phys. {\bf B 387}, 467
 (1992).

\bibitem{chicoupling}
J.~F. Gunion and H. Haber, Nucl. Phys. {\bf B 272}, 1 (1986).

\bibitem{isacode} H.~Baer, F.~E.~Paige, S.~D.~Protopopescu, and X.~Tata, 
Brookhaven report BNL-HET-99-43, hep-ph/0001086.  

\bibitem{lowscaleGMSB} Report of the Working group on Gauge Mediation / 
Low-Scale SUSY Breaking, 
Fermilab workshop in preparation for Run II at the Fermilab Tevatron, 
November, 1998.  

\bibitem{form}
J.~A.~M. Vermaseren, KEK-TH-326 (1992).

\bibitem{naive}
M. Chanowitz, M. Furman, I. Hinchliffe, Nucl. Phys. {\bf B 159}, 225 (1979).

\bibitem{pasvelt}
G. Passarino and M. Veltman, Nucl. Phys. {\bf B 160}, 151 (1979).

\bibitem{msbar}
W.~A. Bardeen, A.~J. Buras, D.~W. Duke, T. Muta, Phys. Rev. {\bf D 18}, 3998
 (1978).


\bibitem{mismatch}
S.~P. Martin and M.~T. Vaughn, Phys. Lett. {\bf B 318}, 331 (1993).

\bibitem{hquark} 
W. Beenakker, H. Kuijf, and W.~L. van Neerven, Phys. Rev. {\bf D 40}, 54 (1989).

\bibitem{facto}
G. Altarelli, R.~K. Ellis, and G. Martinelli, Nucl. Phys. {\bf B 157}, 461
 (1979).

\bibitem{ap}
G.~Altarelli and G.~Parisi, Nucl. Phys. {\bf B 126}, 298 (1977).   

\bibitem{nason}
P. Nason, S. Dawson, and R.~K. Ellis, Nucl. Phys. {\bf B 303}, 607 (1988).

\bibitem{jsmith} 
W. Beenakker, W.~L. van Neerven, R. Meng, G.~A. Schuler, and J. Smith,  
Nucl. Phys. {\bf B 351}, 507 (1991), and references therein.  

\bibitem{cteq5} H.~L.~Lai {\it et al}~[CTEQ Collaboration], Eur. Phys. J. 
{\bf C12}, 375 (2000).

\bibitem{cteq4}
H. L. Lai {\it et al}~[CTEQ Collaboration], Phys. 
Rev. {\bf D 55}, 1280 (1997).

\bibitem{smithwillen} M.~C. Smith and S. Willenbrock,  
Phys.Rev. {\bf D54}, 6696 (1996).   

\bibitem{hopker}
R. H\"opker, Ph.D. Thesis, University of Hamburg, DESY-T-96-02 (1996).

\bibitem{scharf}
A. Denner, U. Nierste, and R. Scharf, Nucl. Phys. {\bf B 367}, 637 (1991).

\end{references}
\end{document}